\documentclass[12pt]{article}
\pdfoutput=1
\usepackage[]{hyperref}
\usepackage{xcolor}
\usepackage[]{putex}
\usepackage[utf8]{inputenc}

\usepackage{amsmath,amsthm,amssymb}
\usepackage{tikz}
\usepackage{tikz-cd}
\usetikzlibrary{decorations.markings}
\usetikzlibrary{calc}
\usepackage{subcaption}
\usepackage{enumerate}
\usepackage{enumitem}
\usepackage{mathtools} 

\DeclarePairedDelimiter{\floor}{\lfloor}{\rfloor}

\numberwithin{equation}{section}

\makeatletter
\DeclareRobustCommand*{\bfseries}{%
  \not@math@alphabet\bfseries\mathbf
  \fontseries\bfdefault\selectfont
  \boldmath
}
\makeatother
\hypersetup{
  colorlinks=true,
  linkcolor=blue,
  citecolor=green!60!black,
  urlcolor=cyan!80!black,
  linktoc=all
  }

\input{glyphtounicode}
\def\musepic#1{\vcenter{\hbox{\usebox{#1}}}}
\newmuskip\pFqmuskip

\newcommand*\pFq[6][8]{%
  \begingroup 
  \pFqmuskip=#1mu\relax
  \mathchardef\normalcomma=\mathcode`,
  \mathcode`\,=\string"8000
  \begingroup\lccode`\~=`\,
  \lowercase{\endgroup\let~}\pFqcomma
  {}_{#2}F_{#3}{\left[\genfrac..{0pt}{}{#4}{#5};#6\right]}%
  \endgroup
}
\newcommand{\pFqcomma}{{\normalcomma}\mskip\pFqmuskip}

\newsavebox{\figFiveCombChannel}
\savebox{\figFiveCombChannel}{%
\begin{tikzpicture}[scale=1]
    \coordinate (x1) at (-5/2,0);
    \coordinate (y1) at (-2,0);
    \coordinate (x2) at (-2,1);
    \coordinate (y2) at (-1,0);
    \coordinate (x3) at (-1,1);
    \coordinate (y3) at (0,0);
    \coordinate (x5) at (0,1);
    \coordinate (x6) at (1/2,0);
	\draw[thick] (x1)--(y1)--(x2);
	\draw[thick] (y1)--(y2)--(x3);
	\draw[thick] (y2)--(y3);
	\draw[thick] (y3)--(x5);
	\draw[thick] (y3)--(x6);
	\draw (x1) node[anchor=east] {${\cal O}_1$};
	\draw (x2) node[anchor=south] {${\cal O}_2$};
	\draw (x3) node[anchor=south] {${\cal O}_3$};
	\draw (x5) node[anchor=south] {${\cal O}_4$};
	\draw (x6) node[anchor=west] {${\cal O}_5$};
	\draw ($(y1)!0.5!(y2)$) node[anchor=north] {${\cal O}_{{\delta_1}}$};
	\draw ($(y2)!0.5!(y3)$) node[anchor=north] {${\cal O}_{{\delta_2}}$};
\end{tikzpicture}}
\newsavebox{\figcalWSample}
\savebox{\figcalWSample}{%
\begin{tikzpicture}[scale=1.2]
    \draw[thick, color=black!30] (0,0) circle[radius=2];
    \coordinate (x1) at (200:2);
    \coordinate (x2) at (160:2);
    \coordinate (x5) at (20:2);
    \coordinate (x6) at (-20:2);
    \coordinate (x3) at (90:2);
    \coordinate (x4) at (-90:2);
    \coordinate (y) at (180:1.3);
    \coordinate (yp) at (0:1.3);
	\draw[thick, color=red, dashed] (x2) to [out =  -15, in = 15, looseness = 1.5] (x1);
	\draw[thick, color=red, dashed] (x5) to [out =  195, in = 165, looseness = 1.5] (x6);
	\draw[blue, thick] (x1)--(y)--(x2);
	\draw[thick, blue] (y)--(x3)--(yp)--(x4)--(y);
	\draw[blue, thick] (x5)--(yp)--(x6);
	\draw[thick, dotted] (y)--(yp);
	\draw[thick, black] (x3)--(x4);
	\draw (x1) node[anchor=north east] {${\cal O}_1$};
	\draw (x2) node[anchor=south east] {${\cal O}_2$};
	\draw (x3) node[anchor=south] {${x}_3$};
	\draw (x4) node[anchor=north] {${x}_4$};
	\draw (x5) node[anchor=south west] {${\cal O}_5$};
	\draw (x6) node[anchor=north west] {${\cal O}_6$};
	\draw[color=blue]  ($(y)!0.55!(x3)$) node[anchor=south east,  align=center] {$\Delta_{L_1}$};
	\draw[color=blue]  ($(yp)!0.55!(x3)$) node[anchor=south west, align=center] {$\Delta_{R_1}$};
	\draw[color=black]  ($(y)!0.35!(yp)$) node[anchor=north, text width=3.5cm, align=center] {$\Delta_{C}$};
	\draw[color=blue]  ($(y)!0.55!(x4)$) node[anchor=north east, align=center] {$\Delta_{L_2}$};
	\draw[color=blue]  ($(yp)!0.55!(x4)$) node[anchor=north west, align=center] {$\Delta_{R_2}$};
	\draw[color=black]  ($(x3)!0.35!(x4)$) node[anchor=west] {$\Delta_D$};
\end{tikzpicture}}
\newsavebox{\figcalWFive}
\savebox{\figcalWFive}{%
\begin{tikzpicture}[scale=1.1]
    \draw[thick, color=black!30] (0,0) circle[radius=2];
    \coordinate (x1) at (200:2);
    \coordinate (x2) at (160:2);
    \coordinate (x4) at (20:2);
    \coordinate (x5) at (-20:2);
    \coordinate (x3) at (90:2);
    \coordinate (y) at (180:1.3);
    \coordinate (yp) at (0:1.3);
	\draw[thick, color=red, dashed] (x2) to [out =  -15, in = 15, looseness = 1.5] (x1);
	\draw[thick, color=red, dashed] (x4) to [out =  195, in = 165, looseness = 1.5] (x5);
	\draw[blue, thick] (x1)--(y)--(x2);
	\draw[blue, thick] (y)--(x3)--(yp);
	\draw[blue, thick] (x4)--(yp)--(x5);
	\draw[thick, dotted] (y)--(yp);
	\draw (x1) node[anchor=north east] {${\cal O}_1$};
	\draw (x2) node[anchor=south east] {${\cal O}_2$};
	\draw (x3) node[anchor=south] {${x}_3$};
	\draw (x4) node[anchor=south west] {${\cal O}_4$};
	\draw (x5) node[anchor=north west] {${\cal O}_5$};
	\draw[color=blue]  ($(y)!0.6!(x3)$) node[anchor=south east,  align=center] {$\Delta_{3\delta_1,\delta_2} +k_{1,2}$};
	\draw[color=blue]  ($(yp)!0.6!(x3)$) node[anchor=south west,  align=center] {$\Delta_{3\delta_2,\delta_1}+k_{2,1}$};
	\draw[color=black]  ($(y)!0.5!(yp)$) node[anchor=north, text width=3cm, align=center] {$\Delta_{\delta_1 \delta_2,3}+k_{12,}$};
\end{tikzpicture}}
\newsavebox{\figcalWFiveVariant}
\savebox{\figcalWFiveVariant}{%
\begin{tikzpicture}[scale=1.2]
    \draw[thick, color=black!30] (0,0) circle[radius=2];
    \coordinate (x1) at (200:2);
    \coordinate (x2) at (160:2);
    \coordinate (x4) at (20:2);
    \coordinate (x5) at (-20:2);
    \coordinate (x3) at (-90:2);
    \coordinate (y) at (180:1.3);
    \coordinate (yp) at (0:1.3);
	\draw[thick, color=red, dashed] (x2) to [out =  -15, in = 15, looseness = 1.5] (x1);
	\draw[thick, color=red, dashed] (x4) to [out =  195, in = 165, looseness = 1.5] (x5);
	\draw[thick] (x1)--(y)--(x2);
	\draw[blue, thick] (y)--(x3)--(yp);
	\draw[thick] (x4)--(yp)--(x5);
	\draw[thick, dotted] (y)--(yp);
	\draw (x1) node[anchor=north east] {${\cal O}_{\delta_1}$};
	\draw (x2) node[anchor=south east] {${\cal O}_3$};
	\draw (x3) node[anchor=north] {${x}_4$};
	\draw (x4) node[anchor=south west] {${\cal O}_5$};
	\draw (x5) node[anchor=north west] {${\cal O}_6$};
	\draw[color=blue]  ($(y)!0.55!(x3)$) node[anchor=north east,  align=center] {$\Delta_{4\delta_2,\delta_3} +k_{2,3}$};
	\draw[color=blue]  ($(yp)!0.55!(x3)$) node[anchor=north west,  align=center] {$\Delta_{4\delta_3,\delta_2}+k_{3,2}$};
	\draw[color=black]  ($(y)!0.5!(yp)$) node[anchor=north, text width=3cm, align=center] {$\Delta_{\delta_2 \delta_3,4}+k_{23,}$};
\end{tikzpicture}}
\newsavebox{\figFourPropBefore}
\savebox{\figFourPropBefore}{%
\begin{tikzpicture}[scale=1.2]
    \draw[thick, color=black!30] (0,0) circle[radius=2];
    \coordinate (x1) at (155:2);
    \coordinate (x2) at (115:2);
    \coordinate (x3) at (-90:2);
    \coordinate (y) at (135:1.3);
    \coordinate (yp) at (0:1.3);
	\draw[thick, color=red, dashed] (x2) to [out =  -60, in = -30, looseness = 1.5] (x1);
	\draw[thick] (x1)--(y)--(x2);
	\draw[thick] (y)--(x3);
	\draw[thick, dotted] (y)--(yp);
	\draw (x1) node[anchor= east] {${\cal O}_{A}(x_1)$};
	\draw (x2) node[anchor=south ] {${\cal O}_B(x_2)$};
	\draw (x3) node[anchor=north] {${\cal O}_D({x}_4)$};
	\draw[color=black]  ($(y)!0.5!(yp)$) node[anchor=north, text width=3cm, align=center] {$\Delta_{C}$};
\end{tikzpicture}}
\newsavebox{\figFourPropAfter}
\savebox{\figFourPropAfter}{%
\begin{tikzpicture}[scale=1.2]
    \draw[thick, color=black!30] (0,0) circle[radius=2];
    \coordinate (x1) at (180:2);
    \coordinate (x3) at (90:2);
    \coordinate (x4) at (-90:2);
    \coordinate (y) at (180:2);
    \coordinate (yp) at (0:1.3);
	\draw[thick] (x4)--(y)--(x3);
	\draw[thick] (x3)--(yp);
	\draw[thick] (y)--(yp);
	\draw[thick, black!40!green] (x3)--(x4);
	\draw (x1) node[anchor=east] {$x_1$};
	\draw (x3) node[anchor=south] {${x}_2$};
	\draw (x4) node[anchor=north] {${x}_4$};
	\draw  ($(y)!0.5!(x3)$) node[anchor=south east,  align=center] {$\Delta_{AB,CD}$};
	\draw  ($(yp)!0.5!(x3)$) node[anchor=south west, align=center] {$\Delta_{BCD,A}+j$};
	\draw  ($(y)!0.5!(yp)$) node[anchor=north,  align=center] {$\Delta_{AC,BD}-j$};
	\draw  ($(y)!0.5!(x4)$) node[anchor=north east,  align=center] {$\Delta_{D}+j$};
	\draw[color=black!40!green]  ($(x3)!0.35!(x4)$) node[anchor=west] {$-j$};
\end{tikzpicture}}
\newsavebox{\figcalWSix}
\savebox{\figcalWSix}{%
\begin{tikzpicture}[scale=1.5]
    \draw[thick, color=black!30] (0,0) circle[radius=2];
    \coordinate (x1) at (200:2);
    \coordinate (x2) at (160:2);
    \coordinate (x5) at (20:2);
    \coordinate (x6) at (-20:2);
    \coordinate (x3) at (90:2);
    \coordinate (x4) at (-90:2);
    \coordinate (y) at (180:1.3);
    \coordinate (yp) at (0:1.3);
	\draw[thick, color=red, dashed] (x2) to [out =  -15, in = 15, looseness = 1.5] (x1);
	\draw[thick, color=red, dashed] (x5) to [out =  195, in = 165, looseness = 1.5] (x6);
	\draw[blue, thick] (x1)--(y)--(x2);
	\draw[thick, blue] (y)--(x3)--(yp)--(x4)--(y);
	\draw[blue, thick] (x5)--(yp)--(x6);
	\draw[thick, dotted] (y)--(yp);
	\draw[thick, black] (x3)--(x4);
	\draw (x1) node[anchor=north east] {${\cal O}_1$};
	\draw (x2) node[anchor=south east] {${\cal O}_2$};
	\draw (x3) node[anchor=south] {${x}_3$};
	\draw (x4) node[anchor=north] {${x}_4$};
	\draw (x5) node[anchor=south west] {${\cal O}_5$};
	\draw (x6) node[anchor=north west] {${\cal O}_6$};
	\draw[color=blue]  ($(y)!0.55!(x3)$) node[anchor=south east,  align=center] {$\Delta_{3\delta_1,\delta_2} +k_{1,2}$};
	\draw[color=blue]  ($(yp)!0.55!(x3)$) node[anchor=south west, align=center] {$\Delta_{3\delta_2,\delta_1} +k_{2,1}+j$};
	\draw[color=black]  ($(y)!0.48!(yp)$) node[anchor=north, text width=3.5cm, align=center] {$\Delta_{\delta_1 \delta_3,34}+k_{13,}-j$};
	\draw[color=blue]  ($(y)!0.55!(x4)$) node[anchor=north east, align=center] {$\Delta_{4\delta_2,\delta_3} +k_{2,3}+j$};
	\draw[color=blue]  ($(yp)!0.55!(x4)$) node[anchor=north west, align=center] {$\Delta_{4\delta_3,\delta_2} +k_{3,2}$};
	\draw[color=black]  ($(x3)!0.35!(x4)$) node[anchor=west] {$-j$};
\end{tikzpicture}}
\newsavebox{\figSixExchThree}
\savebox{\figSixExchThree}{%
\begin{tikzpicture}[scale=.65]
    \tikzstyle{vint}=[draw,scale=0.55,color=teal,fill=green,circle]
    \draw[thick, color=black!30] (0,0) circle[radius=2];
    \coordinate (x1) at (200:2);
    \coordinate (x2) at (160:2);
    \coordinate (x4) at (75:2);
    \coordinate (x5) at (20:2);
    \coordinate (x3) at (105:2);
    \coordinate (x6) at (-20:2);
    \coordinate (y) at (180:1.5);
    \coordinate (yp) at (0:1.5);
    \coordinate (z) at (180:0.5);
    \coordinate (zp) at (0:0.5);
	\draw[thick] (x1)--(y)--(x2);
	\draw[thick] (x5)--(yp)--(x6);
	\draw[thick] (zp)--(x4);
	\draw[thick] (z)--(x3);
	\draw[thick] (y)--(z)--(zp)--(yp);
	\draw (x1) node[anchor=north east] {\footnotesize${\cal O}_1$};
	\draw (x2) node[anchor=south east] {\footnotesize${\cal O}_2$};
	\draw (x3) node[anchor=south] {\footnotesize${\cal O}_3$};
	\draw (x4) node[anchor= south] {\footnotesize${\cal O}_4$};
	\draw (x5) node[anchor= south west] {\footnotesize${\cal O}_5$};
	\draw (x6) node[anchor=north west] {\footnotesize${\cal O}_6$};
 	\draw  ($(y)!0.5!(z)$) node[anchor=north] {\footnotesize$\Delta_{\delta_1}$};
	\draw  ($(z)!0.5!(zp)$) node[anchor=north] {\footnotesize$\Delta_{\delta_2}$};
	\draw  ($(zp)!0.5!(yp)$) node[anchor=north] {\footnotesize$\Delta_{\delta_3}$};
	\draw (y) node[vint] {};
	\draw (yp) node[vint] {};
	\draw (z) node[vint] {};
	\draw (zp) node[vint] {};
\end{tikzpicture}}
\newsavebox{\figSixCombChannel}
\savebox{\figSixCombChannel}{%
\begin{tikzpicture}[scale=.9]
    \coordinate (x1) at (-5/2,0);
    \coordinate (y1) at (-2,0);
    \coordinate (x2) at (-2,1);
    \coordinate (y2) at (-1,0);
    \coordinate (x3) at (-1,1);
    \coordinate (y3) at (0,0);
    \coordinate (x4) at (0,1);
    \coordinate (y4) at (1,0);
    \coordinate (x5) at (1,1);
    \coordinate (x6) at (3/2,0);
	\draw[thick] (x1)--(y1)--(x2);
	\draw[thick] (y1)--(y2)--(x3);
	\draw[thick] (y2)--(y3)--(x4);
	\draw[thick] (y3)--(y4)--(x5);
	\draw[thick] (y4)--(x6);
	\draw (x1) node[anchor=east] {${\cal O}_1$};
	\draw (x2) node[anchor=south] {${\cal O}_2$};
	\draw (x3) node[anchor=south] {${\cal O}_3$};
	\draw (x4) node[anchor=south] {${\cal O}_4$};
	\draw (x5) node[anchor=south] {${\cal O}_5$};
	\draw (x6) node[anchor=west] {${\cal O}_6$};
	\draw ($(y1)!0.5!(y2)$) node[anchor=north] {${\cal O}_{{\delta_1}}$};
	\draw ($(y2)!0.5!(y3)$) node[anchor=north] {${\cal O}_{{\delta_2}}$};
	\draw ($(y3)!0.5!(y4)$) node[anchor=north] {${\cal O}_{{\delta_3}}$};
\end{tikzpicture}}
\newsavebox{\figcalWFivePreAlt}
\savebox{\figcalWFivePreAlt}{%
\begin{tikzpicture}[scale=1.2]
    \draw[thick, color=black!30] (0,0) circle[radius=2];
    \coordinate (x1) at (180:2);
    \coordinate (x5) at (20:2);
    \coordinate (x6) at (-20:2);
    \coordinate (x3) at (90:2);
    \coordinate (x4) at (-90:2);
    \coordinate (y) at (180:2);
    \coordinate (yp) at (0:1.3);
	\draw[thick, color=red, dashed] (x5) to [out =  195, in = 165, looseness = 1.5] (x6);
	\draw[thick, black] (x4)--(y)--(x3);
	\draw[thick, blue] (x3)--(yp)--(x4);
	\draw[blue, thick] (x5)--(yp)--(x6);
	\draw[blue, thick] (y)--(yp);
	\draw[thick, black] (x3)--(x4);
	\draw (x1) node[anchor=east] {$x_1$};
	\draw (x3) node[anchor=south] {${x}_3$};
	\draw (x4) node[anchor=north] {${x}_4$};
	\draw (x5) node[anchor=south west] {${\cal O}_5$};
	\draw (x6) node[anchor=north west] {${\cal O}_6$};
	\draw[color=black]  ($(y)!0.55!(x3)$) node[anchor=south east,  align=center] {$\Delta_{3\delta_1,\delta_2} +k_{1,2}$};
	\draw[color=blue]  ($(yp)!0.55!(x3)$) node[anchor=south west,  align=center] {$\Delta_{3\delta_2,\delta_1} +k_{2,1}+j$};
	\draw[color=blue]  ($(y)!0.5!(yp)$) node[anchor=north, align=center] {$\Delta_{\delta_1 \delta_3,34}+k_{13,}-j$};
	\draw[color=black]  ($(y)!0.55!(x4)$) node[anchor=north east, align=center] {$\Delta_{4\delta_2,\delta_3} +k_{2,3}+j$};
	\draw[color=blue]  ($(yp)!0.55!(x4)$) node[anchor=north west,  align=center] {$\Delta_{4\delta_3,\delta_2} +k_{3,2}$};
	\draw[color=black]  ($(x3)!0.35!(x4)$) node[anchor=west] {$-j$};
\end{tikzpicture}}
\newsavebox{\figcalWFiveAlt}
\savebox{\figcalWFiveAlt}{%
\begin{tikzpicture}[scale=1.2]
    \draw[thick, color=black!30] (0,0) circle[radius=2];
    \coordinate (x1) at (180:2);
    \coordinate (x5) at (20:2);
    \coordinate (x6) at (-20:2);
    \coordinate (x3) at (90:2);
    \coordinate (x4) at (-90:2);
    \coordinate (y) at (180:2);
    \coordinate (yp) at (0:1.3);
	\draw[thick, color=red, dashed] (x5) to [out =  195, in = 165, looseness = 1.5] (x6);
	\draw[thick, black] (x4)--(y)--(x3);
	\draw[thick, blue] (x3)--(yp)--(x4);
	\draw[blue, thick] (x5)--(yp)--(x6);
	\draw[blue, thick] (y)--(yp);
	\draw[thick, black] (x3)--(x4);
	\draw (x1) node[anchor=east] {$x_1$};
	\draw (x3) node[anchor=south] {${x}_3$};
	\draw (x4) node[anchor=north] {${x}_4$};
	\draw (x5) node[anchor=south west] {${\cal O}_5$};
	\draw (x6) node[anchor=north west] {${\cal O}_6$};
	\draw[color=black]  ($(y)!0.55!(x3)$) node[anchor=south east,  align=center] {$\Delta_{3\delta_1,\delta_2}-k_2$};
	\draw[color=blue]  ($(yp)!0.55!(x3)$) node[anchor=south west, align=center] {$\Delta_{3\delta_2,\delta_1}+k_2+j$};
	\draw[color=blue]  ($(y)!0.51!(yp)$) node[anchor=north,  align=center] {$\Delta_{\delta_1 \delta_3,34}+k_3-j$};
	\draw[color=black]  ($(y)!0.55!(x4)$) node[anchor=north east,  align=center] {$\Delta_{4\delta_2,\delta_3} +k_{2,3}+j$};
	\draw[color=blue]  ($(yp)!0.55!(x4)$) node[anchor=north west,  align=center] {$\Delta_{4\delta_3,\delta_2}+k_{3,2}$};
	\draw[color=black]  ($(x3)!0.35!(x4)$) node[anchor=west] {$-j$};
\end{tikzpicture}}
\newsavebox{\figVlimOne}
\savebox{\figVlimOne}{%
\begin{tikzpicture}[scale=1.2]
    \draw[thick, color=black!30] (0,0) circle[radius=2];
    \coordinate (x1) at (150:2);
    \coordinate (x5) at (20:2);
    \coordinate (x6) at (-20:2);
    \coordinate (x4) at (210:2);
    \coordinate (y) at (150:2);
    \coordinate (yp) at (0:1.3);
	\draw[thick, color=red, dashed] (x5) to [out =  195, in = 165, looseness = 1.5] (x6);
	\draw[thick, black] (x4)--(y)--(x1);
	\draw[thick, blue] (x1)--(yp)--(x4);
	\draw[blue, thick] (x5)--(yp)--(x6);
	\draw[thick, black] (x1)--(x4);
	\draw (x1) node[anchor=east] {$x_1$};
	\draw (x4) node[anchor=north east] {${x}_4$};
	\draw (x5) node[anchor=south west] {${\cal O}_5$};
	\draw (x6) node[anchor=north west] {${\cal O}_6$};
	\draw[color=blue]  ($(yp)!0.55!(x1)$) node[anchor=south west, rotate=-20, yshift=.05cm, xshift=-.75cm,  align=center] {$\Delta_{\delta_3\delta_2,4}+k_3$};
	\draw[color=black]  ($(y)!0.5!(x4)$) node[anchor= west,  align=center] {$\Delta_{4\delta_2,\delta_3} -k_{3}$};
	\draw[color=blue]  ($(yp)!0.55!(x4)$) node[anchor=north west,  rotate=20, yshift=-.025cm, xshift=-.75cm,  align=center] {$\Delta_{\delta_3 4,\delta_2}+k_{3}$};
\end{tikzpicture}}
\newsavebox{\figcalWfourXi}
\savebox{\figcalWfourXi}{%
\begin{tikzpicture}[scale=.7]
    \draw[thick, color=black!30] (0,0) circle[radius=2];
    \coordinate (x1) at (200:2);
    \coordinate (x2) at (160:2);
    \coordinate (x6) at (20:2);
    \coordinate (x7) at (-20:2);
    \coordinate (y) at (180:1.3);
    \coordinate (yp) at (0:1.3);
	\draw[thick, color=red, dashed] (x2) to [out =  -15, in = 15, looseness = 1.5] (x1);
	\draw[thick, color=red, dashed] (x6) to [out =  195, in = 165, looseness = 1.5] (x7);
	\draw[thick, blue] (x1)--(y)--(x2);
	\draw[thick, blue] (x6)--(yp)--(x7);
	\draw[thick, dotted] (y)--(yp);
	\draw (x1) node[anchor=north east] {${\cal O}_{4}(x_{4})$};
	\draw (x2) node[anchor=south east] {${\cal O}_{\delta_2}(x_1)$};
	\draw (x6) node[anchor=south west] {${\cal O}_{5}(x_5)$};
	\draw (x7) node[anchor=north west] {${\cal O}_{6}(x_{6})$};
	\draw[color=black]  ($(y)!0.5!(yp)$) node[anchor=north] {\footnotesize $\Delta_{\delta_3}+2k_3$};
\end{tikzpicture}}
\newsavebox{\figVlimTwo}
\savebox{\figVlimTwo}{%
\begin{tikzpicture}[scale=1.2]
    \draw[thick, color=black!30] (0,0) circle[radius=2];
    \coordinate (x1) at (180:2);
    \coordinate (x5) at (0:2);
    \coordinate (x6) at (0:2);
    \coordinate (x3) at (90:2);
    \coordinate (x4) at (-90:2);
    \coordinate (y) at (180:2);
    \coordinate (yp) at (0:2);
	\draw[thick, black] (x4)--(y)--(x3);
	\draw[thick, black] (x3)--(yp)--(x4);
	\draw[thick, black] (y)--(yp);
	\draw[thick, black] (x3)--(x4);
	\draw (x1) node[anchor=east] {$x_1$};
	\draw (x3) node[anchor=south] {${x}_3$};
	\draw (x4) node[anchor=north] {${x}_4$};
	\draw (x5) node[anchor= west] {$x_5$};
	\draw[color=black]  ($(y)!0.5!(x3)$) node[anchor=south east,  align=center] {$\Delta_{3\delta_1,\delta_2}-k_2$};
	\draw[color=black]  ($(yp)!0.5!(x3)$) node[anchor=south west, align=center] {$\Delta_{3\delta_2,\delta_1}+k_2+j$};
	\draw[color=black]  ($(y)!0.42!(yp)$) node[anchor=north,  align=center] {$\Delta_{\delta_1 \delta_3,34}-j$};
	\draw[color=black]  ($(y)!0.5!(x4)$) node[anchor=north east,  align=center] {$\Delta_{4\delta_2,\delta_3} +k_{2}+j$};
	\draw[color=black]  ($(yp)!0.5!(x4)$) node[anchor=north west,  align=center] {$\Delta_{4\delta_3,\delta_2}-k_{2}$};
	\draw[color=black]  ($(x3)!0.30!(x4)$) node[anchor=west] {$-j$};
\end{tikzpicture}}
\newsavebox{\figSevenCombChannel}
\savebox{\figSevenCombChannel}{%
\begin{tikzpicture}[scale=1]
    \coordinate (x1) at (-5/2,0);
    \coordinate (y1) at (-2,0);
    \coordinate (x2) at (-2,1);
    \coordinate (y2) at (-1,0);
    \coordinate (x3) at (-1,1);
    \coordinate (y3) at (0,0);
    \coordinate (x4) at (0,1);
    \coordinate (y4) at (1,0);
    \coordinate (x5) at (1,1);
    \coordinate (y5) at (2,0);
    \coordinate (x6) at (2,1);
    \coordinate (x7) at (5/2,0);
	\draw[thick] (x1)--(y1)--(x2);
	\draw[thick] (y1)--(y2)--(x3);
	\draw[thick] (y2)--(y3)--(x4);
	\draw[thick] (y3)--(y4)--(x5);
	\draw[thick] (y4)--(y5)--(x6);	
	\draw[thick] (y5)--(x7);
	\draw (x1) node[anchor=east] {${\cal O}_1$};
	\draw (x2) node[anchor=south] {${\cal O}_2$};
	\draw (x3) node[anchor=south] {${\cal O}_3$};
	\draw (x4) node[anchor=south] {${\cal O}_4$};
	\draw (x5) node[anchor=south] {${\cal O}_5$};
	\draw (x6) node[anchor=south] {${\cal O}_6$};
	\draw (x7) node[anchor=west] {${\cal O}_7$};
	\draw ($(y1)!0.5!(y2)$) node[anchor=north] {${\cal O}_{{\delta_1}}$};
	\draw ($(y2)!0.5!(y3)$) node[anchor=north] {${\cal O}_{{\delta_2}}$};
	\draw ($(y3)!0.5!(y4)$) node[anchor=north] {${\cal O}_{{\delta_3}}$};
	\draw ($(y4)!0.5!(y5)$) node[anchor=north] {${\cal O}_{{\delta_4}}$};
\end{tikzpicture}}
\newsavebox{\figcalWSeven}
\savebox{\figcalWSeven}{%
\begin{tikzpicture}[scale=1.65]
    \draw[thick, color=black!30] (0,0) circle[radius=2];
    \coordinate (x1) at (200:2);
    \coordinate (x2) at (160:2);
    \coordinate (x6) at (20:2);
    \coordinate (x7) at (-20:2);
    \coordinate (x3) at (120:2);
    \coordinate (x4) at (-90:2);
    \coordinate (x5) at (60:2);
    \coordinate (y) at (180:1.3);
    \coordinate (yp) at (0:1.3);
	\draw[thick, color=red, dashed] (x2) to [out =  -15, in = 15, looseness = 1.5] (x1);
	\draw[thick, color=red, dashed] (x6) to [out =  195, in = 165, looseness = 1.5] (x7);
	\draw[thick] (x1)--(y)--(x2);
	\draw[thick, blue] (y)--(x3)--(yp)--(x4)--(y);
	\draw[thick, orange] (y)--(x5)--(yp);
	\draw[thick] (x6)--(yp)--(x7);
	\draw[thick, dotted] (y)--(yp);
	\draw[thick, black!40!green] (x3)--(x4)--(x5)--(x3);
	\draw (x1) node[anchor=north east] {${\cal O}_1$};
	\draw (x2) node[anchor=south east] {${\cal O}_2$};
	\draw (x3) node[anchor=south] {${x}_3$};
	\draw (x4) node[anchor=north] {${x}_4$};
	\draw (x5) node[anchor=south] {${x}_5$};
	\draw (x6) node[anchor=south west] {${\cal O}_6$};
	\draw (x7) node[anchor=north west] {${\cal O}_7$};
	\draw[color=blue]  ($(y)!0.55!(x3)$) node[anchor=south east, align=center] {\footnotesize $\Delta_{3\delta_1,\delta_2}+k_{1,2}$};
	\draw[color=blue]  ($(yp)!0.5!(x3)$) node[anchor=south west, text width=2.3cm,  rotate=-38, yshift=.05cm, xshift=-.5cm, align=center] {\footnotesize $\Delta_{3\delta_2,\delta_1}$\\$+k_{2,1}+j_{13,}$};
	\draw[color=black]  ($(y)!0.5!(yp)$) node[anchor=north, text width=4cm, align=center] {\footnotesize $\Delta_{\delta_1 \delta_4,345}+k_{14,}-j_{123,}$};
	\draw[color=blue]  ($(y)!0.55!(x4)$) node[anchor=north east,  align=center] {\footnotesize $\Delta_{4\delta_2,\delta_3} +k_{2,3}+j_1$};
	\draw[color=blue]  ($(yp)!0.55!(x4)$) node[anchor=north west, align=center] {\footnotesize $\Delta_{4\delta_3,\delta_2} +k_{3,2}+j_{2}$};
	\draw[color=orange]  ($(yp)!0.55!(x5)$) node[anchor=south west, align=center] {\footnotesize $\Delta_{5\delta_4,\delta_3} +k_{4,3}$};
	\draw[color=orange]  ($(y)!0.5!(x5)$) node[anchor=south west, text width=2.3cm,  rotate=38, yshift=.05cm, xshift=-1.95cm, align=center] {\footnotesize $\Delta_{5\delta_3,\delta_4}$\\$+k_{3,4}+j_{23,}$};
	\draw[color=black!40!green]  ($(x3)!0.65!(x4)$) node[anchor=east] {\footnotesize $-j_1$};
	\draw[color=black!40!green]  ($(x5)!0.65!(x4)$) node[anchor=west] {\footnotesize $-j_2$};
	\draw[color=black!40!green]  ($(x3)!0.5!(x5)$) node[anchor=south] {\footnotesize $-j_3$};
\end{tikzpicture}}
\newsavebox{\figcalWSevenDoubleOPE}
\savebox{\figcalWSevenDoubleOPE}{%
\begin{tikzpicture}[scale=1.65]
    \draw[thick, color=black!30] (0,0) circle[radius=2];
    \coordinate (x1) at (180:2);
    \coordinate (x2) at (180:2);
    \coordinate (x6) at (0:2);
    \coordinate (x7) at (0:2);
    \coordinate (x3) at (120:2);
    \coordinate (x4) at (-90:2);
    \coordinate (x5) at (60:2);
    \coordinate (y) at (180:2);
    \coordinate (yp) at (0:2);
	\draw[thick, blue] (y)--(x3)--(yp)--(x4)--(y);
	\draw[thick, orange] (y)--(x5)--(yp);
	\draw[thick] (y)--(yp);
	\draw[thick, black!40!green] (x3)--(x4)--(x5)--(x3);
	\draw (x1) node[anchor=east] {$x_1$};
	\draw (x3) node[anchor=south] {${x}_3$};
	\draw (x4) node[anchor=north] {${x}_4$};
	\draw (x5) node[anchor=south] {${x}_5$};
	\draw (x6) node[anchor=west] {$x_6$};
	\draw[color=blue]  ($(y)!0.55!(x3)$) node[anchor=south east, align=center] {\footnotesize $\Delta_{3\delta_1,\delta_2}-k_{2}$};
	\draw[color=blue]  ($(yp)!0.5!(x3)$) node[anchor=south west, text width=2.3cm,  rotate=-30, yshift=.05cm, xshift=-.75cm, align=center] {\footnotesize $\Delta_{3\delta_2,\delta_1}$\\$+k_{2}+j_{13,}$};
	\draw[color=black]  ($(y)!0.5!(yp)$) node[anchor=north, text width=4cm, align=center] {\footnotesize $\Delta_{\delta_1 \delta_4,345}-j_{123,}$};
	\draw[color=blue]  ($(y)!0.55!(x4)$) node[anchor=north east,  align=center] {\footnotesize $\Delta_{4\delta_2,\delta_3} +k_{2,3}+j_1$};
	\draw[color=blue]  ($(yp)!0.55!(x4)$) node[anchor=north west, align=center] {\footnotesize $\Delta_{4\delta_3,\delta_2} +k_{3,2}+j_{2}$};
	\draw[color=orange]  ($(yp)!0.55!(x5)$) node[anchor=south west, align=center] {\footnotesize $\Delta_{5\delta_4,\delta_3} -k_{3}$};
	\draw[color=orange]  ($(y)!0.5!(x5)$) node[anchor=south west, text width=2.3cm,  rotate=30, yshift=.05cm, xshift=-1.5cm, align=center] {\footnotesize $\Delta_{5\delta_3,\delta_4}$\\$+k_{3}+j_{23,}$};
	\draw[color=black!40!green]  ($(x3)!0.65!(x4)$) node[anchor=east] {\footnotesize $-j_1$};
	\draw[color=black!40!green]  ($(x5)!0.65!(x4)$) node[anchor=west] {\footnotesize $-j_2$};
	\draw[color=black!40!green]  ($(x3)!0.5!(x5)$) node[anchor=south] {\footnotesize $-j_3$};
\end{tikzpicture}}
\newsavebox{\figNPlusTwoCombChannel}
\savebox{\figNPlusTwoCombChannel}{%
\begin{tikzpicture}[scale=1]
    \coordinate (x1) at (-5/2,0);
    \coordinate (y1) at (-2,0);
    \coordinate (x2) at (-2,1);
    \coordinate (y2) at (-1,0);
    \coordinate (x3) at (-1,1);
    \coordinate (y3) at (0,0);
    \coordinate (z3) at (0,1/2);
    \coordinate (x4) at (0,1);
    \coordinate (y4) at (1,0);
    \coordinate (z4) at (1,1/2);
    \coordinate (x5) at (1,1);
    \coordinate (y5) at (2,0);
    \coordinate (x6) at (2,1);
    \coordinate (x8) at (3,1);
    \coordinate (x7) at (7/2,0);
    \coordinate (y6) at (3,0);
	\draw[thick] (x1)--(y1)--(x2);
	\draw[thick] (y1)--(y2)--(x3);
	\draw[thick] (y2)--(y3)--(x4);
	\draw[very thick, line cap=round, dash pattern=on 0 off 8] (z3)--(z4);
	\draw[thick] (y3)--(y4);
	\draw[thick] (y4)--(x5);
	\draw[thick] (y4)--(y5)--(x6);	
	\draw[thick] (y5)--(y6)--(x8);
	\draw[thick] (y6)--(x7);
	\draw (x1) node[anchor=east] {${\cal O}_{e_1}(x_{1^\prime})$};
	\draw ($(x2)!-0.15!(x3)$) node[anchor=south] {${\cal O}_{e_2}(x_{1})$};
	\draw (x3) node[anchor=south] {${\cal O}_2$};
	\draw (x4) node[anchor=south] {${\cal O}_3$};
	\draw (x5) node[anchor=south] {${\cal O}_{n-2}$};
	\draw (x6) node[anchor=south] {${\cal O}_{n-1}$};
	\draw ($(x8)!-0.25!(x6)$) node[anchor=south] {${\cal O}_{e_3}(x_{n})$};
	\draw (x7) node[anchor=west] {${\cal O}_{e_4}(x_{n^\prime})$};
	\draw ($(y1)!0.5!(y2)$) node[anchor=north] {${\cal O}_{1}$};
	\draw ($(y2)!0.5!(y3)$) node[anchor=north] {${\cal O}_{{\delta_1}}$};
	\draw ($(y4)!0.5!(y5)$) node[anchor=north] {${\cal O}_{{\delta_{n-3}}}$};
	\draw ($(y5)!0.5!(y6)$) node[anchor=north] {${\cal O}_{n}$};
\end{tikzpicture}}
\newsavebox{\figcalWnplustwoA}
\savebox{\figcalWnplustwoA}{%
\begin{tikzpicture}[scale=1.2]
    \draw[thick, color=black!30] (0,0) circle[radius=2];
    \coordinate (x1) at (200:2);
    \coordinate (x2) at (160:2);
    \coordinate (x6) at (20:2);
    \coordinate (x7) at (-20:2);
    \coordinate (y) at (180:1.3);
    \coordinate (yp) at (0:1.3);
	\draw[thick, color=red, dashed] (x2) to [out =  -15, in = 15, looseness = 1.5] (x1);
	\draw[thick, color=red, dashed] (x6) to [out =  195, in = 165, looseness = 1.5] (x7);
	\draw[thick, blue] (x1)--(y)--(x2);
	\draw[thick, blue] (x6)--(yp)--(x7);
	\draw[thick, dotted] (y)--(yp);
	\draw (x1) node[anchor=north east] {${\cal O}_{e_1}(x_{1^\prime})$};
	\draw (x2) node[anchor=south east] {${\cal O}_{e_2}(x_1)$};
	\draw (x6) node[anchor=south west] {${\cal O}_{e_3}(x_n)$};
	\draw (x7) node[anchor=north west] {${\cal O}_{e_4}(x_{n^\prime})$};
	\draw (y) node[anchor=east] {$w$};
	\draw (yp) node[anchor=west] {$w^\prime$};
	\draw[color=black]  ($(y)!0.5!(yp)$) node[anchor=north, text width=4cm, align=center] {\scriptsize $\Delta_{1n,2 \ldots (n-1)}+k_{0(n-2),}$ \\$-\sum_{2 \leq r < s \leq n-1} j_{\langle r|s\rangle}$}; 
\end{tikzpicture}}
\newsavebox{\figcalWnplustwoB}
\savebox{\figcalWnplustwoB}{%
\begin{tikzpicture}[scale=1]
    \draw[thick, color=black!30] (0,0) circle[radius=2];
    \coordinate (x3) at (120:2);
    \coordinate (x4) at (-120:2);
    \coordinate (x5) at (60:2);
    \coordinate (xk) at (90:2);
    \coordinate (xl) at (-90:2);
	\draw[thick, black!100!green] (xk)--(xl);
	\draw (xk) node[anchor=south] {${x}_r$};
	\draw (xl) node[anchor=north] {${x}_s$};
		\draw[color=black!100!green]  ($(xk)!0.5!(xl)$) node[anchor=east,  rotate=0, yshift=.0cm, xshift=-.0cm] {\footnotesize $-j_{\langle r|s\rangle}$};
\end{tikzpicture}}
\newsavebox{\figcalWnplustwoD}
\savebox{\figcalWnplustwoD}{%
\begin{tikzpicture}[scale=1]
    \draw[thick, color=black!30] (0,0) circle[radius=2];
    \coordinate (x4) at (90:2);
    \coordinate (y) at (180:1.3);
    \coordinate (yp) at (0:1.3);
	\draw[thick, blue] (yp)--(x4)--(y);
	\draw (x4) node[anchor=south] {${x}_{t+2}$};
	\draw (y) node[anchor=east] {$w$};
	\draw (yp) node[anchor=west] {$w^\prime$};
	\draw[color=blue]  ($(y)!0.65!(x4)$) node[anchor= east,  align=center] {\footnotesize $\Delta_{(t+2)\delta_{t},\delta_{t+1}} +k_{t,t+1}$\\ \footnotesize$+\sum_{r=2}^{t+1}j_{\langle r|t+2\rangle}$};
	\draw[color=blue]  ($(yp)!0.65!(x4)$) node[anchor=west, align=center, xshift=-0.05cm] {\footnotesize $\Delta_{(t+2)\delta_{t+1},\delta_{t}} +k_{t+1,t}$\\\footnotesize$+\sum_{s=t+3}^{n-1}j_{\langle t+2|s\rangle}$};
\end{tikzpicture}}
\newsavebox{\figNCombChannel}
\savebox{\figNCombChannel}{%
\begin{tikzpicture}[scale=1]
    \coordinate (y1) at (-3/2,0);
    \coordinate (y2) at (-1,0);
    \coordinate (x3) at (-1,1);
    \coordinate (y3) at (0,0);
    \coordinate (z3) at (0,1/2);
    \coordinate (x4) at (0,1);
    \coordinate (y4) at (1,0);
    \coordinate (z4) at (1,1/2);
    \coordinate (x5) at (1,1);
    \coordinate (y5) at (2,0);
    \coordinate (x6) at (2,1);
    \coordinate (y6) at (5/2,0);
	\draw[thick] (y1)--(y2)--(x3);
	\draw[thick] (y2)--(y3)--(x4);
	\draw[very thick, line cap=round, dash pattern=on 0 off 8] (z3)--(z4);
	\draw[thick] (y3)--(y4);
	\draw[thick] (y4)--(x5);
	\draw[thick] (y4)--(y5)--(x6);	
	\draw[thick] (y5)--(y6);
	\draw (y1) node[anchor=east] {${\cal O}_{1}$};
	\draw (x3) node[anchor=south] {${\cal O}_2$};
	\draw (x4) node[anchor=south] {${\cal O}_3$};
	\draw (x5) node[anchor=south] {${\cal O}_{n-2}$};
	\draw (x6) node[anchor=south] {${\cal O}_{n-1}$};
	\draw (y6) node[anchor=west] {${\cal O}_{n}$};
	\draw ($(y2)!0.5!(y3)$) node[anchor=north] {${\cal O}_{{\delta_1}}$};
	\draw ($(y4)!0.5!(y5)$) node[anchor=north] {${\cal O}_{{\delta_{n-3}}}$};
\end{tikzpicture}}
\newsavebox{\figNCombChannelOPE}
\savebox{\figNCombChannelOPE}{%
\begin{tikzpicture}[scale=1]
    \coordinate (y1) at (-3/2,0);
    \coordinate (y2) at (-1,0);
    \coordinate (x3) at (-1,1);
    \coordinate (y3) at (0,0);
    \coordinate (z3) at (0,1/2);
    \coordinate (x4) at (0,1);
    \coordinate (y4) at (1,0);
    \coordinate (z4) at (1,1/2);
    \coordinate (x5) at (1,1);
    \coordinate (y5) at (2,0);
    \coordinate (x6) at (2,1);
    \coordinate (y6) at (5/2,0);
	\draw[thick] (y1)--(y2)--(x3);
	\draw[thick] (y2)--(y3)--(x4);
	\draw[very thick, line cap=round, dash pattern=on 0 off 8] (z3)--(z4);
	\draw[thick] (y3)--(y4);
	\draw[thick] (y4)--(x5);
	\draw[thick] (y4)--(y5)--(x6);	
	\draw[thick] (y5)--(y6);
	\draw (y1) node[anchor=east] {${\cal O}_{1}$};
	\draw (x3) node[anchor=south] {${\cal O}_2$};
	\draw (x4) node[anchor=south] {${\cal O}_3$};
	\draw (x5) node[anchor=south] {${\cal O}_{n-3}$};
	\draw (x6) node[anchor=south] {${\cal O}_{n-2}$};
	\draw (y6) node[anchor=west] {${\cal O}_{\delta_{n-3}}(x_{n-1})$};
	\draw ($(y2)!0.5!(y3)$) node[anchor=north] {${\cal O}_{{\delta_1}}$};
	\draw ($(y4)!0.5!(y5)$) node[anchor=north] {${\cal O}_{{\delta_{n-4}}}$};
\end{tikzpicture}}
\newsavebox{\figWnA}
\savebox{\figWnA}{%
\begin{tikzpicture}[scale=1]
    \draw[thick, color=black!30] (0,0) circle[radius=2];
    \coordinate (y) at (180:2);
    \coordinate (yp) at (0:2);
	\draw[thick, black!100!green] (y)--(yp);
	\draw (y) node[anchor=east] {$x_1$};
	\draw (yp) node[anchor=west] {$x_n$};
	\draw[color= black!100!green]  ($(y)!0.5!(yp)$) node[anchor=north, text width=4cm, align=center] {\footnotesize $\Delta_{1n,2 \ldots (n-1)}$\\$-\sum_{2 \leq r < s \leq n-1} j_{\langle r|s\rangle}$};
\end{tikzpicture}}
\newsavebox{\figWnD}
\savebox{\figWnD}{%
\begin{tikzpicture}[scale=1]
    \draw[thick, color=black!30] (0,0) circle[radius=2];
    \coordinate (x4) at (90:2);
    \coordinate (y) at (180:2);
    \coordinate (yp) at (0:2);
	\draw[thick, black] (yp)--(x4)--(y);
	\draw (x4) node[anchor=south] {${x}_{t+2}$};
	\draw (y) node[anchor=east] {$x_1$};
	\draw (yp) node[anchor=west] {$x_n$};
	\draw[color=black]  ($(y)!0.65!(x4)$) node[anchor= east,  align=center, rotate=45, yshift=1.15cm, xshift=1.0cm] {\footnotesize $\Delta_{(t+2)\delta_{t},\delta_{t+1}} +k_{t,t+1}$\\\footnotesize $+\sum_{r=2}^{t+1}j_{\langle r|t+2\rangle}$};
	\draw[color=black]  ($(yp)!0.65!(x4)$) node[anchor=west, align=center,   rotate=-45, yshift=1.15cm, xshift=-1.1cm] {\footnotesize $\Delta_{(t+2)\delta_{t+1},\delta_{t}} +k_{t+1,t}$\\ \footnotesize$+\sum_{s=t+3}^{n-1}j_{\langle t+2|s\rangle}$};
\end{tikzpicture}}
\newsavebox{\figWnAp}
\savebox{\figWnAp}{%
\begin{tikzpicture}[scale=1]
    \draw[thick, color=black!30] (0,0) circle[radius=2];
    \coordinate (y) at (180:2);
    \coordinate (yp) at (0:2);
	\draw[thick, black!100!green] (y)--(yp);
	\draw (y) node[anchor=east] {$x_1$};
	\draw (yp) node[anchor=west] {$x_{n-1}$};
	\draw[color= black!100!green]  ($(y)!0.5!(yp)$) node[anchor=north, text width=4cm, align=center] {\footnotesize $\Delta_{1\delta_{n-3},2 \ldots (n-2)}$\\$-\sum_{2 \leq r < s \leq n-2} j_{\langle r|s\rangle}$};
\end{tikzpicture}}
\newsavebox{\figWnBp}
\savebox{\figWnBp}{%
\begin{tikzpicture}[scale=1]
    \draw[thick, color=black!30] (0,0) circle[radius=2];
    \coordinate (xk) at (90:2);
    \coordinate (xl) at (-90:2);
	\draw[thick, black!100!green] (xk)--(xl);
	\draw (xk) node[anchor=south] {${x}_r$};
	\draw (xl) node[anchor=north] {${x}_s$};
		\draw[color=black!100!green]  ($(xk)!0.5!(xl)$) node[anchor=east,  rotate=0, yshift=0cm, xshift=0cm] {\footnotesize $-j_{\langle r|s\rangle}$};
\end{tikzpicture}}
\newsavebox{\figWnDp}
\savebox{\figWnDp}{%
\begin{tikzpicture}[scale=1]
    \draw[thick, color=black!30] (0,0) circle[radius=2];
    \coordinate (x4) at (90:2);
    \coordinate (y) at (180:2);
    \coordinate (yp) at (0:2);
	\draw[thick, black] (yp)--(x4)--(y);
	\draw (x4) node[anchor=south] {${x}_{t+2}$};
	\draw (y) node[anchor=east] {$x_1$};
	\draw (yp) node[anchor=west] {$x_{n-1}$};
	\draw[color=black]  ($(y)!0.65!(x4)$) node[anchor= east,  align=center, rotate=45, yshift=1.15cm, xshift=1.1cm] {\footnotesize $\Delta_{(t+2)\delta_{t},\delta_{t+1}} +k_{t,t+1}$\\\footnotesize$+\sum_{r=2}^{t+1}j_{\langle r|t+2\rangle}$};
	\draw[color=black]  ($(yp)!0.65!(x4)$) node[anchor=west, align=center,   rotate=-45, yshift=1.15cm, xshift=-1.1cm] {\footnotesize $\Delta_{(t+2)\delta_{t+1},\delta_{t}} +k_{t+1,t}$\\\footnotesize$+\sum_{s=t+3}^{n-2}j_{\langle t+2|s\rangle}$};
\end{tikzpicture}}
\newsavebox{\figLsqrId}
\savebox{\figLsqrId}{%
\begin{tikzpicture}[scale=.7]
    \draw[thick, color=black!30] (0,0) circle[radius=2];
    \coordinate (x1) at (90:2);
    \coordinate (xa) at (180:2);
    \coordinate (xell) at (-90:2);
	\draw[thick, black] (x1)--(xa)--(xell);
	\draw (xa) node[anchor=east] {${x}$};
	\draw (x1) node[anchor=south] {$x_1$};
	\draw (xell) node[anchor=north] {$x_{\ell}$};
	\draw[very thick, line cap=round, dash pattern=on 0 off 22] (60:2.3) arc (60:-60:2.3);
	\draw[color=black]  ($(x1)!0.5!(xa)$) node[anchor= south,  align=center, rotate=45, yshift=0cm, xshift=0cm] {\footnotesize $\Delta_1$};
	\draw[color=black]  ($(xell)!0.5!(xa)$) node[anchor= south,  align=center, rotate=-45, yshift=0cm, xshift=0cm] {\footnotesize $\Delta_\ell$};
	\draw[color=black]  ($(xell)!0.53!(x1)$) node[anchor= east,  align=center, yshift=0cm, xshift=-.8cm] { $\vdots$};
\end{tikzpicture}}
\newsavebox{\figLcrossIdOne}
\savebox{\figLcrossIdOne}{%
\begin{tikzpicture}[scale=.6]
    \draw[thick, color=black!30] (0,0) circle[radius=2];
    \coordinate (x1) at (90:2);
    \coordinate (xell) at (-90:2);
    \draw[thick] (x1)--(xell);
	\draw (x1) node[anchor=south] {$x_r$};
	\draw (xell) node[anchor=north] {$x_{s}$};
    \draw ($(x1)!0.5!(xell)$) node[anchor=east] {\footnotesize $\Delta_{\langle r|s \rangle}$};
\end{tikzpicture}}
\newsavebox{\figLcrossIdTwo}
\savebox{\figLcrossIdTwo}{%
\begin{tikzpicture}[scale=.65]
    \draw[thick, color=black!30] (0,0) circle[radius=2];
    \coordinate (x1) at (90:2);
    \coordinate (xa) at (180:2);   
    \coordinate (xb) at (0:2);
    \coordinate (xell) at (-90:2);
	\draw[thick, black] (x1)--(xa)--(xell); 
	\draw (xa) node[anchor=east] {${x_{a_i}}$};
	\draw (x1) node[anchor=south] {$x_r$};
	\draw (xell) node[anchor=north] {$x_{s}$};
	\draw[color=black]  ($(x1)!0.5!(xa)$) node[anchor= north,  align=center, rotate=45, yshift=0.1cm, xshift=0cm] {\footnotesize $\Delta_{\langle a_i|r \rangle}$};
	\draw[color=black]  ($(xell)!0.5!(xa)$) node[anchor= south,  align=center, rotate=-45, yshift=-0.1cm, xshift=0cm] {\footnotesize $\Delta_{\langle a_i|s \rangle}$};
\end{tikzpicture}}
\newsavebox{\figWAlt}
\savebox{\figWAlt}{%
\begin{tikzpicture}[scale=.5]
    \draw[thick, color=black!30] (0,0) circle[radius=2];
    \coordinate (x1) at (90:2);
    \coordinate (xell) at (-90:2);
    \draw[thick] (x1)--(xell);
	\draw (x1) node[anchor=south] {$x_u$};
	\draw (xell) node[anchor=north] {$x_{v}$};
    \draw ($(x1)!0.5!(xell)$) node[anchor=east] {\footnotesize $\Delta_{\langle u|v \rangle}$};
\end{tikzpicture}}

\begin{document}

 \title{A multipoint conformal block chain in $d$~dimensions}
\authors{Sarthak Parikh\footnote{\tt sparikh@caltech.edu}}
\institution{Caltech}{Division of Physics, Mathematics and Astronomy, California Institute of Technology,\cr\hskip0.06in Pasadena, CA 91125, USA}

 \abstract{ 
 Conformal blocks play a central role in CFTs as the basic, theory-independent building blocks. However, only limited results are available concerning multipoint blocks associated with the global conformal group. In this paper, we systematically work out the $d$-dimensional $n$-point global conformal blocks (for arbitrary $d$ and $n$) for external and exchanged scalar operators in the so-called comb channel. We use kinematic aspects of holography and previously worked out higher-point AdS propagator identities to first obtain the geodesic diagram representation for the $(n+2)$-point block. Subsequently, upon taking a particular double-OPE limit, we obtain an explicit power series expansion for the $n$-point block expressed in terms of powers of conformal cross-ratios. Interestingly, the expansion coefficient is written entirely in terms of Pochhammer symbols and $(n-4)$ factors of the generalized hypergeometric function~${}_3F_2$, for which we provide a holographic explanation. This generalizes the results previously obtained in the literature for $n=4, 5$. We verify the results explicitly in embedding space using conformal Casimir equations. \\\\
 
\centering{{\it Dedicated to the memory of Steven S.\ Gubser}}
 }

 \maketitle


{\hypersetup{linkcolor=black}
\tableofcontents
}


\section{Introduction}
\label{INTRO}

 Conformal field theories (CFTs) are important for several reasons --- they serve as salient guideposts in the space of quantum field theories, describe a variety of critical phenomena, and help elucidate aspects of quantum gravity via the AdS/CFT correspondence~\cite{Maldacena:1997re,Gubser:1998bc,Witten:1998qj}. 
Conformal blocks play a central role in CFTs. They are the basic kinematic building blocks of local observables, encoding the contribution  of primary operators (and all their descendants)  to any given correlation function.
Given a $d$-dimensional CFT (more precisely the dynamical data in the form of the spectrum of all primary operators and the operator product expansion (OPE) coefficients between them), the knowledge of $d$-dimensional conformal blocks permits the explicit writing of all possible correlators.
Conversely, the  conformal bootstrap program~\cite{Ferrara:1973yt,Polyakov:1974gs,Rattazzi:2008pe} (see also the recent review~\cite{Poland:2018epd} and references therein) provides a non-perturbative approach to reconstructing the full CFT${}_d$ data by exploiting conformal symmetry as well as stringent consistency conditions such as the associativity of the OPE. 
Here as well, conformal blocks are an essential ingredient needed for setting up the bootstrap equations involving conformal correlators.

It is therefore important to understand these basic building blocks in as much detail as possible. In this paper we will be focusing on multipoint $d$-dimensional {\it global} conformal blocks associated with the Euclidean conformal group $SO(d+1,1)$.
While these theory independent objects are in principle fixed entirely from conformal symmetry, working them out presents significant computational challenges, so explicit results are available in only simple cases.
For instance, explicit expressions are known for {\it four-point} scalar conformal blocks  in general spacetime dimensions~\cite{Ferrara:1971vh,Ferrara:1973vz,Ferrara:1974ny,Dolan:2000ut,Dolan:2003hv,Dolan:2011dv}.
A variety of techniques have been developed for computing four-point conformal blocks involving external and internal exchanged operators in arbitrary representations of the Lorentz group  in closed-form, integral or efficient series expansions; a partial list includes various recursive methods~\cite{Dolan:2000ut,Dolan:2011dv,Zamolodchikov:1985ie,Kos:2013tga,Penedones:2015aga,Iliesiu:2015akf,Costa:2016xah,Costa:2016hju,Kravchuk:2017dzd,Erramilli:2019njx}, shadow formalism~\cite{SimmonsDuffin:2012uy}, use of differential operators~\cite{Costa:2011mg,Costa:2011dw,Echeverri:2015rwa,Echeverri:2016dun,Karateev:2017jgd,Cuomo:2017wme,Isono:2017grm,Fortin:2016lmf,Fortin:2019fvx,Fortin:2019dnq,Fortin:2019gck},\footnote{See also ref.~\cite{Costa:2018mcg} for an AdS interpretation and refs.~\cite{Arkani-Hamed:2018kmz,Baumann:2019oyu} in the context of cosmological bootstrap in dS backgrounds.} Wilson line constructions~\cite{Besken:2016ooo,Bhatta:2016hpz,Bhatta:2018gjb}, integrability methods~\cite{Isachenkov:2016gim,Schomerus:2016epl,Buric:2019rms,Buric:2019dfk} and holographic geodesic diagram techniques~\cite{Hijano:2015zsa,Nishida:2016vds,Castro:2017hpx,Dyer:2017zef,Chen:2017yia,Gubser:2017tsi,Kraus:2017ezw,Tamaoka:2017jce,Nishida:2018opl,Das:2018ajg}.
The situation is disproportionately dire for {\it higher-point} global conformal blocks. Recent work in the shadow formalism has led to explicit series expansions for $n$-point scalar conformal blocks in dimensions one and two in a specific channel called the comb channel, for arbitrary $n$~\cite{Rosenhaus:2018zqn}. In higher dimensions, a series expansion was obtained for the scalar five-point block restricted to the exchange of scalar representations~\cite{Rosenhaus:2018zqn} (see also ref.~\cite{Goncalves:2019znr}).
Geodesic diagram representations have also been obtained for the same five-point block in general spacetime dimensions~\cite{Parikh:2019ygo}, as well as for the $d$-dimensional  six-point scalar conformal block involving scalar exchanges in a different channel called the OPE channel~\cite{Jepsen:2019svc}. 

While one can recursively reduce any higher-point conformal correlator into a combination of two- and three-point functions via repeated use of the OPE (or equivalently, reduce to a combination of four-point correlation functions),  higher-point correlators and conformal blocks are important in their own right for a number of reasons.
For one, knowledge of higher-point blocks allows immediately an efficient writing of conformal correlators and repackaging of higher-point AdS diagrams directly in  position space.
Moreover, higher-point diagrams involve exchange of multi-twist exchanges in their conformal block decomposition, which can provide a new window into understanding  multi-twist exchanges  appearing in the setting of light-cone bootstrap of four-point functions~\cite{Komargodski:2012ek,Fitzpatrick:2012yx,Fitzpatrick:2014vua,Kaviraj:2015cxa,Kaviraj:2015xsa,Alday:2015ewa,Alday:2016njk,Simmons-Duffin:2016wlq,Caron-Huot:2017vep,Albayrak:2019gnz}.
Additionally, knowledge of higher-point scalar conformal blocks opens up the possibility of setting up an equivalent but possibly more efficient alternative to the conventional bootstrap program. In the conventional approach, typically one must study crossing equations of four-point correlators of {\it all} operators in the spectrum including those in non-trivial representations of the Lorentz group.  
As a potential alternative, one can instead aim to solve crossing equations for {\it scalar} $n$-point functions, but for all $n$~\cite{Gadde2018,Rosenhaus:2018zqn}.
Clearly, $n$-point blocks will play a crucial role here.

\begin{figure}
    \centering
    \[ {W}_{\Delta_{\delta_1};\:\ldots ;\:\Delta_{\delta_{n-3}}}^{\Delta_{1},\Delta_2,\ldots,\Delta_{n-1},\Delta_{n}}(x_1,x_2,\ldots,x_{n-1},x_n) 
\equiv \musepic{\figNCombChannel} \equiv W^{(n)}(x_i)\]
    \caption{The comb channel $n$-point global conformal block for external scalar operators ${\cal O}_1(x_1), \ldots, {\cal O}_n(x_n)$ with conformal dimensions $\Delta_1,\ldots,\Delta_n$ and insertion coordinates $x_1,\ldots,x_n$ respectively, and exchanged scalar operators ${\cal O}_{\delta_1},\ldots,{\cal O}_{\delta_{n-3}}$ with conformal dimensions $\Delta_{\delta_1},\ldots,\Delta_{\delta_{n-3}}$, respectively. When there is no scope for confusion, we will often abbreviate it  as $W^{(n)}(x_i)$ or simply $W^{(n)}$.}
    \label{fig:nblock}
\end{figure}

Motivated by these considerations, in this paper we will compute {\it all higher-point} $d$-dimensional global conformal blocks  in a channel which ref.~\cite{Rosenhaus:2018zqn} referred to as the {\it comb channel}, for external and exchanged scalar operators (see figure~\ref{fig:nblock}). This will generalize the series expansion for the $d$-dimensional five-point block computed in ref.~\cite{Rosenhaus:2018zqn} to $n$-point blocks for any $n$.
The main techniques we will be employing are the AdS propagator identities and geodesic diagram techniques of refs.~\cite{Hijano:2015zsa,Gubser:2017tsi,Parikh:2019ygo,Jepsen:2019svc}. 
 Our strategy will be to obtain the holographic geodesic diagram representation of an $(n+2)$-point block in the comb channel with the help of various recently derived AdS propagator identities~\cite{Jepsen:2019svc}. Higher-point geodesic diagrams, like the four-point case~\cite{Hijano:2015zsa}, are higher-point AdS diagrams where all bulk integrations are restricted to geodesic integrals, and they are related to higher-point conformal blocks~\cite{Parikh:2019ygo,Jepsen:2019svc}.
 For comb channel blocks, these geodesic diagram representations involve precisely two geodesic integrals. Taking a particular double-OPE limit gets rid of these geodesic integrals, producing an $n(n-3)/2$-fold power series expansion of the $n$-point conformal block. We also verify our result via a proof by conformal Casimir equations.

This paper is organized as follows. In section~\ref{FIVESIX} we illustrate the main computational strategy in the simplest non-trivial example. Particularly, in section~\ref{SIXCOMB} we derive the holographic geodesic diagram representation of the six-point comb channel block, and in section~\ref{SIXOPE} we reproduce the well-known series expansion of the four-point block by taking a double-OPE limit. A second example is provided in appendix~\ref{SEVENCOMB}, where we briefly discuss the holographic seven-point block and its double-OPE limit which recovers the series expansion of the five-point block.
In section~\ref{NCOMB}, with the help of these examples, we propose a holographic representation of the general $(n+2)$-point block, whose double-OPE limit leads to the power series expansion of the $n$-point block. Sections~\ref{NCOMBOPE}-\ref{CASIMIR} are concerned with proving our claim via conformal Casimir equations.
We conclude in section~\ref{DISCUSSION} with comments on extending the results to multipoint scalar blocks in {\it arbitrary channels}, brief remarks on the comparison with analogous results in the framework of $p$-adic AdS/CFT, and  a proposal for an alternate, more rapidly convergent series expansion for $n$-point blocks in the comb channel. Further computational details can be found in appendix~\ref{TECH}.

\vspace{.5em}

We end this section with a presentation of the main technical result of this paper, which is an explicit power series expansion for the comb channel $n$-point global conformal block in $d$ spacetime dimensions for external and exchanged scalar operators (see figure~\ref{fig:nblock}):\footnote{For convenience, this result is also included  with the arXiv submission in an ancillary {\tt Mathematica} notebook.}
\eqn{nCombBlock}{
& {W}^{(n)}(x_i)  = 
 { \prod_{t=0}^{n-3} \Gamma(1-\Delta_{\delta_t \delta_{t+1},(t+2)} ) \over \Gamma(1-\Delta_{1n,2\ldots (n-1)})\prod_{i=1}^{n-3} \Gamma(1-\Delta_{\delta_i})} \:
  W_0^{(n)}(x_i) \left( \prod_{i=1}^{n-3} u_i^{\Delta_{\delta_i} \over 2} \right)  
  \cr 
 &\qquad \times  \sum_{\substack{k_1,\ldots,k_{n-3},\\j_{\langle 2|3\rangle},\:j_{\langle 2|4\rangle},\:\ldots,\:j_{\langle n-2|n-1\rangle}=0}}^\infty \left[
\left( \prod_{i=1}^{n-3} {u_i^{k_i} \over k_i!} \right) \left( \prod_{2\leq r < s \leq n-1} {(-w_{r;s})^{j_{\langle r|s\rangle}} \over j_{\langle r|s\rangle}!} \right) 
\right.\cr 
&\qquad  \times  
 \left( \Delta_{1n,2\ldots(n-1)} \right)_{-\sum_{2 \leq r < s \leq n-1} j_{\langle r|s\rangle}}
 \left( \prod_{t=0}^{n-3}  { \left(1-\Delta_{(t+1)\delta_{t-1},\delta_{t}}\right)_{k_{t}} \left(1-\Delta_{(t+2)\delta_{t+1},\delta_{t}}\right)_{k_{t}} \over  \left(\Delta_{\delta_{t}} - d/2+1 \right)_{k_{t}} } \right.
 \cr 
& \qquad  \times 
\left(\Delta_{(t+2)\delta_t,\delta_{t+1}} \right)_{k_{t,t+1}+\sum_{2 \leq r<t+2 }j_{\langle r|t+2\rangle}}
\left(\Delta_{(t+2)\delta_{t+1},\delta_{t}} \right)_{k_{t+1,t}+\sum_{t+2 < s\leq n-1 }j_{\langle t+2|s\rangle}} 
 \cr 
 & \qquad   \times  {}_3F_2\left[\left\{-k_t,-k_{t+1},\Delta_{\delta_t\delta_{t+1} (t+2),}-{d\over 2}\right\}; \left\{\Delta_{(t+2)\delta_{t+1},\delta_t}-k_t, \Delta_{(t+2)\delta_t,\delta_{t+1}}-k_{t+1}\right\};1\right] \Bigg) \Bigg],
}
where $(a)_b \equiv \Gamma(a+b)/\Gamma(a)$ is the Pochhammer symbol and we are using the notation
\eqn{DeltaijkDef}{
\Delta_{i_1\ldots i_\ell,i_{\ell+1} \ldots i_k} \equiv {1\over 2} \left( \Delta_{i_1} + \cdots + \Delta_{i_\ell} - \Delta_{i_{\ell+1}} - \cdots - \Delta_{i_k} \right)
}
for conformal dimensions $\Delta_i$, whereas
 \eqn{kabc}{
k_{i_1 \ldots i_\ell, i_{\ell+1} \ldots i_{n}} &\equiv k_{i_1} + \cdots + k_{i_\ell} - k_{i_{\ell+1}} - \cdots - k_{i_n} 
}
for the  integral parameters $k_i$, as well as the additional definitions\footnote{The apparent dependence  of the conformal block~\eno{nCombBlock} on the undefined dimension $\Delta_{\delta_{-1}}$ is spurious, since  $\Delta_{\delta_{-1}}$ always appears inside Pochhammer symbols of the form $(a)_0$ for some non-zero $a$, which evaluate identically to unity.}
\eqn{kdeltaExtraDef}{ 
k_0 \equiv 0 \qquad k_{n-2} \equiv 0 \qquad
\Delta_{\delta_0} \equiv \Delta_1 \qquad  \Delta_{\delta_{n-2}}\equiv \Delta_{n}\,,
}
so that there are precisely $(n-3)$ independent $k_i$ parameters to be summed over in~\eno{nCombBlock}.
The other set of integral parameters is denoted $j_{\langle r|s\rangle}$, for $2 \leq r < s \leq n-1$, where we use the notation $\langle \cdot|\cdot\rangle$ in the subscript to index the $\binom{n-2}{2}$ independent $j$ parameters.\footnote{This notation should not be confused with the bra-ket notation of quantum mechanics which will not play any role in this paper.} Combined, this leads to an $n(n-3)/2$-fold sum.

The coordinate dependence of the conformal block~\eno{nCombBlock} is factorized into a ``leg factor'', which depends  solely on external dimensions and is given by
\eqn{W0nDef}{
{W}_{0}^{\Delta_{1},\ldots,\Delta_{n}}(x_1,\ldots,x_n) \equiv \left( { x_{2n}^2 \over x_{1n}^2 x_{12}^2} \right)^{\Delta_1 \over 2} \left( { x_{1(n-1)}^2 \over x_{1n}^2 x_{(n-1)n}^2} \right)^{\Delta_n \over 2} \prod_{i=2}^{n-1} \left( { x_{1n}^2 \over x_{1i}^2 x_{in}^2} \right)^{\Delta_i \over 2}  \equiv W_0^{(n)}(x_i) ,
}
where $x_{ij} = x_i -x_j$, while the rest of the dependence  is expressible as an $n(n-3)/2$-fold power series expansion entirely in terms of  a set of $n(n-3)/2$ independent\footnote{There are $n(n-3)/2$ independent cross-ratios as long as the spacetime dimension $d$ is sufficiently large, more precisely $d+1 \geq n$. In lower number of dimensions, some of the cross-ratios defined in~\eno{uwDef} become dependent, such that there are only $nd - (d+2)(d+1)/2$ independent cross-ratios.} cross-ratios $0 \leq u_i, w_{r;s} \leq 1$ defined as follows,\footnote{The preferred appearance of $x_1, x_n$ in the choice of cross-ratios, and the related asymmetry between the external scaling dimensions $\Delta_1, \Delta_n$ and the remaining ones $\Delta_2,\ldots,\Delta_{n-1}$ in the power series expansion~\eno{nCombBlock},  arises naturally in the derivation of  the $n$-point conformal block as a particular double-OPE limit of the holographic representation of an $(n+2)$-point conformal block, discussed in section~\ref{NCOMB}, which preferentially identifies $x_1$ and $x_n$. }
\eqn{uwDef}{
u_i \equiv { x_{1(i+1)}^2 x_{(i+2)n}^2 \over x_{(i+1)n}^2 x_{1(i+2)}^2} \qquad 1 \leq i \leq n-3\,, \qquad
w_{ r;s } \equiv { x_{1n}^2 x_{rs}^2 \over x_{rn}^2 x_{1s}^2} \qquad 2 \leq r < s \leq n-1   \,.
}

 The conformal block is uniquely determine based on the following conditions~\cite{SimmonsDuffin:2012uy}. They satisfy the multipoint conformal Casimir eigenvalue equations
\eqn{NptCasimirEqns}{
\left({\cal L}^{(1)} + \cdots + {\cal L}^{(K)}\right)^2
 {W}_{\Delta_{\delta_1};\:\ldots ;\:\Delta_{\delta_{n-3}}}^{\Delta_{1},\Delta_2,\ldots,\Delta_{n-1},\Delta_{n}}(x_1,\dots,x_n) &= C_2(\Delta_{\delta_{K-1}})  {W}_{\Delta_{\delta_1};\:\ldots ;\:\Delta_{\delta_{n-3}}}^{\Delta_{1},\Delta_2,\ldots,\Delta_{n-1},\Delta_{n}}(x_1,\ldots,x_n)
}
for all $2 \leq K \leq n-2$, where ${\cal L}_{AB}^{(i)}$ are the generators of the Euclidean conformal group $SO(d+1,1)$, realized as differential operators built out of and acting on the coordinate $x_i$, with the quadratic Casimir operator defined as $({\cal L}^{(i)})^2 \equiv {1\over 2} {\cal L}_{AB}^{(i)} {\cal L}^{AB(i)}$ (no sum over $i$). The eigenvalues are given by $C_2(\Delta) = \Delta(\Delta-d)$ for scalar exchange operators, which will be the case throughout this paper.
Moreover, the blocks satisfy the following OPE limit
\eqn{nptOPE}{
 \lim_{x_n \to x_{n-1}} \!\! {W}_{\Delta_{\delta_1};\:\ldots ;\:\Delta_{\delta_{n-3}}}^{\Delta_{1},\Delta_2,\ldots,\Delta_{n-1},\Delta_{n}}(x_1,\ldots,x_n)  
 = (x_{(n-1)n}^2)^{\Delta_{\delta_{n-3},(n-1)n}} 
{W}_{\Delta_{\delta_1};\:\ldots ;\:\Delta_{\delta_{n-4}}}^{\Delta_{1},\Delta_2,\ldots,\Delta_{n-2},\Delta_{\delta_{n-3}}}(x_1,\ldots,x_{n-1}),
 }
or pictorially,
\eqn{nptOPEpic}{
\lim_{x_n \to x_{n-1}} \!\!\! \musepic{\figNCombChannel} \!\!\!\!\!\! = \! (x_{(n-1)n}^2)^{\Delta_{\delta_{n-3},(n-1)n}} \!\!\!\!\!\!\!\!\!\!\!\! \musepic{\figNCombChannelOPE}
}
and symmetrically an analogous limit when $x_2 \to x_1$.
Further, the blocks have been normalized such that for $u_i \ll 1$ for all $1 \leq i \leq n-3$, and $w_{r;s} \approx 1$ for all $2 \leq r \leq s\leq n-1$, the $n$-point block has the leading behavior 
\eqn{Leading}{
 {W}_{\Delta_{\delta_1};\:\ldots ;\:\Delta_{\delta_{n-3}}}^{\Delta_{1},\ldots,\Delta_{n}}(x_i) \approx  W_0^{(n)}(x_i) \left( \prod_{i=1}^{n-3} u_i^{\Delta_{\delta_i} \over 2} \right) .
 }
 The final claim follows trivially from an alternate, rapidly convergent power series expansion of the conformal blocks presented in section~\ref{DISCUSSION}.

\section{Low-point examples of holographic duals}
\label{FIVESIX}
In this section we provide the simplest non-trivial demonstration of the new techniques.
First we will obtain the holographic dual of the six-point block in the comb channel, from which we shall recover the well-known purely boundary power series expansion of the four-point block.
A second non-trivial example is provided in appendix~\ref{SEVENCOMB} --- it focuses on the holographic dual for the comb channel seven-point block, from which a power series expansion is obtained for the five-point block.
In the next section, we will generalize these results to obtain the $n$-point comb channel conformal block.

\subsection{Holographic dual of the six-point block}
\label{SIXCOMB}

\begin{figure}
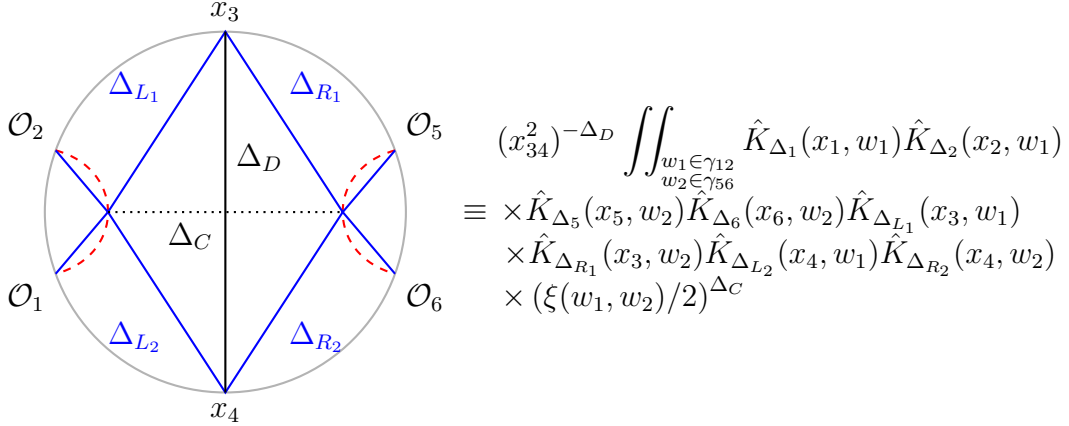

     \centering
     \[\musepic{\figcalWSample} \equiv  \begin{matrix} \displaystyle{  (x_{34}^2)^{-\Delta_D} \iint_{\substack{w_1 \in \gamma_{12}\\ w_2 \in \gamma_{56}}} \hat{K}_{\Delta_1}(x_1,w_1) \hat{K}_{\Delta_2}(x_2,w_1)} \cr
     \hspace{-.5cm} \times \hat{K}_{\Delta_5}(x_5,w_2) \hat{K}_{\Delta_6}(x_6,w_2)  \hat{K}_{\Delta_{L_1}}(x_3,w_1)  \cr 
      \times \hat{K}_{\Delta_{R_1}}(x_3,w_2)  \hat{K}_{\Delta_{L_2}}(x_4,w_1)  \hat{K}_{\Delta_{R_2}}(x_4,w_2) \cr 
     \hspace{-4.15cm} \times \left({\xi(w_1,w_2) / 2}\right)^{\Delta_C} 
     \end{matrix} \]
     \caption{{\it How to read comb-channel geodesic bulk diagrams.} Geodesic bulk diagrams (also referred to as geodesic Witten diagrams) are AdS Feynman diagrams  except with all bulk integrations restricted to boundary-anchored geodesics. Throughout this paper, boundary-anchored geodesics over which bulk points are to be integrated will be shown as {\it red-dashed} lines.
     In the diagram above, they represent the geodesics $\gamma_{12}$ and $\gamma_{56}$ joining $x_1$ to $x_2$ and $x_5$ to $x_6$, respectively.
    Bulk-to-boundary propagators $\hat{K}_\Delta(x,z)$ will be shown with {\it solid blue} lines, and whenever the conformal dimension $\Delta$ associated with it is not clear from the figure, it will be mentioned explicitly.
    For example, in the six-point geodesic diagram above, the four bulk-to-boundary propagators incident on bulk points to be integrated over boundary anchored geodesics are associated with the conformal dimensions $\Delta_i$ of the operator insertions ${\cal O}_i$ as marked. The remaining four bulk-to-boundary propagators emanating from the coordinates $x_3$ and $x_4$ have conformal dimensions as displayed next to the blue lines. Unless stated otherwise, the operator ${\cal O}_i$ is understood to be located at boundary coordinate $x_i$.
    {\it Solid black} lines will refer to purely boundary contractions; for example in the diagram above the solid black line joining $x_3$ to $x_4$ corresponds to a factor of $(x_{34}^2)^{-\Delta_D}$.
    Finally {\it dotted black} lines will stand for factors of chordal distance $(\xi(w_1,w_2)/2)^{\Delta}$ where $\xi(w_1,w_2)^{-1} = \cosh \sigma(w_1,w_2)$ where $\sigma(w_1,w_2)$ is the geodesic distance between bulk points $w_1$ and $w_2$. We will be using the same propagator normalizations as in ref.~\cite{Parikh:2019ygo}; see in particular~\cite[sec.~2]{Parikh:2019ygo} for the normalization of the bulk-to-boundary propagator as well as the relation between the bulk-to-bulk propagator and the chordal distance factor above.
 }
     \label{fig:geodesicdiag}
 \end{figure}

Before discussing the six-point block, let us first establish some notation by reviewing recent results for the five-point block.
In ref.~\cite{Parikh:2019ygo} a holographic geodesic diagram representation was worked out for the $d$-dimensional global scalar five-point conformal block $W_{\Delta_{\delta_1};\Delta_{\delta_2}}^{\Delta_1,\ldots,\Delta_5}(x_i)$. Such a block corresponds to external scalar insertions of dimensions $\Delta_1,\ldots,\Delta_5$, and represents the contribution coming from the exchange of scalar representations (and their conformal families) labelled by dimensions $\Delta_{\delta_1}$ and $\Delta_{\delta_2}$. The precise relation is,
\eqn{calW5confblock}{
{W}_{\Delta_{\delta_1};\:\Delta_{\delta_2}}^{\Delta_1,\ldots,\Delta_5}(x_i) &= \musepic{\figFiveCombChannel} \cr 
&= {4 \over  
B(\Delta_{\delta_1 1,2},\Delta_{\delta_1 2,1})\: B(\Delta_{\delta_2 4,5},\Delta_{\delta_2 5,4})} \:
{\cal W}_{\Delta_{\delta_1};\:\Delta_{\delta_2}}^{\Delta_1,\ldots,\Delta_5}(x_i) \,,
}
where $B(s,t) = \Gamma(s)\Gamma(t)/\Gamma(s+t)$ is the Euler Beta function, and ${\cal W}$ is a linear combination of five-point geodesic diagrams (see figure~\ref{fig:geodesicdiag} for notation and definition),
\eqn{calW5}{
{\cal W}_{\Delta_{\delta_1};\:\Delta_{\delta_2}}^{\Delta_1,\ldots,\Delta_5}(x_i) &= \sum_{k_1,k_2=0}^\infty c_{k_1,\:k_2}^{\Delta_{\delta_1},\:\Delta_{\delta_2};\:\Delta_3} \musepic{\figcalWFive},
}
with the coefficients $c_{k_1,\:k_2}^{\Delta_{\delta_1},\:\Delta_{\delta_2};\:\Delta_3}$ given by, 
\eqn{c5Def}{
c_{k_1,\:k_2}^{\Delta_{\delta_1},\:\Delta_{\delta_2};\:\Delta_3} &\equiv  {(-1)^{k_1+k_2} \over k_1!k_2!} 
{ \left(1-\Delta_{3\delta_2,\delta_1}\right)_{k_1} \left(1-\Delta_{3\delta_1,\delta_2}\right)_{k_2} \over
\left(\Delta_{\delta_1}-d/2+1\right)_{k_1} 
\left(\Delta_{\delta_2}-d/2+1\right)_{k_2}
} \cr 
& \times 
\left( \Delta_{3\delta_1,\delta_2} \right)_{k_{1,2}} 
\left(\Delta_{3\delta_2,\delta_1}\right)_{k_{2,1}} 
\left(\Delta_{\delta_1\delta_2,3}\right)_{k_{12,}} 
\cr 
& \times {}_3F_2\left[\{-k_1,-k_2,\Delta_{\delta_1\delta_2 3,}-d/2\}; \{\Delta_{3\delta_2,\delta_1}-k_1, \Delta_{3\delta_1,\delta_2}-k_2\};1\right]  .
 }
 The geodesic bulk diagram in~\eno{calW5} is a generalization of the geodesic Witten diagrams first obtained in the case of the four-point block~\cite{Hijano:2015zsa}.
 
The (holographic representation of the) conformal block  satisfies the Casimir equations~\eno{NptCasimirEqns} (for $n=5$ and $K=2,3$) and has the expected leading behavior in the two OPE limits.
Later in this section we will obtain an alternate geodesic diagram representation for the five-point block involving a single geodesic integral.
Moreover, in appendix~\ref{SEVENCOMB}, we will provide a  power series (purely CFT) representation of the five-point block obtained by taking a so-called double-OPE limit of the holographic dual of the seven-point block. 
In the rest of this section, we illustrate this procedure for the case of the six-point block where we recover the four-point block in the double-OPE limit.

\vspace{0.5em}
We first briefly discuss how to obtain the holographic representation of the six-point comb channel block. 
This discussion is light on technical details; we refer to reader to refs.~\cite{Parikh:2019ygo,Jepsen:2019svc} where the systematic procedure is described and several examples are worked out in detail.
The first step involves partially evaluating a particular AdS diagram involving cubic scalar couplings, in this case the diagram
\eqn{SixExchThree}{
\musepic{\figSixExchThree}\,,
}
where all internal (green) cubic vertices are to be integrated over all of ${\rm AdS}_{d+1}$.
In its direct-channel conformal block decomposition, the term corresponding to the exchange of three single-trace primaries takes the form
\eqn{SixExchThreeCBD}{
\musepic{\figSixExchThree} \!\!\!\supset C_{\Delta_1 \Delta_2 \Delta_{\delta_1}} C_{\Delta_{\delta_1} \Delta_3 \Delta_{\delta_2}} C_{\Delta_{\delta_2} \Delta_4 \Delta_{\delta_3}} C_{\Delta_{\delta_3} \Delta_5 \Delta_6}\!\!\! \musepic{\figSixCombChannel},
}
where $C_{\Delta_i \Delta_j \Delta_k}$ are the OPE coefficients associated with the dual generalized free-field CFT.
This is the six-point conformal block we are after. The trick to extracting the block is to use the split representation for the central bulk-to-bulk propagator as well as certain powerful two-propagator~\cite{Hijano:2015zsa} and three-propagator identities~\cite{Jepsen:2019svc}. These propagator identities are especially helpful in evaluating AdS diagrams in a way which makes their direct-channel conformal block decomposition manifest~\cite{Hijano:2015zsa,Jepsen:2019svc}. Performing the remaining the boundary and spectral integrals arising from the split representation, one manifestly isolates precisely the term shown above, proportional to the right combination of OPE coefficients.
The object multiplying this factor of OPE coefficients produces a candidate for the holographic geodesic diagram representation of the block.
Technically, the method outlined above is a heuristic derivation; the object obtained   must be independently checked to be the claimed conformal block, for example  via conformal Casimir equations.
We will comment on certain interesting implications of this derivation in section~\ref{DISCUSSION} in the context of obtaining multipoint conformal blocks in arbitrary channels.

The procedure outlined above leads  to the following 
 geodesic diagram representation, which we claim to be the six-point conformal block in the comb channel,
\eqn{calW6confblock}{
{W}_{\Delta_{\delta_1};\:\Delta_{\delta_2};\:\Delta_{\delta_3}}^{\Delta_1,\ldots,\Delta_6}(x_i) &= \musepic{\figSixCombChannel} \cr 
&= {4 \over  
B(\Delta_{\delta_1 1,2},\Delta_{\delta_1 2,1})\: B(\Delta_{\delta_3 5,6},\Delta_{\delta_3 6,5})} \:
{\cal W}_{\Delta_{\delta_1};\:\Delta_{\delta_2};\:\Delta_{\delta_3}}^{\Delta_1,\ldots,\Delta_6}(x_i) \,,
}
where the ${\cal W}$ is a linear combination of six-point geodesic diagrams (see figure~\ref{fig:geodesicdiag}), 
\eqn{calW6}{
{\cal W}_{\Delta_{\delta_1};\:\Delta_{\delta_2};\:\Delta_{\delta_3}}^{\Delta_1,\ldots,\Delta_6}(x_i) &= \sum_{k_1,k_2,k_3,j=0}^\infty \!\!\!\! c_{k_1,\:k_2,\:k_3;\:j}^{\Delta_{\delta_1},\: \Delta_{\delta_2},\:\Delta_{\delta_3};\:\Delta_3,\:\Delta_4}\!\!\!\!\!\!\!\!\!\!\!\! \musepic{\figcalWSix},
}
with the coefficients given by
\eqn{c6Def}{
c_{k_1,\:k_2,\:k_3;\:j}^{\Delta_{\delta_1},\: \Delta_{\delta_2},\:\Delta_{\delta_3};\:\Delta_3,\:\Delta_4} &\equiv  \lambda_6 {(-1)^{k_1+k_3+j} \over k_1!k_2!k_3!j!} 
{ \left(1-\Delta_{3\delta_2,\delta_1}\right)_{k_1}\! \left(1-\Delta_{3\delta_1,\delta_2}\right)_{k_2}\! 
\left(1-\Delta_{4\delta_3,\delta_2}\right)_{k_2} \! \left(1-\Delta_{4\delta_2,\delta_3}\right)_{k_3} 
\over
\left(\Delta_{\delta_1}-d/2+1\right)_{k_1} 
\left(\Delta_{\delta_2}-d/2+1\right)_{k_2}
\left(\Delta_{\delta_3}-d/2+1\right)_{k_3}
} \cr 
& \times 
\left( \Delta_{3\delta_1,\delta_2} \right)_{k_{1,2}} 
\left(\Delta_{3\delta_2,\delta_1}\right)_{k_{2,1}+j} 
\left(\Delta_{4\delta_2,\delta_3}\right)_{k_{2,3}+j} 
\left( \Delta_{4\delta_3,\delta_2} \right)_{k_{3,2}} 
\left(\Delta_{\delta_1\delta_3,34}\right)_{k_{13,}-j} 
\cr 
& \times {}_3F_2\left[\{-k_1,-k_2,\Delta_{\delta_1\delta_2 3,}-d/2\}; \{\Delta_{3\delta_2,\delta_1}-k_1, \Delta_{3\delta_1,\delta_2}-k_2\};1\right] \cr 
& \times {}_3F_2\left[\{-k_2,-k_3,\Delta_{\delta_2\delta_3 4,}-d/2\}; \{\Delta_{4\delta_3,\delta_2}-k_2, \Delta_{4\delta_2,\delta_3}-k_3\};1\right] ,
 }
and  
\eqn{lambda6Def}{
\lambda_6 \equiv {\Gamma(1-\Delta_{\delta_1 \delta_2,3}) \Gamma(1-\Delta_{\delta_2 \delta_3,4}) \over \Gamma(1-\Delta_{\delta_1 \delta_3,34}) \Gamma(1-\Delta_{\delta_2}) }\,. 
}
It is worth pointing out the close structural similarity of the geodesic bulk diagram as well as the functional form of the coefficients with the five-point example. In particular, the five-point geodesic diagram had one ``triangle'' formed by two blue lines (bulk-to-boundary propagators) and a single dotted line (factor of chordal distance), and precisely one factor of the hypergeometric ${}_3F_2$ function in the coefficient, while the six-point block has two triangles and two factors of ${}_3F_2$.\footnote{The four-point case has no triangles and no factors of ${}_3F_2$ in its double power series expansion.} We will return to this observation in section~\ref{DISCUSSION}.

To prove the claim above we need to  show that the conformal block satisfies the right differential equations~\eno{NptCasimirEqns} (for $n=6$, $K=2,3,4$) 
with the right boundary conditions, expressed in terms of the OPE limits~\eno{nptOPE}, reproduced below for convenience
\eqn{OPElim6}{
{W}_{\Delta_{\delta_1};\:\Delta_{\delta_2};\:\Delta_{\delta_3}}^{\Delta_1,\ldots,\Delta_6}(x_1,x_2,x_3,x_4,x_5,x_6) &\stackrel{x_2 \to x_1}{\longrightarrow} (x_{12}^2)^{\Delta_{\delta_1,12}}\: W_{\Delta_{\delta_2};\:\Delta_{\delta_3}}^{\Delta_{\delta_1},\Delta_3,\Delta_4,\Delta_5,\Delta_6}(x_1,x_3,x_4,x_5,x_6) \cr
{W}_{\Delta_{\delta_1};\:\Delta_{\delta_2};\:\Delta_{\delta_3}}^{\Delta_1,\ldots,\Delta_6}(x_1,x_2,x_3,x_4,x_5,x_6) &\stackrel{x_6 \to x_5}{\longrightarrow} (x_{56}^2)^{\Delta_{\delta_3,56}}\: W_{\Delta_{\delta_1};\:\Delta_{\delta_2}}^{\Delta_1,\Delta_2,\Delta_3,\Delta_4,\Delta_{\delta_3}}(x_1,x_2,x_3,x_4,x_5)\,.
}
The conformal Casimir check is set up in embedding space and  the calculation proceeds identically to the conformal Casimir checks for the holographic representations of the five-point block~\cite{Parikh:2019ygo} and the six-point block in the OPE channel~\cite{Jepsen:2019svc}. 
Since no new ingredients are required, we refrain from including the somewhat lengthy albeit straightforward proof and simply point the reader to Refs.~\cite{Parikh:2019ygo,Jepsen:2019svc} for reference.\footnote{The Casimir check relies on a particular non-trivial functional identity obeyed by the hypergeometric function~${}_3F_2$ which was also previously employed in the five-point case~\cite{Parikh:2019ygo},
\eqn{3F2id}{
   E(D-B)\: \pFq{3}{2}{A+1,B,C}{D+1,E}{1} + B(E-C)\: \pFq{3}{2}{A+1,B+1,C}{D+1,E+1}{1} - D E\: \pFq{3}{2}{A,B,C}{D,E}{1} = 0\,.
}
The only difference is that due to two factors of the hypergeometric function in~\eno{c6Def}, this identity must now be applied twice, once for each factor.} 
In lieu of that, in the next section we will provide a conformal Casimir proof for the {\it power series} expansion of the six-point block which will be obtained later.

In the remainder of this section, we would like to focus on the OPE limits~\eno{OPElim6} of the six-point block~\eno{calW6confblock}-\eno{calW6}.
If this is indeed the six-point conformal block as claimed, taking a single OPE limit should produce an (alternate) holographic representation for the five-point block involving a single geodesic integral, and taking a further OPE limit should reproduce the power series expansion of the four-point block purely in terms of boundary coordinates. 
We will utilize this strategy in the next section to go from the holographic representation of the $(n+2)$-point conformal block in the comb channel to produce an explicit power series expansion of the $n$-point comb channel block. 
For illustrative purposes, a seven-point to five-point block example is provided in appendix~\ref{SEVENCOMB}.

\subsection{OPE limit of the six-point block}
\label{SIXOPE}

Generically, all OPE limits in this paper will involve integrals of the following kind:
\eqn{OPElimGeneral}{
& \lim_{x_2 \to x_1} \int_{w\in \gamma_{12}} \hat{K}_{\Delta_1}(x_1,w) \hat{K}_{\Delta_2}(x_2,w) \left(\prod_{i=1}^j \hat{K}_{\Delta_{a_i}}(x_{a_i},w) \right) \left(\prod_{i=1}^k \left({\xi(w_{b_i},w) \over 2} \right)^{\Delta_{b_i}} \right) \cr 
 &= { B(\Delta_{a_1 \ldots a_j b_1 \ldots b_k 1,2},\Delta_{a_1 \ldots a_j b_1 \ldots b_k 2,1}) \over 2} { (x_{12}^2)^{\Delta_{a_1 \ldots a_j b_1 \ldots b_k,12}} \over \prod_{i=1}^j (x_{1a_i}^2)^{\Delta_{a_i}}} \left(\prod_{i=1}^k \hat{K}_{\Delta_{b_i}}(x_1, w_{b_i}) \right),
}
 which is straightforward to derive (assuming convergence of the integral).
In this section we focus on the OPE limit $x_2 \to x_1$ of the six-point block. Due to the symmetry of the six-point block, we need only check one of the OPE limits in~\eno{OPElim6}; the other one follows immediately from a simple relabeling. 
Using~\eno{OPElimGeneral}, we have
\eqn{x2x1OPE}{
{W}_{\Delta_{\delta_1};\:\Delta_{\delta_2};\:\Delta_{\delta_3}}^{\Delta_1,\ldots,\Delta_6}(x_i) \stackrel{x_2 \to x_1}{\longrightarrow} & \!\!\!\!\!  \sum_{k_1,k_2,k_3,j=0}^\infty  \!\!\!\!\! (x_{12}^2)^{\Delta_{\delta_1,12}+k_1} {2B(\Delta_{\delta_1 1,2}+k_1,\Delta_{\delta_1 2,1}+k_1) \over  
B(\Delta_{\delta_1 1,2},\Delta_{\delta_1 2,1})\: B(\Delta_{\delta_3 5,6},\Delta_{\delta_3 6,5})}  c_{k_1,\:k_2,\:k_3;\:j}^{\Delta_{\delta_1},\: \Delta_{\delta_2},\:\Delta_{\delta_3};\:\Delta_3,\:\Delta_4} \cr 
& \times \musepic{\figcalWFivePreAlt}\,.
}
We are interested in the leading behavior as $x_2 \to x_1$, so we set the non-negative integral parameter $k_1=0$ above. Comparing with the first line of~\eno{OPElim6}, it is sufficient to show that  this holographic representation is indeed the five-point  conformal block expected to be obtained in this limit, that is
\eqn{W5altToShow}{
{W}_{\Delta_{\delta_2};\:\Delta_{\delta_3}}^{\Delta_{\delta_1},\Delta_3,\Delta_4,\Delta_5,\Delta_6}(x_1,x_3,x_4,x_5,x_6) & \stackrel{!}{=} {2 \over B(\Delta_{\delta_3 5, 6}, \Delta_{\delta_3 6,5})} \sum_{k_2,k_3,j=0}^\infty c_{0,\:k_2,\:k_3;\:j}^{\Delta_{\delta_1},\: \Delta_{\delta_2},\:\Delta_{\delta_3};\:\Delta_3,\:\Delta_4} \cr 
 &\times \musepic{\figcalWFiveAlt} \equiv V \,.
}
This is easily checked by showing  that the RHS above, which we call $V$, satisfies the differential equations and boundary conditions of a five-point block. Specifically,
\eqn{CasimirV}{
\left({\cal L}^{(1)} + {\cal L}^{(3)}\right)^2
V &= C_2(\Delta_{\delta_2})\: V \cr 
\left({\cal L}^{(1)} + {\cal L}^{(3)} + {\cal L}^{(4)}\right)^2
V &= C_2(\Delta_{\delta_3})\: V\,,
}
with $V$ reducing to four-point blocks  in the OPE limit,
\eqn{OPElimV}{
V &\stackrel{x_3 \to x_1}{\longrightarrow} (x_{13}^2)^{\Delta_{\delta_2,\delta_1 3}}\: W_{\Delta_{\delta_3}}^{\Delta_{\delta_2},\Delta_4,\Delta_5,\Delta_6}(x_1,x_4,x_5,x_6) \cr
V &\stackrel{x_6 \to x_5}{\longrightarrow} (x_{56}^2)^{\Delta_{\delta_3,56}}\: W_{\Delta_{\delta_2}}^{\Delta_{\delta_1},\Delta_3,\Delta_4,\Delta_{\delta_3}}(x_1,x_3,x_4,x_5)\,.
}
Just as discussed near the end of  section~\ref{SIXCOMB}, the conformal Casimir check~\eno{CasimirV} is once again straightforward to show using directly the techniques of ref.~\cite{Parikh:2019ygo} and the functional identity~\eno{3F2id}.
In the interest of keeping the length of this paper reasonable, we will refrain from presenting the technical details.

The OPE limits~\eno{OPElimV} are also straightforward to work out. Interestingly, the first OPE limit in~\eno{OPElimV}  leads to the (holographic) geodesic bulk diagram representation of the four-point block, and the second furnishes a (boundary) power series expansion. We discuss these limits next.

\subsubsection{Recovering the four-point block from the six-point block}

Let's first discuss the limit $x_3 \to x_1$. In this limit, using~\eno{OPElimGeneral} we find that $V$ is proportional to $(x_{13}^2)^{\Delta_{\delta_2,\delta_1 3}+k_2}$ where $k_2$ is summed over non-negative integers. Thus the leading order contribution comes from setting $k_2=0$, which gives
\eqn{Vx3x1}{
V &\stackrel{x_3 \to x_1}{\longrightarrow}  {2\:  (x_{13}^2)^{\Delta_{\delta_2,\delta_1 3}} \over B(\Delta_{\delta_3 5, 6}, \Delta_{\delta_3 6,5})} \sum_{k_3,j=0}^\infty c_{0,\:0,\:k_3;\:j}^{\Delta_{\delta_1},\: \Delta_{\delta_2},\:\Delta_{\delta_3};\:\Delta_3,\:\Delta_4} \musepic{\figVlimOne} \,.
}
The free sum over $j$ is easily performed to give
\eqn{Vx3x1Again}{
&V \stackrel{x_3 \to x_1}{\longrightarrow}  {2\:  (x_{13}^2)^{\Delta_{\delta_2,\delta_1 3}} \over B(\Delta_{\delta_3 5, 6}, \Delta_{\delta_3 6,5})} \sum_{k_3=0}^\infty  \frac{(\Delta_{\delta_3\delta_2,4})_{k_3} (\Delta_{\delta_34,\delta_2})_{k_3}}{k_3! (\Delta_{\delta_3} - d/2+1)_{k_3}}
\musepic{\figVlimOne} \cr 
 &= {4(x_{13}^2)^{\Delta_{\delta_2,\delta_1 3}} \over B(\Delta_{\delta_3 5, 6}, \Delta_{\delta_3 6,5}) B(\Delta_{\delta_3\delta_2,4}, \Delta_{\delta_34,\delta_2})}\: \sum_{k_3=0}^\infty  \frac{ (\Delta_{\delta_3})_{2k_3}}{k_3! (\Delta_{\delta_3} - d/2+1)_{k_3}} \!\!\!\!\!\!\!\!\!
\musepic{\figcalWfourXi} \cr
 &= (x_{13}^2)^{\Delta_{\delta_2,\delta_1 3}}\: W_{\Delta_{\delta_3}}^{\Delta_{\delta_2},\Delta_4,\Delta_5,\Delta_6}(x_1,x_4,x_5,x_6) \,,
}
where to get to the second equality we used ref.~\cite[eqn.~(A.4)]{Parikh:2019ygo} which re-expresses a particular combination of bulk-to-boundary propagators in the first equality as a geodesic integral. 
Performing the $k_3$ sum by using the relation between the chordal distance measure $\xi$ and the bulk-to-bulk propagator (see e.g.\ ref.~\cite[eqn.~(2.8)]{Parikh:2019ygo} for the precise relation in our conventions), one immediately recovers the original geodesic diagram representation of the four-point block~\cite{Hijano:2015zsa}.

Now let's consider the other OPE limit in~\eno{OPElimV}. As $x_6 \to x_5$, the leading contribution is given by
\eqn{Vx6x5}{
V &\stackrel{x_6 \to x_5}{\longrightarrow}   (x_{56}^2)^{\Delta_{\delta_3,56}}\sum_{k_2,j=0}^\infty c_{0,\:k_2,\:0;\:j}^{\Delta_{\delta_1},\: \Delta_{\delta_2},\:\Delta_{\delta_3};\:\Delta_3,\:\Delta_4}  \musepic{\figVlimTwo} \,,
}
with the subleading contributions suppressed by higher positive powers of $x_{56}^2$. Recalling from figure~\ref{fig:geodesicdiag} that lines joining points on the boundary of the Poincar\'{e} disk represent factors of the form $(x_{34}^2)^{j}$ etc., this limit can be written explicitly as
\eqn{Vx6x5Again}{
V &\stackrel{x_6 \to x_5}{\longrightarrow}   
{(x_{56}^2)^{\Delta_{\delta_3,56}} \over (x_{13}^2)^{\Delta_{\delta_1 3,}} (x_{45}^2)^{\Delta_{4\delta_3,}}} 
\left({x_{35}^2 \over x_{15}^2 }\right)^{\Delta_{\delta_1,3}}
\left({x_{15}^2 \over x_{14}^2 }\right)^{\Delta_{4,\delta_3}}
u^{\Delta_{\delta_2} \over 2}
\sum_{k_2,j=0}^\infty c_{0,\:k_2,\:0;\:j}^{\Delta_{\delta_1},\:
\Delta_{\delta_2},\:\Delta_{\delta_3};\:\Delta_3,\:\Delta_4}   u^{k_2} v^{j} \,,
}
where the conformal cross-ratios are defined to be
\eqn{uvVlim}{
u \equiv {x_{13}^2 x_{45}^2 \over x_{14}^2 x_{35}^2 } \qquad v \equiv { x_{15}^2 x_{34}^2 \over x_{14}^2 x_{35}^2 }\,.
}
Up to the expected overall factor of $ (x_{56}^2)^{\Delta_{\delta_3,56}}$,~\eno{Vx6x5Again} is of the same form as~\eno{nCombBlock} except for a different set of coordinate labels and conformal dimensions. More precisely,
\eqn{}{
V &\stackrel{x_6 \to x_5}{\longrightarrow}  (x_{56}^2)^{\Delta_{\delta_3,56}} {W}_{\Delta_{\delta_2}}^{\Delta_{\delta_1},\Delta_3,\Delta_4,\Delta_{\delta_3}}(x_1,x_3,x_4,x_5)\,,
}
in the notation described after~\eno{nCombBlock}.
To show that~\eno{Vx6x5Again} is indeed the power series expansion of the global four-point block as claimed, it helps to bring it to a more familiar form, by rewriting the series expansion in terms of powers of $(1-v)$,\footnote{\label{fn:rearrange}The standard trick to do that is as follows: We expand
\eqn{}{
v^j = (1+v-1)^j = \sum_{\ell=0}^j {j \choose \ell} (-1)^\ell (1-v)^\ell\,.
}
Then extending the upper limit of the binomial sum above to infinity with impunity and switching the order of $\ell$ and $j$ sums,  perform the summation over $j$  in~\eno{Vx6x5Again} to obtain a power series expansion in powers of $(1-v)$.}
\eqn{Vx6x5Final}{
V &\stackrel{x_6 \to x_5}{\longrightarrow}   (x_{56}^2)^{\Delta_{\delta_3,56}} \:
{W}_0^{\Delta_{\delta_1},\Delta_3,\Delta_4,\Delta_{\delta_3}}(x_1,x_3,x_4,x_5) \cr 
&\qquad \times u^{\Delta_{\delta_2} \over 2}
\sum_{k_2,\ell=0}^\infty  { u^{k_2}  (1-v)^{\ell} \over k_2! \ell!} 
{(\Delta_{\delta_2 \delta_1,3})_{k_2} (\Delta_{\delta_2 \delta_3,4})_{k_2} \over (\Delta_{\delta_2} -d/2+1)_{k_2}} {(\Delta_{\delta_2 3,\delta_1})_{k_2+\ell} (\Delta_{\delta_2 4,\delta_3})_{k_2+\ell} \over (\Delta_{\delta_2})_{2k_2+\ell}}\,,
}
where the leg-factor $W_0$ was defined in~\eno{W0nDef}.
Up to the overall factor of  $(x_{56}^2)^{\Delta_{\delta_3,56}}$, this is precisely the power series expansion of the appropriate global scalar four-point block~\cite{Dolan:2003hv}, thus confirming the second line of~\eno{OPElimV}.

To conclude, we emphasize the main results of this section: We obtained and verified a \emph{holographic} representation of the six-point block in the comb channel~\eno{calW6confblock}-\eno{lambda6Def}, and in the double-OPE limit $x_2 \to x_1, x_n\to x_{n-1}$ for $n=6$, we recovered the explicit \emph{power series} expansion of the four-point block~\eno{Vx6x5}.
In appendix~\ref{SEVENCOMB}, we provide another example of this --- the holographic dual of the seven-point block leading to the power series expansion of the five-point block.
In the next section, we will generalize this result to obtain the power series expansion of the $n$-point comb channel block from  a similar double-OPE limit of the holographic dual of the $(n+2)$-point comb channel block.

\section{Multipoint block in the comb channel}
\label{NCOMB}

\subsection{Holographic dual of the $(n+2)$-point block and its double-OPE limit}

The low-point examples of five- and six-point blocks in the previous section, along with the seven-point example in appendix~\ref{SEVENCOMB} allow us to make a guess for the {\it holographic} representation of any multipoint block in the comb channel. 
We conjecture  the holographic representation of the $(n+2)$-point comb channel block ($n \geq 3$), labelled as follows,
\eqn{calWnPlus2confblock}{
{W}_{\Delta_1;\:\Delta_{\delta_1};\:\ldots ;\:\Delta_{\delta_{n-3}};\:\Delta_{n}}^{\Delta_{e_1},\Delta_{e_2},\Delta_2,\ldots,\Delta_{n-1},\Delta_{e_3},\Delta_{e_4}}(x_i) &\equiv \musepic{\figNPlusTwoCombChannel} \cr 
&= {4\: {\cal W}_{\Delta_1;\:\Delta_{\delta_1};\:\ldots ;\:\Delta_{\delta_{n-3}};\:\Delta_{n}}^{\Delta_{e_1},\Delta_{e_2},\Delta_2,\ldots,\Delta_{n-1},\Delta_{e_3},\Delta_{e_4}}(x_i) \over  
B(\Delta_{1 e_1,e_2},\Delta_{1 e_2,e_1})\: B(\Delta_{n e_3,e_4},\Delta_{n e_4,e_3})} \,, 
}
to be given by ${\cal W}$, which is the following linear combination of geodesic bulk diagrams (see figure~\ref{fig:geodesicdiag} for the graphical notation and below for more notes),
\eqn{calWnPlus2}{
& {\cal W}_{\Delta_1;\:\Delta_{\delta_1};\:\ldots ;\:\Delta_{\delta_{n-3}};\:\Delta_{n}}^{\Delta_{e_1},\Delta_{e_2},\Delta_2,\ldots,\Delta_{n-1},\Delta_{e_3},\Delta_{e_4}}(x_i) = \sum_{\substack{k_0,k_1,k_2,\ldots,k_{n-3},k_{n-2},\\j_{\langle 2|3\rangle},\:j_{\langle 2|4\rangle},\:\ldots,\:j_{\langle n-2|n-1\rangle}=0}}^\infty \!\!\!\!\!\!\!\!
c_{k_0,\:k_1,\:k_2,\:\ldots,\:k_{n-3},\:k_{n-2};\:j_{\langle 2|3\rangle},\:j_{\langle 2|4\rangle},\:\ldots,\:j_{\langle n-2|n-1\rangle}}^{\Delta_1,\:\Delta_{\delta_1},\: \Delta_{\delta_2},\:\ldots,\:\Delta_{\delta_{n-3}},\:\Delta_n;\:\Delta_2,\:\Delta_3,\:\ldots,\:\Delta_{n-1}} \cr  
& \quad \times \musepic{\figcalWnplustwoA} \times \left(\prod_{2 \leq r < s \leq n-1} \musepic{\figcalWnplustwoB} \right)\cr 
& \quad  
\times \left(\prod_{t=0}^{n-3} \musepic{\figcalWnplustwoD} \right) ,
}
where the indices ${\langle r|s\rangle}$ represent the $\binom{n-2}{2}$ values the subscript on $j$ can take. 
The coefficients are given by (for $n\geq 3$)
\eqn{cnPlus2Def}{
& c_{k_0,\:k_1,\:\ldots,\:k_{n-3},\:k_{n-2};\:j_{\langle 2|3\rangle},\:j_{\langle 2|4\rangle},\:\ldots,\:j_{\langle n-2|n-1\rangle}}^{\Delta_1,\:\Delta_{\delta_1},\:\ldots,\:\Delta_{\delta_{n-3}},\:\Delta_n;\:\Delta_2,\:\ldots,\:\Delta_{n-1}} \equiv \lambda_{n+2}   \left( \prod_{2 \leq r < s \leq n-1} {(-1)^{j_{\langle r|s\rangle}} \over j_{\langle r|s\rangle}!} \right) \left( \prod_{i=0}^{n-2} {1 \over k_i!} \right) \cr 
 & \times \left( \prod_{i=1}^{n-3}  { \left(1-\Delta_{(i+1)\delta_{i-1},\delta_{i}}\right)_{k_{i}} \left(1-\Delta_{(i+2)\delta_{i+1},\delta_{i}}\right)_{k_{i}} \over  \left(\Delta_{\delta_{i}} - d/2+1 \right)_{k_{i}} } \right) 
 {\left(1-\Delta_{2\delta_1,1}\right)_{k_0} \left(1-\Delta_{(n-1)\delta_{n-3},n}\right)_{k_{n-2}} \over \left(\Delta_1 -d/2+1\right)_{k_0} \left(\Delta_n-d/2+1\right)_{k_{n-2}} } \cr 
 &\times
 (-1)^{k_0 + k_{n-2}} \left( \Delta_{1n,2\ldots(n-1)} \right)_{k_{0(n-2),}-\sum_{2 \leq r < s \leq n-1} j_{\langle r|s\rangle}} \cr 
& \times \left( \prod_{t=0}^{n-3}
\left(\Delta_{(t+2)\delta_t,\delta_{t+1}} \right)_{k_{t,t+1}+\sum_{r=2}^{t+1}j_{\langle r|t+2\rangle}}
\left(\Delta_{(t+2)\delta_{t+1},\delta_{t}} \right)_{k_{t+1,t}+\sum_{s=t+3}^{n-1}j_{\langle t+2|s\rangle}} 
\right. \cr 
& \quad  \times  {}_3F_2\left[\left\{-k_t,-k_{t+1},\Delta_{\delta_t\delta_{t+1} (t+2),}-{d\over 2}\right\}; \left\{\Delta_{(t+2)\delta_{t+1},\delta_t}-k_t, \Delta_{(t+2)\delta_t,\delta_{t+1}}-k_{t+1}\right\};1\right] \Bigg) ,
}
with
\eqn{lambdanPlus2Def}{
\lambda_{n+2} \equiv  {  \prod_{t=0}^{n-3} \Gamma(1-\Delta_{\delta_t \delta_{t+1},(t+2)} ) \over \Gamma(1-\Delta_{1n,2\ldots (n-1)})\prod_{i=1}^{n-3} \Gamma(1-\Delta_{\delta_i})} \,,
}
and the additional definitions
\eqn{}{
\Delta_{\delta_0} \equiv \Delta_1 \qquad \Delta_{\delta_{n-2}} \equiv \Delta_n\,.
}
A few remarks are in order regarding the conjecture for the $(n+2)$-point block above:
\begin{itemize}
    \item For clarity we have split up a single geodesic bulk diagram in~\eno{calWnPlus2} into a chain of constituent factors. The first factor should be familiar to the reader from the geodesic diagram representation of a four-point block~\cite{Hijano:2015zsa} (see also~\eno{Vx3x1Again}). The second factor, corresponding to $\binom{n-2}{2}$ contractions (in fact, a perfect graph) between boundary points $x_2, \ldots, x_{n-1}$, appears in the holographic dual for all $n \geq 4$ (i.e.\ for the six-point block and higher). The third factor, which represents a product over $(n-2)$ pairs of bulk-to-boundary propagators, is present in the holographic dual for all $n \geq 3$ (i.e.\ five-point block and higher). 
    \item All these constituent factors are to be merged together and should be understood as having been drawn on the same Poincar\'{e} disk. The bulk points $w, w^\prime$, which are to be integrated over the two boundary anchored geodesics have been marked explicitly to emphasize that the bulk-to-boundary propagators are incident on precisely the same bulk points; thus there are only two geodesic integrals to be performed.
    \item It is straightforward to check that this conjecture reduces to the already established holographic duals for the five-point block~\eno{calW5confblock}-\eno{c5Def}, the six-point block~\eno{calW6confblock}-\eno{lambda6Def}, and the seven-point block~\eno{calW7confblock}-\eno{lambda7Def}.
\end{itemize}

If this conjecture is true, its double-OPE limit, $x_{1^\prime} \to x_1, x_{n^\prime} \to x_n$   should lead to the $n$-point block (for $n \geq 4$) described in figure~\ref{fig:nblock}, up to some expected overall scaling factors,
\eqn{nPlus2ptOPE}{
 \lim_{\substack{x_{1^\prime} \to x_1\\ x_{n^\prime} \to x_{n}}} \!\!  &{W}_{\Delta_1;\:\Delta_{\delta_1};\:\ldots ;\:\Delta_{\delta_{n-3}};\:\Delta_{n}}^{\Delta_{e_1},\Delta_{e_2},\Delta_2,\ldots,\Delta_{n-1},\Delta_{e_3},\Delta_{e_4}}(x_{1^\prime}, x_1,\ldots,x_n, x_{n^\prime})  \cr 
 &= (x_{11^\prime}^2)^{\Delta_{e_1 e_2,1}} (x_{nn^\prime}^2)^{\Delta_{e_3 e_4,n}} 
{W}_{\Delta_{\delta_1};\:\ldots ;\:\Delta_{\delta_{n-3}}}^{\Delta_{1},\Delta_2,\ldots,\Delta_{n-1},\Delta_{n}}(x_1,\ldots,x_{n})\,,
 }
 where, using~\eno{OPElimGeneral},\footnote{In particular, the integer parameters $k_0$ and $k_{n-2}$ are set to zero to obtain the leading contribution on the RHS of~\eno{nPlus2ptOPE}.} the explicit representation can be worked out to be (for $n \geq 4$)
\eqn{nCombSeries}{
&{W}_{\Delta_{\delta_1};\:\ldots ;\:\Delta_{\delta_{n-3}}}^{\Delta_{1},\Delta_2,\ldots,\Delta_{n-1},\Delta_{n}}(x_i)  = \sum_{\substack{k_1,k_2,\ldots,k_{n-3},\\j_{\langle 2|3\rangle},\:j_{\langle 2|4\rangle},\:\ldots,\:j_{\langle n-2|n-1\rangle}=0}}^\infty c_{0,k_1,\:k_2,\:\ldots,\:k_{n-3},\:0;\:j_{\langle 2|3\rangle},\:j_{\langle 2|4\rangle},\:\ldots,\:j_{\langle n-2|n-1\rangle}}^{\Delta_1,\:\Delta_{\delta_1},\: \Delta_{\delta_2},\:\ldots,\:\Delta_{\delta_{n-3}},\:\Delta_n;\:\Delta_2,\:\Delta_3,\:\ldots,\:\Delta_{n-1}} \cr  
& \qquad \times \musepic{\figWnA} \times\;\left(\prod_{2 \leq r < s \leq n-1} \musepic{\figcalWnplustwoB} \right) \cr 
& \qquad \times 
\left( \prod_{t=0}^{n-3} \musepic{\figWnD} \right),
}
where we now additionally impose the identifications~\eno{kdeltaExtraDef}.
Once again the chain of Poincar\'{e} disks above is interpreted as explained for the geodesic diagram representation. 
The coefficients themselves simplify to (for $n \geq 4$),
\eqn{cn}{
& c_{0,k_1,\:\ldots,\:k_{n-3},\:0;\:j_{\langle 2|3\rangle},\:j_{\langle 2|4\rangle},\:\ldots,\:j_{\langle n-2|n-1\rangle}}^{\Delta_1,\:\Delta_{\delta_1},\:\ldots,\:\Delta_{\delta_{n-3}},\:\Delta_n;\:\Delta_2,\:\ldots,\:\Delta_{n-1}} = \lambda_{n+2} \left( \prod_{2 \leq r < s \leq n-1} {(-1)^{j_{\langle r|s\rangle}} \over j_{\langle r|s\rangle}!} \right) 
\left( \prod_{i=1}^{n-3} {1 \over k_i!} \right) \cr 
& \times 
  \left( \Delta_{1n,2\ldots(n-1)} \right)_{-\sum_{2 \leq r < s \leq n-1} j_{\langle r|s\rangle}} 
  \left( \prod_{t=0}^{n-3}  { \left(1-\Delta_{(t+1)\delta_{t-1},\delta_{t}}\right)_{k_{t}} \left(1-\Delta_{(t+2)\delta_{t+1},\delta_{t}}\right)_{k_{t}} \over  \left(\Delta_{\delta_{t}} - d/2+1 \right)_{k_{t}} } \right.
 \cr 
&  \times
\left(\Delta_{(t+2)\delta_t,\delta_{t+1}} \right)_{k_{t,t+1}+\sum_{r=2}^{t+1}j_{\langle r|t+2\rangle}}
\left(\Delta_{(t+2)\delta_{t+1},\delta_{t}} \right)_{k_{t+1,t}+\sum_{s=t+3}^{n-1}j_{\langle t+2|s\rangle}}  \cr 
 &  \times   {}_3F_2\left[\left\{-k_t,-k_{t+1},\Delta_{\delta_t\delta_{t+1} (t+2),}-{d\over 2}\right\}; \left\{\Delta_{(t+2)\delta_{t+1},\delta_t}-k_t, \Delta_{(t+2)\delta_t,\delta_{t+1}}-k_{t+1}\right\};1\right] \Bigg) , 
}
where $\lambda_{n+2}$ is given in~\eno{lambdanPlus2Def} and we employed~\eno{kdeltaExtraDef} to write the coefficient above compactly.\footnote{For example, for $n=4$ one obtains the simplified coefficients
\eqn{c4}{
 c_{0,k_1,\:0;\:j_{\langle 2|3 \rangle}}^{\Delta_1,\:\Delta_{\delta_1},\:\Delta_4;\:\Delta_2,\:\Delta_{3}} &= \lambda_{6}    {(-1)^{j_{\langle 2|3 \rangle}} \over j_{\langle 2|3 \rangle}!}   {1 \over k_1!} 
{  \left(\Delta_{2\delta_1,1} \right)_{k_1+j_{\langle 2|3 \rangle}} 
 \left(\Delta_{3\delta_{1},4} \right)_{k_{1} +j_{\langle 2|3 \rangle}} 
  \left( \Delta_{14,23} \right)_{-j_{\langle 2|3 \rangle}}   \over  \left(\Delta_{\delta_1} - d/2+1 \right)_{k_1} } 
  \,.
}
}

The coordinate dependence of the putative conformal block is captured pictorially in~\eno{nCombSeries}. Recalling the notation from figure~\ref{fig:geodesicdiag}, one can easily convert this to conformal cross-ratios as follows,
\eqn{nCombSeriesExpressAlt}{
{W}_{\Delta_{\delta_1};\:\ldots ;\:\Delta_{\delta_{n-3}}}^{\Delta_{1},\ldots,\Delta_{n}}(x_i) & = W_0^{(n)}(x_i) \left( \prod_{i=1}^{n-3} u_i^{\Delta_{\delta_i} \over 2} \right) \!\!\! \sum_{\substack{k_1,\ldots,k_{n-3},\\j_{\langle 2|3\rangle},\:j_{\langle 2|4\rangle},\:\ldots,\:j_{\langle n-2|n-1\rangle}=0}}^\infty \!\!\!\!\!\!\!\!\!\!\!\!\!\!\! c_{0,k_1,\:\ldots,\:k_{n-3},\:0;\:j_{\langle 2|3\rangle},\:j_{\langle 2|4\rangle},\:\ldots,\:j_{\langle n-2|n-1\rangle}}^{\Delta_1,\:\Delta_{\delta_1},\:\ldots,\:\Delta_{\delta_{n-3}},\:\Delta_n;\:\Delta_2,\:\ldots,\:\Delta_{n-1}} \cr 
& \qquad \qquad \qquad \qquad \qquad \qquad \qquad \qquad \times \left( \prod_{i=1}^{n-3} u_i^{k_i} \right) \left( \prod_{2\leq r < s \leq n-1} w_{r;s}^{j_{\langle r|s\rangle}} \right),
}
where the ``leg factor'' $W_0^{(n)}(x_i)$ is given by~\eno{W0nDef} 
and the cross-ratios $u_i, w_{r;s}$ are defined in~\eno{uwDef}. 
This is precisely the power series expansion~\eno{nCombBlock}.\footnote{Alternately, one can write it as
\eqn{nCombSeriesExpress}{
{W}_{\Delta_{\delta_1};\:\ldots ;\:\Delta_{\delta_{n-3}}}^{\Delta_{1},\ldots,\Delta_{n}}(x_i) & = W_0^{(n)}(x_i) \left( \prod_{i=1}^{n-3} u_i^{\Delta_{\delta_i} \over 2} \right) \!\!\! \sum_{\substack{k_1,\ldots,k_{n-3},\\j_{\langle 2|3\rangle},\:j_{\langle 2|4\rangle},\:\ldots,\:j_{\langle n-2|n-1\rangle}=0}}^\infty \!\!\!\!\!\!\!\!\!\!\!\!\!\!\! c_{0,k_1,\:\ldots,\:k_{n-3},\:0;\:j_{\langle 2|3\rangle},\:j_{\langle 2|4\rangle},\:\ldots,\:j_{\langle n-2|n-1\rangle}}^{\Delta_1,\:\Delta_{\delta_1},\:\ldots,\:\Delta_{\delta_{n-3}},\:\Delta_n;\:\Delta_2,\:\ldots,\:\Delta_{n-1}} \cr 
& \qquad\qquad\quad\qquad \qquad \qquad \qquad \qquad \times \left( \prod_{i=1}^{n-3} u_i^{k_i} v_i^{j_{\langle i+1|i+2\rangle}} \right) \left( \prod_{\substack{2\leq r < s \leq n-1\\ s\neq r+1}} w_{r;s}^{j_{\langle r|s\rangle}} \right),
}
where we disallowed $s=r+1$ in the $w_{r;s}$ cross-ratios, and collect them separately into the cross-ratios
\eqn{vDef}{
 v_i  \equiv w_{i+1;i+2} = { x_{1n}^2 x_{(i+1)(i+2)}^2 \over x_{(i+1)n}^2 x_{1(i+2)}^2}  \qquad\qquad 1 \leq i \leq n-3\,.
}
}

While we do not directly prove the conjecture~\eno{calWnPlus2confblock}-\eno{cnPlus2Def} for the holographic representation of the $(n+2)$-point block as this is not the main focus of this paper, in the remainder of this section, we will prove via conformal Casimir equations~\eno{NptCasimirEqns}-\eno{nptOPE} that~\eno{nCombSeriesExpressAlt} is indeed the desired power series expansion of the $n$-point comb channel block.

\subsection{OPE limit of the $n$-point block}
\label{NCOMBOPE}

Let's first verify the boundary conditions. 
Due to the symmetrical nature of the conjectural conformal block~\eno{nCombSeries}, we need only check one of the two OPE limits. We choose to work out the limit $x_{n}\to x_{n-1}$. It is easily gleaned from the diagrammatic representation~\eno{nCombSeries}  that the power series expansion is proportional to a factor of $(x_{(n-1)n}^2)^{k_{n-3}-\Delta_{(n-1)n,\delta_{n-3}}}$, so we set $k_{n-3}=0$ to obtain the leading contribution in the OPE limit. 
It is straightforward to work out the full expansion,
\eqn{nCombSeriesOPE}{
& \lim_{x_n \to x_{n-1}} {W}_{\Delta_{\delta_1};\:\ldots ;\:\Delta_{\delta_{n-3}}}^{\Delta_{1},\ldots,\Delta_{n}}(x_i)  \to (x_{(n-1)n}^2)^{\Delta_{\delta_{n-3},(n-1)n}} 
\!\!\!\!\!\!\!\!\!\!\!\!\!\!\!\!\!\!\!\!\! \sum_{\substack{k_1,k_2,\ldots,k_{n-4},\\j_{\langle 2|3\rangle},\:j_{\langle 2|4\rangle},\:\ldots,\:j_{\langle n-2|n-1\rangle}=0}}^\infty
\!\!\!\!\!\!\!\!\!\!\!\!\!\!\!\!\!\!\!\!\! c_{0,k_1,\:k_2,\:\ldots,\:k_{n-4},\:0,\:0;\:j_{\langle 2|3\rangle},\:j_{\langle 2|4\rangle},\:\ldots,\:j_{\langle n-2|n-1\rangle}}^{\Delta_1,\:\Delta_{\delta_1},\: \Delta_{\delta_2},\:\ldots,\:\Delta_{\delta_{n-3}},\:\Delta_n;\:\Delta_2,\:\Delta_3,\:\ldots,\:\Delta_{n-1}} \cr  
& \qquad  \times \musepic{\figWnAp} \times \left(\prod_{2 \leq r < s \leq n-2} \musepic{\figWnBp} \right) \cr 
& \qquad  \times \left(\prod_{t=0}^{n-4} \musepic{\figWnDp} \right) ,
}
where remember that $\Delta_{\delta_0} = \Delta_1$ and $k_0=k_{n-3}=0$ as well. 
Comparing with~\eno{nCombSeries}, we observe that~\eno{nCombSeriesOPE} has precisely the right position space dependence of the $(n-1)$-point conformal block obtained in the OPE limit $x_n \to x_{n-1}$. 
Indeed, when $k_{n-3}=0$, it is straightforward to check that the coefficient in~\eno{nCombSeries} can be expressed in terms of the coefficient associated with the $(n-1)$-point block times some ``extraneous factors'' of Euler Gamma function and Pochhammer symbol, as shown:
\eqn{cnOPE}{
& c_{0,k_1,\:\ldots,\:k_{n-4},\:0,\:0;\:j_{\langle 2|3\rangle},\: j_{\langle 2|4 \rangle},\:\ldots,\:j_{\langle n-2|n-1\rangle}}^{\Delta_1,\:\Delta_{\delta_1},\:\ldots,\:\Delta_{\delta_{n-3}},\:\Delta_n;\:\Delta_2,\:\ldots,\:\Delta_{n-1}} \cr 
&=  c_{0,k_1,\:\ldots,\:k_{n-4},\:0;\:j_{\langle 2|3\rangle},\: j_{\langle 2|4 \rangle},\:\ldots,\:j_{\langle n-3|n-2\rangle}}^{\Delta_1,\:\Delta_{\delta_1},\:\ldots,\:\Delta_{\delta_{n-4}},\:\Delta_{\delta_{n-3}};\:\Delta_2,\:\ldots,\:\Delta_{n-2}} 
{\Gamma\left(1- \Delta_{\delta_{n-3}n,(n-1)}\right) \Gamma\left(1-\Delta_{1\delta_{n-3},2\ldots (n-2)}\right)  \over \Gamma\left(1-\Delta_{1n,2\ldots (n-1)}\right) \Gamma\left(1-\Delta_{\delta_{n-3}}\right) } 
\cr 
 & \times  \left( \prod_{\ell=2}^{n-2} {(-1)^{j_{\langle \ell|n-1\rangle}} \over j_{\langle \ell|n-1\rangle}!} \right)
  \left( \prod_{t=0}^{n-4} \left(\Delta_{(t+2)\delta_{t+1},\delta_{t}} + k_{t+1,t}+\sum_{s=t+3}^{n-2}j_{\langle t+2|s\rangle} \right)_{j_{\langle t+2|n-1 \rangle}}  \right)
 \cr 
 &\times  \left(\Delta_{(n-1)\delta_{n-3},n} \right)_{\sum_{r=2}^{n-2}j_{\langle r|n-1\rangle}} 
 {   \left( \Delta_{1n,2\ldots(n-1)} \right)_{-\sum_{2 \leq r < s \leq n-1} j_{\langle r|s\rangle}}  \over   \left( \Delta_{1\delta_{n-3},2\ldots(n-2)} \right)_{-\sum_{2 \leq r < s \leq n-2} j_{\langle r|s\rangle}}  } 
   \,,
}
where we remember to set $\delta_0 = \Delta_1$ and $k_0=k_{n-3}=0$.
Thus, we obtain the correct OPE limit~\eno{nptOPE}-\eno{nptOPEpic},
provided the following $(n-3)$-dimensional coordinate independent sum over the indices $j_{\langle 2|n-1\rangle}, \ldots, j_{\langle n-2|n-1\rangle}$, which involves  the ``extraneous factors'' in~\eno{cnOPE}, evaluates to unity:
\eqn{ToShowSum}{
  1 &\stackrel{!}{=} 
 {\Gamma\left(1- \Delta_{\delta_{n-3}n,(n-1)}\right) \Gamma\left(1-\Delta_{1\delta_{n-3},2\ldots (n-2)}\right)  \over \Gamma\left(1-\Delta_{1n,2\ldots (n-1)}\right) \Gamma\left(1-\Delta_{\delta_{n-3}}\right) } 
\cr 
 & \times
  \sum_{\substack{j_{\langle 2|n-1\rangle}, \ldots,\\j_{\langle n-2|n-1\rangle}=0}}^\infty  \!\!\!\!\!
   \left( \prod_{\ell=2}^{n-2} {(-1)^{j_{\langle \ell|n-1\rangle}} \over j_{\langle \ell|n-1\rangle}!} \right)
  \left( \prod_{t=0}^{n-4} \left(\Delta_{(t+2)\delta_{t+1},\delta_{t}} + k_{t+1,t}+\sum_{s=t+3}^{n-2}j_{\langle t+2|s\rangle} \right)_{j_{\langle t+2|n-1 \rangle}}  \right)
 \cr 
 &\times  \left(\Delta_{(n-1)\delta_{n-3},n} \right)_{\sum_{r=2}^{n-2}j_{\langle r|n-1\rangle}} 
 {   \left( \Delta_{1n,2\ldots(n-1)} \right)_{-\sum_{2 \leq r < s \leq n-1} j_{\langle r|s\rangle}}  \over   \left( \Delta_{1\delta_{n-3},2\ldots(n-2)} \right)_{-\sum_{2 \leq r < s \leq n-2} j_{\langle r|s\rangle}}  } 
 \,.
}
Indeed, this turns out to be the case, as shown in appendix~\ref{USEFUL2F1ID}.

\subsection{Conformal Casimir check}
\label{CASIMIR}

Having established the OPE limits for the $n$-point conjecture, we now turn our attention to the multipoint Casimir equations~\eno{NptCasimirEqns}, repeated below:
\eqn{NptCasimirEqnsRpt}{
\left({\cal L}^{(1)} + \cdots + {\cal L}^{(K)}\right)^2
 {W}_{\Delta_{\delta_1};\:\ldots ;\:\Delta_{\delta_{n-3}}}^{\Delta_{1},\Delta_2,\ldots,\Delta_{n-1},\Delta_{n}}(x_1,\dots,x_n) &= C_2(\Delta_{\delta_{K-1}})  {W}_{\Delta_{\delta_1};\:\ldots ;\:\Delta_{\delta_{n-3}}}^{\Delta_{1},\Delta_2,\ldots,\Delta_{n-1},\Delta_{n}}(x_1,\ldots,x_n)
}
for all $2 \leq K \leq n-2$, where ${\cal L}_{AB}^{(r)}$ are the generators of the Euclidean conformal group acting on and built out of the coordinate $x_r$, $A,B$ are $SO(d+1,1)$ indices, and $({\cal L}^{(r)})^2 = 1/2 {\cal L}^{(r)}_{AB} {\cal L}^{(r)AB}$. In embedding space $\mathbb{R}^{d+1,1}$, these generators act linearly as Lorentz group generators in $d+2$ dimensions,
\eqn{}{
{\cal L}_{AB}^{(r)} = -i \left( X_{A}^r {\partial \over \partial X^{rB}} - X_{B}^r {\partial \over \partial X^{rA}} \right)\,,
}
where $X^r \in \mathbb{R}^{d+1,1}$ are embedding space coordinates which upon taking an appropriate  section give the corresponding  Poincar\'{e} coordinates $x_r$.  In the rest of this section,  $x_r$ will refer to a boundary coordinate so that $X^r$ will be a null coordinate in embedding space with $X^r \cdot X^r = 0$ (no sum over $r$).
While we will not explicitly state it every time, all calculations are worked out in embedding space because of the convenience of working with linear differential operators. Various useful formulae relevant for such computations can be found in ref.~\cite[sec.~2.2]{Parikh:2019ygo}.

Expand the multipoint Casimir operator
\eqn{CasimirExpand}{
\left({\cal L}^{(1)} + \cdots + {\cal L}^{(K)}\right)^2 = \sum_{i=1}^K \left({\cal L}^{(i)} \right)^2 + \sum_{1 \leq r <s \leq K} {\cal L}^{(r)}_{AB} {\cal L}^{(s)AB}\,.
}
First, the following action of the quadratic Casimir on the conjectural conformal block~\eno{nCombSeries} is easily checked:
\eqn{Lsqr}{
\left({\cal L}^{(i)} \right)^2  {W}_{\Delta_{\delta_1};\:\ldots ;\:\Delta_{\delta_{n-3}}}^{\Delta_{1},\Delta_2,\ldots,\Delta_{n-1},\Delta_{n}}(x_1,\ldots,x_n) = m_{\Delta_i}^2  {W}_{\Delta_{\delta_1};\:\ldots ;\:\Delta_{\delta_{n-3}}}^{\Delta_{1},\Delta_2,\ldots,\Delta_{n-1},\Delta_{n}}(x_1,\ldots,x_n)  \,,
}
for all $1 \leq i \leq n$ and for all $n$. This follows from the explicit conformal dimension assignment of~\eno{nCombSeries} and the following obvious identity
\eqn{LsqrId}{
\left({\cal L}^{(x)} \right)^2 \musepic{\figLsqrId} = m^2_{\sum_{i=1}^\ell \Delta_i} \musepic{\figLsqrId}\,,
}
where as explained in figure~\ref{fig:geodesicdiag}, the solid lines joining together boundary points are to be interpreted as a boundary contractions of the form $(x_{ij}^2)^{-\Delta}$.
The action of the cross-term in~\eno{CasimirExpand} on the putative conformal block~\eno{nCombSeries} involves terms of the form
\eqn{TypicalTerm}{
I \equiv {\cal L}^{(r)}_{AB} {\cal L}^{(s)AB} \left[\musepic{\figLcrossIdOne} \left(\prod_{i=1}^{n-2} \musepic{\figLcrossIdTwo}\right)\right],
}
with  $1 \leq r < s \leq K$ fixed, and $1 \leq a_1 < a_2 <\ldots < a_{n-2} \leq n$ with $a_1,a_2,\ldots, a_{n-2} \notin \{r, s\}$. Like in the previous subsections, the pictorial representation above is split into a product of Poincar\'{e} disks for clarity, but should be understood as merged into a single diagram. We are using the symbols $\langle a|b \rangle$ as indices labelling conformal dimensions with $\Delta_{\langle a|b \rangle} = \Delta_{\langle b|a \rangle}$.
Based on convenience, we will sometimes index $a_i$ using the subscripts $i=1,\ldots,n-2$, and at other times directly as $a_i = 1,\ldots, n$ with the restriction that $a_i \neq r,s$.

To evaluate $I$, we decompose it as
\eqn{IDef}{
I = I_1 + I_2 + I_3\,,
 }
where we have defined
\begingroup
\allowdisplaybreaks
\begin{align*}
I_1 &\equiv \left(\prod_{i=1}^{n-2}\musepic{\figLcrossIdTwo}\right)  {\cal L}^{(r)}_{AB} {\cal L}^{(s)AB} \musepic{\figLcrossIdOne} \\
I_2 &\equiv \musepic{\figLcrossIdOne}\quad  {\cal L}^{(r)}_{AB} {\cal L}^{(s)AB} \left(\prod_{i=1}^{n-2}\musepic{\figLcrossIdTwo}\right) \\ 
I_3 &\equiv \left( {\cal L}^{(r)}_{AB} \musepic{\figLcrossIdOne} \right) \left( {\cal L}^{(s)AB} - {\cal L}^{(r)AB} \right) \left(\prod_{i=1}^{n-2}\musepic{\figLcrossIdTwo}\right) .
\stepcounter{equation}\tag{\theequation}\label{IiDef}
\end{align*}
\endgroup
In writing $I_3$ we used the obvious fact that
\eqn{}{
{\cal L}^{(s)}_{AB} \musepic{\figLcrossIdOne} = - {\cal L}^{(r)}_{AB} \musepic{\figLcrossIdOne}\,.
}
In fact, using this it is straightforward to evaluate $I_1$, since
\eqn{}{
 {\cal L}^{(r)}_{AB} {\cal L}^{(s)AB} \musepic{\figLcrossIdOne} = -2m_{\Delta_{\langle r|s\rangle}}^2 \musepic{\figLcrossIdOne}\,.
}
To evaluate $I_2$ and $I_3$, we use the identity
\eqn{UsefulId}{
{\cal L}_{AB}^{(s)} & \left(\prod_{i=1}^{n-2}\musepic{\figLcrossIdTwo}\right) \cr 
&= \left( \sum_{\ell=1}^{n-2} {-i\Delta_{\langle a_\ell|s \rangle} \left(2X^s_A X^{a_\ell}_B - 2X^s_B X^{a_\ell}_A \right) \over (-2X^s \cdot X^{a_\ell}) } \right) \left(\prod_{i=1}^{n-2}\musepic{\figLcrossIdTwo}\right),
}
which is easily checked. 
Here we have made explicit reference to the embedding space coordinates in the prefactor.\footnote{\label{fn:embedding}At the end of the calculation, we will re-express all expressions in terms of $d$-dimensional boundary coordinates, using the simple identification $(-2X^i \cdot X^j) = x_{ij}^2$. }
Then, it follows
\eqn{}{
I_2 &=  -2\left( \sum_{u,v=1}^{n-2} {\Delta_{\langle a_u|s \rangle}\Delta_{\langle a_v|r\rangle} \left((-2X^s \cdot X^r)(-2 X^{a_u}\cdot X^{a_v}) - ((-2X^s\cdot X^{a_v}) (-2X^r \cdot X^{a_u}) \right) \over (-2X^s \cdot X^{a_u}) (-2X^r\cdot X^{a_v})} \right) \cr 
&\quad \times \left[\musepic{\figLcrossIdOne} \left(\prod_{i=1}^{n-2} \musepic{\figLcrossIdTwo}\right)\right] \cr 
I_3 &= -2 \Delta_{\langle r|s\rangle} \sum_{\ell=1}^{n-2}(\Delta_{\langle a_\ell|s\rangle}+\Delta_{\langle a_\ell|r\rangle}) \left[\musepic{\figLcrossIdOne} \left(\prod_{i=1}^{n-2} \musepic{\figLcrossIdTwo}\right)\right] .
}

Now rewrite the putative $n$-point conformal block~\eno{nCombSeries} as
\eqn{nCombSeriesAlt}{
{W}_{\Delta_{\delta_1};\:\ldots ;\:\Delta_{\delta_{n-3}}}^{\Delta_{1},\Delta_2,\ldots,\Delta_{n-1},\Delta_{n}}(x_i)  &= \sum_{\substack{k_1,k_2,\ldots,k_{n-3},\\j_{\langle 2|3\rangle},\:j_{\langle 2|4\rangle},\:\ldots,\:j_{\langle n-2|n-1\rangle}=0}}^\infty \!\!\!\!\!\!\!\!\!\! c_{0,k_1,\:k_2,\:\ldots,\:k_{n-3},\:0;\:j_{\langle 2|3\rangle},\:j_{\langle 2|4\rangle},\:\ldots,\:j_{\langle n-2|n-1\rangle}}^{\Delta_1,\:\Delta_{\delta_1},\: \Delta_{\delta_2},\:\ldots,\:\Delta_{\delta_{n-3}},\:\Delta_n;\:\Delta_2,\:\ldots,\:\Delta_{n-1}} \cr 
&\quad\times \left(\prod_{1 \leq u < v \leq n} \musepic{\figWAlt}\right),
}
where the conformal dimensions $\Delta_{\langle u|v\rangle}$, which depend on $k_1,\ldots,k_{n-3}$, and various $j_{\langle \cdot|\cdot \rangle}$ can be read-off of~\eno{nCombSeries}. Then, using the computations above, we can evaluate
\eqn{}{
& {\cal L}^{(r)}_{AB} {\cal L}^{(s)AB} 
 {W}_{\Delta_{\delta_1};\:\ldots ;\:\Delta_{\delta_{n-3}}}^{\Delta_{1},\Delta_2,\ldots,\Delta_{n-1},\Delta_{n}}(x_i)   = -2\!\!\!\!\!\!\!\!\!\!\! \sum_{\substack{k_1,k_2,\ldots,k_{n-3},\\j_{\langle 2|3\rangle},\:j_{\langle 2|4\rangle},\:\ldots,\:j_{\langle n-2|n-1\rangle}=0}}^\infty \!\!\!\!\!\!\!\!\!\!\!\!\!\!\!\!\!\! c_{0,k_1,\:k_2,\:\ldots,\:k_{n-3},\:0;\:j_{\langle 2|3\rangle},\:j_{\langle 2|4\rangle},\:\ldots,\:j_{\langle n-2|n-1\rangle}}^{\Delta_1,\:\Delta_{\delta_1},\: \Delta_{\delta_2},\:\ldots,\:\Delta_{\delta_{n-3}},\:\Delta_n;\:\Delta_2,\:\ldots,\:\Delta_{n-1}} \cr 
 &\times \left(  m_{\Delta_{\langle r|s\rangle}}^2 + \Delta_{\langle r|s\rangle}\sum_{\ell=1}^{n-2} (\Delta_{ \langle a_\ell|s\rangle}+\Delta_{\langle a_\ell|r\rangle})  + \sum_{u,v=1}^{n-2} \Delta_{\langle a_u|s \rangle}\Delta_{\langle a_v|r\rangle}{ \left( x_{rs}^2 x_{a_u a_v}^2  - x_{sa_v}^2 x_{ra_u}^2 \right) \over x_{sa_u}^2 x_{ra_v}^2 }   \right) \cr 
 &\times \left(\prod_{1 \leq u < v \leq n} \musepic{\figWAlt}\right),
}
where as before $1 \leq r < s \leq K$ for a fixed $K$ satisfying $2 \leq K \leq  n-2$, and $1 \leq a_1 < a_2 < \ldots < a_{n-3} < a_{n-2} \leq n$ with $a_1,a_2,\ldots, a_{n-2} \notin \{r, s\}$. 

Combining all the computations (see also footnote~\ref{fn:embedding}), the left hand side of~\eno{NptCasimirEqnsRpt} evaluates to
\eqn{}{
 LHS &= \sum_{\substack{k_1,k_2,\ldots,k_{n-3},\\j_{\langle 2|3\rangle},\:j_{\langle 2|4\rangle},\:\ldots,\:j_{\langle n-2|n-1\rangle}=0}}^\infty \!\!\!\!\!\!\!\!\!\!\!\!\!\!\!\!\!\!  c_{0,k_1,\:\ldots,\:k_{n-3},\:0;\:j_{\langle 2|3\rangle},\:j_{\langle 2|4\rangle},\:\ldots,\:j_{\langle n-2|n-1\rangle}}^{\Delta_1,\:\Delta_{\delta_1},\:\ldots,\:\Delta_{\delta_{n-3}},\:\Delta_n;\:\Delta_2,\:\ldots,\:\Delta_{n-1}} 
 \left(\prod_{1 \leq u < v \leq n} \musepic{\figWAlt}\right) \cr 
 & \times \left[ 
 \sum_{i=1}^K m_{\Delta_i}^2 
 -2 \sum_{1 \leq r<s \leq K} \left( m_{\Delta_{\langle r|s\rangle}}^2  +   \Delta_{\langle r|s\rangle}\sum_{\ell=1}^{n-2}(\Delta_{\langle a_\ell|s\rangle }+\Delta_{\langle a_\ell|r\rangle})  \right. \right. \cr
 & \left. \left. \quad +\sum_{u,v=1}^{n-2} \Delta_{\langle a_u|s \rangle}\Delta_{\langle a_v|r\rangle}{ \left( x_{rs}^2 x_{a_u a_v}^2  - x_{sa_v}^2 x_{ra_u}^2 \right) \over x_{sa_u}^2 x_{ra_v}^2 }  \right) 
 \right],
}
where we used $\sum_{\substack{u=1\\ u \neq v}}^{n}  \Delta_{\langle u|v \rangle} = \Delta_v$ for all $1 \leq v \leq n$, which can be easily checked by reading~$ \Delta_{\langle u|v \rangle}$ off of~\eno{nCombSeries}.
Conveniently, a particular partial sum above vanishes,\footnote{More precisely, the following partial sum vanishes,
\eqn{PartialSum}{
 \sum_{1 \leq r<s \leq K}\sum_{\substack{a_u,a_v=1 \\  a_u, a_v \notin \{r,s\},a_u \neq a_v}}^{K} \Delta_{\langle a_u|s \rangle}\Delta_{\langle a_v|r\rangle}{ \left( x_{rs}^2 x_{a_u a_v}^2  - x_{sa_v}^2 x_{ra_u}^2 \right) \over x_{sa_u}^2 x_{ra_v}^2 } = 0 \,,
}
which is proven in appendix~\ref{APP:PARTIALSUM}.} leaving us with
\eqn{}{
LHS &= \sum_{\substack{k_1,k_2,\ldots,k_{n-3},\\j_{\langle 2|3\rangle},\:j_{\langle 2|4\rangle},\:\ldots,\:j_{\langle n-2|n-1\rangle}=0}}^\infty \!\!\!\!\!\!\!\!\!\!\!\!\!\!\!\!\!\!  c_{0,k_1,\:\ldots,\:k_{n-3},\:0;\:j_{\langle 2|3\rangle},\:j_{\langle 2|4\rangle},\:\ldots,\:j_{\langle n-2|n-1\rangle}}^{\Delta_1,\:\Delta_{\delta_1},\:\ldots,\:\Delta_{\delta_{n-3}},\:\Delta_n;\:\Delta_2,\:\ldots,\:\Delta_{n-1}} 
 \left(\prod_{1 \leq u < v \leq n} \musepic{\figWAlt}\right) \cr
 &\times \left[ 
 \sum_{i=1}^K m_{\Delta_i}^2 -2 \sum_{1 \leq r<s \leq K} \left( m_{\Delta_{\langle r|s\rangle}}^2  +   \Delta_{\langle r|s\rangle}\sum_{\ell=1}^{n-2}(\Delta_{\langle a_\ell|s\rangle} + \Delta_{ \langle a_\ell|r\rangle})  - \sum_{\ell=1}^{n-2} \Delta_{\langle a_\ell|s\rangle} \Delta_{\langle a_\ell|r\rangle} \right.
 \right. \cr
 &\quad + \left. \left. 
 \sum_{\substack{a_u,a_v=K+1 \\ a_u \neq a_v}}^{n} \Delta_{\langle a_u|s \rangle}\Delta_{\langle a_v|r\rangle}{ \left( x_{rs}^2 x_{a_u a_v}^2  - x_{sa_v}^2 x_{ra_u}^2 \right) \over x_{sa_u}^2 x_{ra_v}^2 }  \right) 
 \right] .
}
The position space dependence above can be re-expressed in terms of the cross-ratios defined in~\eno{uwDef} by employing the form~\eno{nCombSeriesExpressAlt} of the $n$-point block and observing that
\eqn{}{
 {x_{sr}^2 x_{a_u a_v}^2 \over x_{s a_u}^2 x_{ra_v}^2} &= 
\begin{cases} \displaystyle{u_{s-1} u_s \ldots u_{a_u-2} {w_{r;s} w_{a_u;a_v} \over w_{s;a_u} w_{r;a_v}}} & \qquad K+1 \leq a_u < a_v \leq n-1  \\ 
\displaystyle{u_{s-1} u_s \ldots u_{a_v-2} {w_{r;s} w_{a_v;a_u} \over w_{s;a_u} w_{r;a_v}}} & \qquad K+1 \leq a_v < a_u \leq n-1  \\ 
    \displaystyle{u_{s-1} u_s \ldots u_{a_u-2} {w_{r;s} \over  w_{s;a_u}}}        & \qquad K+1 \leq a_u \leq n-1 , a_v = n\\ 
    \displaystyle{u_{s-1} u_s \ldots u_{a_v-2} {w_{r;s} \over  w_{r;a_v}}}        & \qquad  K+1 \leq a_v \leq n-1, a_u = n
\end{cases}\cr 
  {x_{sa_v}^2 x_{ra_u}^2 \over x_{sa_u}^2 x_{ra_v}^2} &=  {w_{s;a_v} w_{r;a_u} \over w_{s;a_u} w_{r;a_v}}\,,
}
for $1 \leq r<s \leq K$, and a fixed $K$ satisfying $2 \leq K \leq n-2$.
Substituting this back in the left hand side, we get
\eqn{LHSuw}{
LHS&= W_0^{(n)}(x_i) \left( \prod_{i=1}^{n-3} u_i^{\Delta_{\delta_i} \over 2} \right) \!\!\! \sum_{\substack{k_1,\ldots,k_{n-3},\\j_{\langle 2|3\rangle},\:j_{\langle 2|4\rangle},\:\ldots,\:j_{\langle n-2|n-1\rangle}=0}}^\infty \!\!\!\!\!\!\!\!\!\!\!\!\!\!\! c_{0,k_1,\:\ldots,\:k_{n-3},\:0;\:j_{\langle 2|3\rangle},\:j_{\langle 2|4\rangle},\:\ldots,\:j_{\langle n-2|n-1\rangle}}^{\Delta_1,\:\Delta_{\delta_1},\:\ldots,\:\Delta_{\delta_{n-3}},\:\Delta_n;\:\Delta_2,\:\ldots,\:\Delta_{n-1}} \cr 
  &\times \Bigg\{
 \sum_{i=1}^K m_{\Delta_i}^2 -2 \sum_{1 \leq r<s \leq K} \Bigg[ m_{\Delta_{\langle r|s\rangle}}^2  +   \Delta_{\langle r|s\rangle}\sum_{\ell=1}^{n-2}(\Delta_{\langle r|a_\ell\rangle} + \Delta_{ \langle s|a_\ell\rangle})  - \sum_{\ell=1}^{n-2} \Delta_{\langle r|a_\ell\rangle} \Delta_{\langle s|a_\ell\rangle} 
 \cr
 &\quad +
 \sum_{K+1 \leq a_u < a_v \leq n-1} \Delta_{\langle r|a_v\rangle}\Delta_{\langle s|a_u \rangle} \left(  u_{s-1} u_s \ldots u_{a_u-2} {w_{r;s} w_{a_u;a_v} \over w_{s;a_u} w_{r;a_v}}  -  {w_{s;a_v} w_{r;a_u} \over w_{s;a_u} w_{r;a_v}} \right)  \cr 
  &\quad +
 \sum_{K+1 \leq a_v < a_u \leq n-1} \Delta_{\langle r|a_v\rangle}\Delta_{\langle s|a_u \rangle} \left(  u_{s-1} u_s \ldots u_{a_v-2} {w_{r;s} w_{a_v;a_u} \over w_{s;a_u} w_{r;a_v}}  -  {w_{s;a_v} w_{r;a_u} \over w_{s;a_u} w_{r;a_v}} \right)  \cr
 &\quad +
 \sum_{a_u=K+1}^{n-1} \Delta_{\langle r|n\rangle}\Delta_{\langle s|a_u \rangle} \left(  u_{s-1} u_s \ldots u_{a_u-2} {w_{r;s}  \over w_{s;a_u} }  -  {w_{r;a_u} \over w_{s;a_u}} \right) \cr 
  &\quad +
 \sum_{a_v=K+1}^{n-1} \Delta_{\langle r|a_v\rangle}\Delta_{\langle s|n \rangle} \left(  u_{s-1} u_s \ldots u_{a_v-2} {w_{r;s}  \over w_{r;a_v} }  -  {w_{s;a_v} \over w_{r;a_v}} \right)
 \Bigg]  \Bigg\}
 \left( \prod_{i=1}^{n-3} u_i^{k_i} \right) \left( \prod_{2\leq r < s \leq n-1} w_{r;s}^{j_{\langle r|s\rangle}} \right).
}
This must be shown to equal the right hand side of~\eno{NptCasimirEqnsRpt}, which can be written as
\eqn{RHSuw}{
RHS &= m_{\Delta_{\delta_{K-1}}}^2  W_0^{(n)}(x_i) \left( \prod_{i=1}^{n-3} u_i^{\Delta_{\delta_i} \over 2} \right) \!\!\! \sum_{\substack{k_1,\ldots,k_{n-3},\\j_{\langle 2|3\rangle},\:j_{\langle 2|4\rangle},\:\ldots,\:j_{\langle n-2|n-1\rangle}=0}}^\infty  \!\!\!\!\!\!\!\!\!\!\!\!\!\!\!  c_{0,k_1,\:\ldots,\:k_{n-3},\:0;\:j_{\langle 2|3\rangle},\:j_{\langle 2|4\rangle},\:\ldots,\:j_{\langle n-2|n-1\rangle}}^{\Delta_1,\:\Delta_{\delta_1},\:\ldots,\:\Delta_{\delta_{n-3}},\:\Delta_n;\:\Delta_2,\:\ldots,\:\Delta_{n-1}} \cr 
& \times    \left( \prod_{i=1}^{n-3} u_i^{k_i} \right) \left( \prod_{2\leq r < s \leq n-1} w_{r;s}^{j_{\langle r|s\rangle}} \right),
}
after substituting $C_2(\Delta_{\delta_{K-1}}) = m^2_{\Delta_{\delta_{K-1}}}$. 

By integer shifting appropriate integral parameters $k_i, j_{\langle \cdot|\cdot \rangle}$, we can make the position space dependence of the combination $(LHS-RHS)$ identical across individual terms. For instance, if a term  contains an extra factor of $u_i$, then shift $k_i \to k_i - 1$. Similarly, if a term has an extra factor of $w_{r;s}$, shift $j_{\langle r|s\rangle} \to j_{\langle r|s\rangle} -1$.
Subsequently, after some further simplifications (see appendix~\ref{APP:CASIMIR} for calculational details), we end up with
\eqn{LHSmRHS}{
LHS-RHS &=   W_0^{(n)}(x_i) \left( \prod_{i=1}^{n-3} u_i^{\Delta_{\delta_i} \over 2} \right) \!\!\! \sum_{\substack{k_1,\ldots,k_{n-3},\\j_{\langle 2|3\rangle},\:j_{\langle 2|4\rangle},\:\ldots,\:j_{\langle n-2|n-1\rangle}=0}}^\infty  \!\!\!\!\!\!\!\!\!\!\!\!\!\!\!\Bigg( 4\: k_{{K-1}}(\delta_{K-1}+k_{K-1} -d/2)\: c_{(\cdot)}  \cr 
&  \quad -4 \sum_{\substack{1 \leq r<s \leq K \\  \\ K+1 \leq a_u < a_v \leq n}} \!\!\!
    \widetilde{c}^{r,s; a_u,a_v}_{(k_{s-1}-1, k_s -1, \ldots, k_{a_u-2}-1,j_{\langle r|s\rangle}-1, j_{\langle a_u|a_v\rangle}-1,j_{\langle s|a_u\rangle}+1,j_{\langle r|a_v\rangle}+1) }  \Bigg) \cr 
& \times    \left( \prod_{i=1}^{n-3} u_i^{k_i} \right) \left( \prod_{2\leq r < s \leq n-1} w_{r;s}^{j_{\langle r|s\rangle}} \right),
}
where we have defined the scaled (tilded) coefficient
\eqn{ctDef}{
\widetilde{c}_{0,k_1,\:\ldots,\:k_{n-3},\:0;\:j_{\langle 2|3\rangle},\:j_{\langle 2|4\rangle},\:\ldots,\:j_{\langle n-2|n-1\rangle}}^{r,s; a_u,a_v}  \equiv c_{0,k_1,\:\ldots,\:k_{n-3},\:0;\:j_{\langle 2|3\rangle},\:j_{\langle 2|4\rangle},\:\ldots,\:j_{\langle n-2|n-1\rangle}}^{\Delta_1,\:\Delta_{\delta_1},\:\ldots,\:\Delta_{\delta_{n-3}},\:\Delta_n;\:\Delta_2,\:\ldots,\:\Delta_{n-1}}  \Delta_{\langle r|a_v\rangle} \Delta_{\langle s|a_u \rangle}\,,
}
and for brevity, we are using the short-hands $c_{( \cdot )}, \widetilde{c}_{( k_i \pm 1, j_{\langle r|s\rangle} \pm 1,\ldots)}^{r,s; a_u,a_v}$ to stand for 
\eqn{cctShort}{
c_{( \cdot )} &\equiv c_{0,k_1,\:\ldots,\:k_{n-3},\:0;\:j_{\langle 2|3\rangle},\:j_{\langle 2|4\rangle},\:\ldots,\:j_{\langle n-2|n-1\rangle}}^{\Delta_1,\:\Delta_{\delta_1},\:\ldots,\:\Delta_{\delta_{n-3}},\:\Delta_n;\:\Delta_2,\:\ldots,\:\Delta_{n-1}}  \cr 
 \widetilde{c}_{( k_i \pm 1, j_{\langle r|s\rangle} \pm 1,\ldots)}^{r,s; a_u,a_v} &\equiv  \widetilde{c}_{0,k_1,\:\ldots,\:k_{n-3},\:0;\:j_{\langle 2|3\rangle},\:j_{\langle 2|4\rangle},\:\ldots,\:j_{\langle n-2|n-1\rangle}}^{r,s; a_u,a_v} \Big|_{k_i \to k_i \pm 1, j_{\langle r|s\rangle} \to j_{\langle r|s\rangle} \pm 1, \ldots}\,,
 }
that is, coefficients with (un)shifted integral parameters.

Thus to show that the conformal Casimir equations are satisfied, we need to show~\eno{LHSmRHS} vanishes identically. Since it must vanish irrespective of the choice of boundary coordinates, each individual term in the sum must vanish. That is, we need to show 
\eqn{ShowIdAgain2}{
&  4\: k_{{K-1}}(\delta_{K-1}+k_{K-1} -d/2)\: c_{(\cdot)} \cr 
&-4 \sum_{\substack{1 \leq r<s \leq K \\  \\ K+1 \leq a_u < a_v \leq n}} \!\!\!
    \widetilde{c}^{r,s; a_u,a_v}_{(k_{s-1}-1, k_s -1, \ldots, k_{a_u-2}-1,j_{\langle r|s\rangle}-1, j_{\langle a_u|a_v\rangle}-1,j_{\langle s|a_u\rangle}+1,j_{\langle r|a_v\rangle}+1) }  
\stackrel{!}{=} 0\,,
}
for all $2 \leq K \leq n-2$, and for all $k_1,\ldots, k_{n-3}, j_{\langle 2|3 \rangle}, j_{\langle 2|4 \rangle}, \ldots, j_{\langle n-2|n-1\rangle} \in \mathbb{Z}^{\geq 0}$.\footnote{Due to the symmetry of the conjectural conformal block, we need only show~\eno{ShowIdAgain2} for $2 \leq K \leq \floor{n/2}$ where $\floor{\cdot}$ is the floor function; the rest follow after a simple relabelling.} We proved this analytically for all $n \leq 7$ (and all admissible values of $K$). The calculations are straightforward, though lengthy and are provided in an ancillary {\tt Mathematica} notebook with the arXiv submission. The key tool useful for proving~\eno{ShowIdAgain2} is an $(n-4)$-fold application of the hypergeometric identity~\eno{3F2id}, once for each factor of ${}_3F_2$ in the original expansion coefficient~\eno{cn}. For higher $n$, the analytics become particularly lengthy and unwieldy, but we cannot rule out a simple proof may exist. Nevertheless, it is straightforward to check~\eno{ShowIdAgain2} numerically for arbitrary values of $n$ and $K$ to arbitrary numerical precision; such a check is also included in the same {\tt Mathematica} notebook. Thus while~\eno{ShowIdAgain2} remains to be established fully analytically for $n\geq 8$, in our view the highly non-trivial numerical checks provide convincing supporting evidence.

\section{Discussion}
\label{DISCUSSION}

In this paper, we used the holographic principle, particularly its theory-independent kinematic aspects, to obtain for the first time explicit expressions for a class of multipoint conformal blocks.
We started by establishing the holographic geodesic diagram representations of $d$-dimensional comb channel six-point  (\eno{calW6confblock}-\eno{lambda6Def}) and seven-point (\eno{calW7confblock}-\eno{lambda7Def})  scalar conformal blocks involving scalar exchanges.  From them we recovered power series expansions for respectively the four- and five-point blocks via a double-OPE limit.
The explicit low-point examples, along with the holographic dual of the five-point block~\cite{Parikh:2019ygo}, allowed a generalization to the holographic dual of the $(n+2)$-point block for arbitrary $n$ (\eno{calWnPlus2confblock}-\eno{lambdanPlus2Def}). Like the low-point examples, it was expressible in terms of a linear combination of $(n+2)$-point geodesic diagrams involving two bulk geodesic integrals.
Its double-OPE limit then led us to a power series expansion for a scalar $n$-point block in~\eno{nCombSeries} (given also in~\eno{nCombBlock}-\eno{kdeltaExtraDef}), which is the main technical result of this paper and proven in sections~\ref{NCOMBOPE}-\ref{CASIMIR}.
Obtaining the holographic representations of low-point but non-trivial comb channel blocks served a crucial purpose in this paper. The low-point examples were simple enough that we needed only existing technology~\cite{Jepsen:2019svc} to obtain them, but non-trivial enough for us to recognize the pattern and guess the form of the holographic dual of an arbitrary-point block.

There are various avenues for further investigation.
In this paper, we restricted ourselves to studying $n$-point comb channel scalar conformal blocks involving solely scalar exchanges. To be able to set up an alternative $n$-point conformal bootstrap involving scalar $n$-point functions for all $n$, one must also have in hand higher-point scalar blocks involving exchange of other non-trivial representations of the Lorentz group. It should  be possible to generate $n$-point comb channel blocks involving both external and internal spin exchanges by, for example, operating on the blocks obtained in this paper via weight-shifting operators~\cite{Karateev:2017jgd}. It should also be possible to generate higher-point spinning geodesic diagram representations using the AdS differential operators of ref.~\cite{Costa:2018mcg}. Various recursive techniques, when supplemented with the results of this paper, may also turn out to be fruitful.

For setting up the $n$-point conformal bootstrap, one also needs $n$-point blocks in channels other than the comb channel. 
The number of topologically distinct channels, not related via conformal transformation or simple relabeling, grows quickly with $n$. Thus it is likely inefficient to compute multipoint conformal blocks one specific channel at a time.
On the other hand, it is conceivable there exist some version of ``Feynman-like'' rules for writing out conformal blocks, akin to Feynman rules for Mellin amplitudes~\cite{Fitzpatrick:2011ia,Paulos:2011ie,Nandan:2011wc}, which can be worked out once and for all.
 We hope the explicit expressions for the $n$-point comb channel block we obtained in this paper will help elucidate these Feynman-like rules.
 
For instance, consider the diagrammatic representation of the comb channel conformal block in figure~\ref{fig:nblock}. 
It has two internal cubic vertices where exactly two external legs and one internal leg are incident, and $(n-4)$ internal cubic vertices at which exactly one external leg and two internal legs are incident. 
In this paper we saw that both the holographic representation as well as power series expansion of the $n$-point comb channel block have precisely $(n-4)$ factors of the  hypergeometric ${}_3 F_2$ function in the explicit expansion.\footnote{One might be tempted to assert  from~\eno{nCombBlock} that there are in fact $(n-2)$ factors of the hypergeometric function, but the precise count is $(n-4)$ since $k_0$ and $k_{n-2}$ vanish by definition~\eno{kdeltaExtraDef}. This is a simple generalization of the previously known cases for scalar blocks with scalar exchanges --- the four-point block doesn't have any ${}_3F_2$ functions in its double power series expansion~\cite{Dolan:2003hv}, while the power series expansion for the $d$-dimensional five-point block obtained in ref.~\cite{Rosenhaus:2018zqn} (as well as its holographic representation~\cite{Parikh:2019ygo}) carries precisely one factor of the hypergeometric ${}_3F_2$ function.}  This is not a coincidence, and is in fact reminiscent of Feynman rules for scalar Mellin amplitudes. Our work suggests a holographic origin for this. Given a conformal block, as argued previously~\cite{Parikh:2019ygo,Jepsen:2019svc}, a general strategy for extracting its explicit holographic dual (and consequently a power series expansion via a double-OPE limit as discussed in this paper) is to start with a canonical tree-level AdS diagram.\footnote{The canonical choice for the AdS diagram for extracting the six-point comb channel block was provided in~\eno{SixExchThree}.} The AdS diagram should have solely cubic couplings and its direct-channel conformal block decomposition should include the given block. Performing the bulk integrals in the diagram carefully using various two- and three-propagator AdS identities helps extract an explicit representation for the block~\cite{Hijano:2015zsa,Parikh:2019ygo,Jepsen:2019svc}. Indeed for comb channel blocks, the $(n-4)$~factors of~${}_3F_2$ arise due to the presence of $(n-4)$~cubic bulk integrals  involving exactly one bulk-to-boundary propagator and  two bulk-to-bulk  propagators. This integral was fully worked out in ref.~\cite{Jepsen:2019svc} and it involves precisely the right factor of the hypergeometric function~${}_3F_2$ with precisely the right arguments.\footnote{The remaining two bulk integrals involving integration over a product of precisely two bulk-to-boundary propagators and one bulk-to-bulk propagator contribute simply factors of Gamma functions or Pochhammer symbols~\cite{Jepsen:2019svc}.} In fact, this general argument should extend to any arbitrary-point scalar conformal block with scalar exchanges in an arbitrary channel. All the necessary three-propagator integrals appearing in such a derivation were worked out in ref.~\cite{Jepsen:2019svc}. This should make it tractable to work out the putative Feynman rules for all scalar blocks.

It is interesting to compare our results  with the parallel, albeit considerably simpler story in the framework of $p$-adic AdS/CFT, where the conformal group is ${\rm PGL}(2,\mathbb{Q}_{p^d})$~\cite{Gubser:2016guj,Heydeman:2016ldy}.\footnote{Here $\mathbb{Q}_{p^d}$ is the unique unramified degree $d$ field extension of the $p$-adic numbers $\mathbb{Q}_p$ (see e.g.\ the book~\cite{gouvea1997p}).}
Due to the lack of descendant operators in $p$-adic CFTs~\cite{Melzer:1988he}, the conformal blocks are simply scaling blocks. 
Nevertheless, analogous to the real conformal blocks, the $p$-adic blocks admit holographic duals, written as geodesic bulk diagrams on the Bruhat--Tits tree~\cite{Gubser:2017tsi,Parikh:2019ygo,Jepsen:2019svc}. 
 In fact, all results presented in this paper also admit a $p$-adic counterpart --- various recent accounts of comparison and translation between  objects in the usual (real) and $p$-adic holographic settings can be found in refs.~\cite{Gubser:2016guj,Heydeman:2016ldy,Gubser:2016htz,Gubser:2017vgc,Gubser:2017tsi,Gubser:2017qed,Dutta:2018qrv,Bhowmick:2018bmn,Gubser:2018bpe,Gubser:2018cha,Gubser:2018ath,Jepsen:2018dqp,Jepsen:2018ajn,Qu:2018ned,Heydeman:2018qty,Hung:2018mcn,Hung:2019zsk,Huang:2019nog,Parikh:2019ygo,Jepsen:2019svc,Bentsen:2019rlr,Gubser:2019uyf,Ebert:2019src}. 
 Essentially, to recover the $p$-adic result, one can truncate all power series expansions featured in this paper to their respective first terms, since the infinite multi-fold series expansions in real CFTs sum up descendant contributions which do not exist in $p$-adic CFTs.\footnote{For example, the holographic dual of the comb channel $p$-adic block is given by~\eno{calWnPlus2confblock}-\eno{calWnPlus2} except with all integral parameters being summed over set to zero.}
 Conversely, it is practical to work out holographic duals of blocks in the simpler $p$-adic setting first.
 This is because the conformal dimension dependence of propagators appearing in the associated $p$-adic geodesic diagrams is identical to that for geodesic diagrams in real CFTs (more precisely, the ``primary contribution'' is identical). So the simpler $p$-adic technology can be used to figure out in advance the expected primary contribution to conformal blocks in a real CFT; the full block, which sums up also the descendant contributions, can then in principle be determined from conformal invariance.

 Finally, a potential practical concern about the power series expansion~\eno{nCombBlock} could be that it is not rapidly convergent for operator insertions in the ``OPE regime'' of the comb channel, i.e.\ for cross-ratios $u_i \ll 1$ and $w_{r;s} \approx 1$ defined in~\eno{uwDef} (for all allowed values of the subscripts). 
However, one can remedy this slow convergence by a simple transformation which re-expresses the series expansion in powers of $w_{r;s}$ as an expansion in powers of $(1-w_{r;s})$.
 This is straightforward to work out,  though increasingly tedious as $n$ grows.
More precisely, to transform~\eno{nCombBlock} to a more efficient and rapidly convergent power series expansion for any fixed $n$, the following identity is useful (see e.g.\ footnote~\ref{fn:rearrange}):
\eqn{RearrangementID}{
\sum_{j=0}^\infty {(-w)^j \over j!}  (a_1)_{b_1+j} (a_2)_{b_2+j}  (a_3)_{b_3-j} &=  \sum_{j=0}^\infty {(-1)^{b_3} w^j \over j!} { (a_1)_{b_1+j} (a_2)_{b_2+j}  \over (1-a_3)_{-b_3+j} } \cr 
  &= { (-1)^{b_1+b_2} \Gamma(1-a_3) \Gamma\left(1-(a_1+a_2+a_3)\right) \over \Gamma\left(1-(a_1+a_3+b_1+b_3)\right) \Gamma\left(1-(a_2+a_3+b_2+b_3)\right) } \cr 
& \times \sum_{j=0}^\infty { (1-w)^j \over j!} {(a_1)_{b_1+j} (a_2)_{b_2+j}  \over (a_1+a_2+a_3)_{b_1+b_2+b_3+j}}
}
for integers $b_1, b_2, b_3$ and  the parameter space where the power series in $w$ and $(1-w)$ are simultaneously convergent; we analytically continue to extend the result outside this convergent regime.

For the four-, five-, six-, and seven-point blocks, we can explicitly check that a repeated application of this identity leads to the following alternate representation (for $n=4,5,6,7$)
\eqn{ConvergentConjecture}{
 {W}_n & = 
  W_0^{(n)}(x_i) \left( \prod_{i=1}^{n-3} u_i^{\Delta_{\delta_i} \over 2} \right) \!\!\! \sum_{\substack{k_1,\ldots,k_{n-3},\\j_{\langle 2|3\rangle},\:j_{\langle 2|4\rangle},\:\ldots,\:j_{\langle n-2|n-1\rangle}=0}}^\infty \!\!\! \Bigg[
\left( \prod_{i=1}^{n-3} {u_i^{k_i} \over k_i!} \right) \left( \prod_{2\leq r < s \leq n-1} {(1-w_{r;s})^{j_{\langle r|s\rangle}} \over j_{\langle r|s\rangle}!} \right) \cr 
&  \times  
 \Bigg( \prod_{t=0}^{n-3}  { \left(1-\Delta_{(t+1)\delta_{t-1},\delta_{t}}\right)_{k_{t}} \left(1-\Delta_{(t+2)\delta_{t+1},\delta_{t}}\right)_{k_{t}} \over  \left(\Delta_{\delta_{t}} - d/2+1 \right)_{k_{t}}  \left(\Delta_{\delta_t} \right)_{2k_t + \sum_{2 \leq r < t+2 \leq s \leq n-1} j_{\langle r|s \rangle} } } 
 \left(\Delta_{(t+2)\delta_t,\delta_{t+1}} \right)_{k_{t,t+1}+\sum_{2 \leq r<t+2 }j_{\langle r|t+2\rangle}}
 \cr 
& \quad \times
\left(\Delta_{(t+2)\delta_{t+1},\delta_{t}} \right)_{k_{t+1,t}+\sum_{t+2 < s \leq n-1}j_{\langle t+2|s\rangle}} 
 \left(\Delta_{\delta_t \delta_{t+1},(t+2)} \right)_{k_{t(t+1),}+\sum_{2 \leq r < t+2 < s \leq n-1}j_{\langle r|s\rangle}}  \cr  
 &  \quad \times  {}_3F_2\left[\left\{-k_t,-k_{t+1},\Delta_{\delta_t\delta_{t+1} (t+2),}-{d\over 2}\right\}; \left\{\Delta_{(t+2)\delta_{t+1},\delta_t}-k_t, \Delta_{(t+2)\delta_t,\delta_{t+1}}-k_{t+1}\right\};1\right] \Bigg) \Bigg].
}
Recently, a similar power series expansion was obtained for the same five-point block in general spacetime dimensions~\cite{Rosenhaus:2018zqn}. There, a different set of conformal cross-ratios were used.
We expect it to be possible to  establish analytically the equality between the result of ref.~\cite{Rosenhaus:2018zqn} and~\eno{ConvergentConjecture} for $n=5$, but we have not found any simple transformations which achieve this.
However, we have  verified numerically that the power series expansion~\eno{ConvergentConjecture}  matches the one in ref.~\cite{Rosenhaus:2018zqn} to desired numerical precision in the mutual regime of convergence of the two series expansions, as expected.
While we have not worked out the alternate representation for general $n$, we conjecture this to also hold for all $n \geq 8$.

\subsection*{Acknowledgements}
I thank Robert Clemenson for valuable discussions and collaboration in the early stages, and Christian B.\ Jepsen for useful discussions. I would also like to thank Steve Gubser, my Ph.D.\ advisor. Our joint work from about two years ago~\cite{Gubser:2017tsi} served as an important stepping stone and a springboard for many subsequent ideas, including the present work. Steve was an inspiring mentor, an extraordinary physicist and an exceptional human being. I am deeply grateful to Steve Gubser for everything, and would like to dedicate this paper to his fond memory.

\appendix

\section{Five- and seven-point examples}
\label{SEVENCOMB}

In this appendix we briefly discuss the holographic representation of the seven-point comb channel block and its double-OPE limit, which leads to a power series expansion for the five-point block. 
The holographic representations~\eno{calW5} and~\eno{calW6} for the five- and six-point cases inform the following conjecture for the seven-point block (which is consistent with the general conjecture~\eno{calWnPlus2confblock}-\eno{lambdanPlus2Def}):
\eqn{calW7confblock}{
{W}_{\Delta_{\delta_1};\:\Delta_{\delta_2};\:\Delta_{\delta_3};\:\Delta_{\delta_4}}^{\Delta_1,\ldots,\Delta_7}(x_i) &= \musepic{\figSevenCombChannel} \cr 
&= {4 \over  
B(\Delta_{\delta_1 1,2},\Delta_{\delta_1 2,1})\: B(\Delta_{\delta_4 6,7},\Delta_{\delta_4 7,6})} \:
{\cal W}_{\Delta_{\delta_1};\:\Delta_{\delta_2};\:\Delta_{\delta_3};\:\Delta_{\delta_4}}^{\Delta_1,\ldots,\Delta_7}(x_i) \,,
}
where the linear combination of seven-point geodesic bulk diagrams ${\cal W}$ is given by
\eqn{calW7}{
{\cal W}_{\Delta_{\delta_1};\:\Delta_{\delta_2};\:\Delta_{\delta_3};\:\Delta_{\delta_4}}^{\Delta_1,\ldots,\Delta_7}(x_i) &= \sum_{\substack{k_1,k_2,k_3,k_4,\\j_1,j_2,j_3=0}}^\infty c_{k_1,\:k_2,\:k_3,\:k_4;\:j_1,\:j_2,\:j_3}^{\Delta_{\delta_1},\: \Delta_{\delta_2},\:\Delta_{\delta_3},\:\Delta_{\delta_4};\:\Delta_3,\:\Delta_4,\:\Delta_5} \cr   & \times \musepic{\figcalWSeven},
}
with the coefficients
\eqn{c7Def}{
c_{k_1,\:k_2,\:k_3,\:k_4;\:j_1,\:j_2,\:j_3}^{\Delta_{\delta_1},\: \Delta_{\delta_2},\:\Delta_{\delta_3},\:\Delta_{\delta_4};\:\Delta_3,\:\Delta_4,\:\Delta_5} &\equiv  \lambda_7\,  {(-1)^{k_1+k_4+j_1+j_2+j_3} \over k_1!k_2!k_3!k_4!j_1!j_2!j_3!} 
{ \left(1-\Delta_{3\delta_2,\delta_1}\right)_{k_1} \left(1-\Delta_{3\delta_1,\delta_2}\right)_{k_2}  
\over
\left(\Delta_{\delta_1}-d/2+1\right)_{k_1} 
} \cr 
& \times {\left(1-\Delta_{4\delta_3,\delta_2}\right)_{k_2} \left(1-\Delta_{4\delta_2,\delta_3}\right)_{k_3}
\left(1-\Delta_{5\delta_4,\delta_3}\right)_{k_3} \left(1-\Delta_{5\delta_3,\delta_4}\right)_{k_4} 
\over
\left(\Delta_{\delta_2}-d/2+1\right)_{k_2}
\left(\Delta_{\delta_3}-d/2+1\right)_{k_3}
\left(\Delta_{\delta_4}-d/2+1\right)_{k_4}
} \cr 
& \times 
\left( \Delta_{3\delta_1,\delta_2} \right)_{k_{1,2}} 
\left(\Delta_{3\delta_2,\delta_1}\right)_{k_{2,1}+j_{13,}} 
\left(\Delta_{4\delta_2,\delta_3}\right)_{k_{2,3}+j_1} 
\left( \Delta_{4\delta_3,\delta_2} \right)_{k_{3,2}+j_2} \cr 
& \times \left(\Delta_{5\delta_3,\delta_4} \right)_{k_{3,4}+j_{23,}}
\left(\Delta_{5\delta_4,\delta_3} \right)_{k_{4,3}}
\left(\Delta_{\delta_1\delta_4,345}\right)_{k_{14,}-j_{123,}} 
\cr 
& \times {}_3F_2\left[\{-k_1,-k_2,\Delta_{\delta_1\delta_2 3,}-d/2\}; \{\Delta_{3\delta_2,\delta_1}-k_1, \Delta_{3\delta_1,\delta_2}-k_2\};1\right] \cr 
& \times {}_3F_2\left[\{-k_2,-k_3,\Delta_{\delta_2\delta_3 4,}-d/2\}; \{\Delta_{4\delta_3,\delta_2}-k_2, \Delta_{4\delta_2,\delta_3}-k_3\};1\right] \cr 
& \times {}_3F_2\left[\{-k_3,-k_4,\Delta_{\delta_3\delta_4 5,}-d/2\}; \{\Delta_{5\delta_4,\delta_3}-k_3, \Delta_{5\delta_3,\delta_4}-k_4\};1\right],
 }
where
\eqn{lambda7Def}{
\lambda_7 \equiv   {\Gamma(1-\Delta_{\delta_1 \delta_2,3}) \Gamma(1-\Delta_{\delta_2 \delta_3,4}) \Gamma(1-\Delta_{\delta_3 \delta_4,5}) \over  \Gamma(1-\Delta_{\delta_1 \delta_4,345}) \Gamma(1-\Delta_{\delta_2}) \Gamma(1-\Delta_{\delta_3})}\,.
}
 Here we are using the subscript convention~\eno{kabc} for both $k$ and $j$ integral parameters.
In drawing the geodesic diagram above,  we have disobeyed the strict color-coding prescribed in figure~\ref{fig:geodesicdiag} to better help guide the eye. 
It should be clear from the diagram which lines represent a bulk-to-boundary propagator, and which lines represent purely boundary contractions of the form $(x_{ij}^2)^{-\Delta}$. The factor of chordal distance continues to be shown as a dotted black line.

To prove this conjecture, we must show that~\eno{calW7confblock} satisfies the conformal Casimir equations~\eno{NptCasimirEqns} for $n=7$ and $K=2,3,4,5$.
Due to the symmetry of the object, we need only check the cases $K=2,3$; the remaining two cases follow trivially after relabelling.
Further, we must show the OPE limit~\eno{nptOPE}. 
As remarked in section~\ref{SIXCOMB}, the conformal Casimir check is in fact straightforward though lengthy to work out, but the procedure is identical to the ones described in refs.~\cite{Parikh:2019ygo,Jepsen:2019svc} in the context of five- and six-point blocks. No new ingredients are needed, except for a triple-application of the hypergeometric identity~\eno{3F2id}, once for each factor of the ${}_3F_2$ function in~\eno{c7Def}.
Similarly, the object obtained in the OPE limit can be shown to be an alternate holographic representation of the six-point block involving a single geodesic integral, via a similar proof by Casimir.
So to keep the paper to a reasonable length, we will refrain from providing the somewhat lengthy details here.
In lieu of this, we provide a conformal Casimir check of the {\it series expansion} of the seven-point block in section~\ref{NCOMB}.
The OPE limits of the six-point block obtained above themselves lead to two different representations of the five-point block --- one corresponding to a holographic representation involving a single geodesic integral similar to~\eno{W5altToShow}, with the other more interesting limit producing a power series expansion for the five-point block as discussed next.

\subsection{Double-OPE limit and the five-point block}

Consider the following double-OPE limit of the seven-point block,
\eqn{W7doubleOPE}{
{W}_{\Delta_{\delta_1};\:\Delta_{\delta_2};\:\Delta_{\delta_3};\:\Delta_{\delta_4}}^{\Delta_1,\ldots,\Delta_7}(x_1,\ldots,x_7) &\stackrel{\substack{x_2 \to x_1\\ x_7 \to x_6}}{\longrightarrow} (x_{12}^2)^{\Delta_{\delta_1,12}}  (x_{67}^2)^{\Delta_{\delta_4,67}} \:
\sum_{\substack{k_2,k_3,\\j_1,j_2,j_3=0}}^\infty c_{0,\:k_2,\:k_3,\:0;\:j_1,\:j_2,\:j_3}^{\Delta_{\delta_1},\: \Delta_{\delta_2},\:\Delta_{\delta_3},\:\Delta_{\delta_4};\:\Delta_3,\:\Delta_4,\:\Delta_5} \cr 
& \times \musepic{\figcalWSevenDoubleOPE} \cr 
 &\equiv  (x_{12}^2)^{\Delta_{\delta_1,12}}  (x_{67}^2)^{\Delta_{\delta_4,67}}\: \widetilde{V}\,.
}
The claim is that $\widetilde{V}$ is a (power series expansion of the) five-point block. Explicitly, we may write it as
\eqn{Vtilde}{
\widetilde{V} &= W_0^{\Delta_{\delta_1}, \Delta_3,\Delta_4,\Delta_5,\Delta_{\delta_4}}(x_1,x_3,x_4,x_5,x_6) \cr  
&\quad \times u_1^{\Delta_{\delta_2} \over 2} u_2^{\Delta_{\delta_3}\over 2} 
\sum_{\substack{k_2,k_3,\\j_1,j_2,j_3=0}}^\infty c_{0,\:k_2,\:k_3,\:0;\:j_1,\:j_2,\:j_3}^{\Delta_{\delta_1},\: \Delta_{\delta_2},\:\Delta_{\delta_3},\:\Delta_{\delta_4};\:\Delta_3,\:\Delta_4,\:\Delta_5} 
u_1^{k_2} u_2^{k_3} v_1^{j_1} v_2^{j_2} w^{j_3}\,,
}
where the leg factor $W_0^{\Delta_{\delta_1}, \Delta_3,\Delta_4,\Delta_5,\Delta_{\delta_4}}(x_1,x_3,x_4,x_5,x_6)$ was defined in~\eno{W0nDef},\footnote{An alternate way to write the leg factor is as follows:
\eqn{W05Def}{
W_0^{\Delta_{\delta_1}, \Delta_3,\Delta_4,\Delta_5,\Delta_{\delta_4}}(x_1,x_3,x_4,x_5,x_6) = {1 \over (x_{13}^2)^{\Delta_{\delta_1 3,}} (x_{46}^2)^{\Delta_4 \over 2} (x_{56}^2)^{\Delta_{\delta_4 5,}} } 
\left( {x_{16}^2 \over x_{36}^2} \right)^{\Delta_{3,\delta_1}}
\left( {x_{16}^2 \over x_{14}^2} \right)^{\Delta_{4} \over 2}
\left( {x_{15}^2 \over x_{16}^2} \right)^{\Delta_{\delta_4,5}}.
}
}
and we have defined the cross-ratios
\eqn{FiveCrossRatios}{
u_1 \equiv  {x_{13}^2 x_{46}^2 \over x_{14}^2 x_{36}^2 } \qquad v_1 \equiv  {x_{16}^2 x_{34}^2 \over x_{14}^2 x_{36}^2 } \qquad u_2 \equiv  {x_{14}^2 x_{56}^2 \over x_{15}^2 x_{46}^2 } \qquad v_2 \equiv  {x_{16}^2 x_{45}^2 \over x_{15}^2 x_{46}^2 } \qquad w \equiv   {x_{16}^2 x_{35}^2 \over x_{15}^2 x_{36}^2 }\,.
}
Writing out the coefficients explicitly, we get
\eqn{c7Special}{
c_{0,\:k_2,\:k_3,\:0;\:j_1,\:j_2,\:j_3}^{\Delta_{\delta_1},\: \Delta_{\delta_2},\:\Delta_{\delta_3},\:\Delta_{\delta_4};\:\Delta_3,\:\Delta_4,\:\Delta_5} &=  \lambda_7\,  {(-1)^{k_2+k_3+j_1+j_2+j_3} \over k_2!k_3!j_1!j_2!j_3!} 
  { 
\left(1-\Delta_{4\delta_3,\delta_2}\right)_{k_2} 
\left(1-\Delta_{4\delta_2,\delta_3}\right)_{k_3}
\over
\left(\Delta_{\delta_2}-d/2+1\right)_{k_2}
\left(\Delta_{\delta_3}-d/2+1\right)_{k_3}
} \cr 
& \times 
\left(\Delta_{3\delta_2,\delta_1}\right)_{k_{2}+j_{13,}} 
\left(\Delta_{4\delta_2,\delta_3}\right)_{k_{2,3}+j_1} 
\left( \Delta_{4\delta_3,\delta_2} \right)_{k_{3,2}+j_2} \cr 
& \times \left(\Delta_{5\delta_3,\delta_4} \right)_{k_{3}+j_{23,}}
\left(\Delta_{\delta_1\delta_4,345}\right)_{-j_{123,}} 
\cr 
& \times {}_3F_2\left[\{-k_2,-k_3,\Delta_{\delta_2\delta_3 4,}-d/2\}; \{\Delta_{4\delta_3,\delta_2}-k_2, \Delta_{4\delta_2,\delta_3}-k_3\};1\right] ,
 }
where $\lambda_7$ was given in~\eno{lambda7Def}. It can be checked that this is just a rewriting of~\eno{nCombBlock} for $n=5$ in different variables.
The proof that this is indeed a five-point block is given in section~\ref{NCOMB}.

\section{Further technical details}
\label{TECH}

\subsection{Proof of~\eno{ToShowSum}}
\label{USEFUL2F1ID}

 In this section we will prove~\eno{ToShowSum}. Consider first the $j_{\langle 2|n-1\rangle}$ sum on the right hand side. We can use the following identity  to evaluate this sum:
\eqn{Useful2F1id}{
\sum _{k=0}^{\infty } \frac{(-1)^k}{k!} (a)_k (b)_{\ell_1+k} (c)_{-\ell_2-k} = \frac{(-1)^{\ell_2} \Gamma (1-c) \Gamma (b+\ell_1) \Gamma (-a-b-c-\ell_1+\ell_2+1)}{\Gamma (b) \Gamma (-a-c+\ell_2+1) \Gamma (-b-c-\ell_1+\ell_2+1)} \,,
}
which holds for integers $\ell_1,\ell_2$, and we assume convergence of the sum. Performing the sum using the identity above, the right hand side of~\eno{ToShowSum} becomes
\eqn{RHS}{
RHS &=    {\Gamma\left(1- \Delta_{\delta_{n-3}n,(n-1)}\right) \Gamma\left(1-\Delta_{\delta_1\delta_{n-3},3\ldots (n-2)}\right)  \over \Gamma\left(1-\Delta_{\delta_1 n,3\ldots (n-1)}\right) \Gamma\left(1-\Delta_{\delta_{n-3}}\right) }
 \cr 
 &\times \sum_{\substack{j_{\langle 3|n-1\rangle}, \ldots,\\j_{\langle n-2|n-1\rangle}=0}}^\infty  \!\!\!\!\!
 \left( \prod_{\ell=3}^{n-2} {(-1)^{j_{\langle \ell|n-1\rangle}} \over j_{\langle \ell|n-1\rangle}!} \right) 
  \left( \prod_{t=1}^{n-4} \left(\Delta_{(t+2)\delta_{t+1},\delta_{t}} + k_{t+1,t}+\sum_{s=t+3}^{n-2}j_{\langle t+2|s\rangle} \right)_{j_{\langle t+2|n-1 \rangle}}  \right) \cr 
  &\times \left(\Delta_{(n-1)\delta_{n-3},n} \right)_{\sum_{r=3}^{n-2}j_{\langle r|n-1\rangle}}
  {   \left( \Delta_{\delta_1n,3\ldots(n-1)} +k_1\right)_{-\sum_{3 \leq r < s \leq n-1} j_{\langle r|s\rangle}}  \over   \left( \Delta_{\delta_1\delta_{n-3},3\ldots(n-2)}+k_1 \right)_{-\sum_{3 \leq r < s \leq n-2} j_{\langle r|s\rangle}}  } 
\,,
}
where $\delta_0=\Delta_1, k_0=0$ and $k_{n-3}=0$ as before.
Notice that~\eno{RHS} is of precisely the same form as the original sum on the right hand side of~\eno{ToShowSum}, except with one fewer sum. Thus one can iteratively apply~\eno{Useful2F1id} to systematically reduce $RHS$ further and further. For instance, performing the $j_{\langle 3|n-1\rangle}$ sum next yields
\eqn{RHS2}{
RHS &=    {\Gamma\left(1- \Delta_{\delta_{n-3}n,(n-1)}\right) \Gamma\left(1-\Delta_{\delta_2\delta_{n-3},4\ldots (n-2)}\right)  \over \Gamma\left(1-\Delta_{\delta_2 n,4\ldots (n-1)}\right) \Gamma\left(1-\Delta_{\delta_{n-3}}\right) }
 \cr 
 &\times \sum_{\substack{j_{\langle 4|n-1\rangle}, \ldots,\\j_{\langle n-2|n-1\rangle}=0}}^\infty  \!\!\!\!\!
 \left( \prod_{\ell=4}^{n-2} {(-1)^{j_{\langle \ell|n-1\rangle}} \over j_{\langle \ell|n-1\rangle}!} \right) 
  \left( \prod_{t=2}^{n-4} \left(\Delta_{(t+2)\delta_{t+1},\delta_{t}} + k_{t+1,t}+\sum_{s=t+3}^{n-2}j_{\langle t+2|s\rangle} \right)_{j_{\langle t+2|n-1 \rangle}}  \right) \cr 
  &\times \left(\Delta_{(n-1)\delta_{n-3},n} \right)_{\sum_{r=4}^{n-2}j_{\langle r|n-1\rangle}}
  {   \left( \Delta_{\delta_2n,4\ldots(n-1)} +k_2\right)_{-\sum_{4 \leq r < s \leq n-1} j_{\langle r|s\rangle}}  \over   \left( \Delta_{\delta_2\delta_{n-3},4\ldots(n-2)}+k_2 \right)_{-\sum_{4 \leq r < s \leq n-2} j_{\langle r|s\rangle}}  } 
\,,
}
where the pattern should be clear by now.
Reducing $RHS$ iteratively by performing the $j_{\langle 4|n-1\rangle}, j_{\langle 5|n-1\rangle}, \ldots$ sums in this order, it is straightforward to check that $RHS$ reduces to unity.

\subsection{Proof of~\eno{PartialSum}}
\label{APP:PARTIALSUM}

Define,
\eqn{}{
J&\equiv \sum_{1 \leq r<s \leq K}\sum_{\substack{a_u,a_v=1 \\  a_u, a_v \notin \{r,s\},a_u \neq a_v}}^{K} \Delta_{\langle a_u|s \rangle}\Delta_{\langle a_v|r\rangle}{ \left( x_{rs}^2 x_{a_u a_v}^2  - x_{sa_v}^2 x_{ra_u}^2 \right) \over x_{sa_u}^2 x_{ra_v}^2 } \,.
}
We will show that $J=0$, which proves~\eno{PartialSum}. 
First expand $J$ as
\eqn{}{ 
J&=  \sum_{1 \leq r<s \leq K} \left(\sum_{\substack{1 \leq a_u < a_v \leq K \\  a_u, a_v \notin \{r,s\}}} +   \sum_{\substack{1 \leq a_v < a_u \leq K  \\  a_u, a_v \notin \{r,s\}}}  \right) \Delta_{\langle a_u|s \rangle}\Delta_{\langle a_v|r\rangle}{ \left( x_{rs}^2 x_{a_u a_v}^2  - x_{sa_v}^2 x_{ra_u}^2 \right) \over x_{sa_u}^2 x_{ra_v}^2 }  \cr 
&=  \left(\sum_{1 \leq r<s \leq K} + \sum_{1 \leq s<r \leq K}\right)  \sum_{\substack{1 \leq a_u < a_v \leq K \\  a_u, a_v \notin \{r,s\}}} \Delta_{\langle a_u|s \rangle}\Delta_{\langle a_v|r\rangle}{ \left( x_{rs}^2 x_{a_u a_v}^2  - x_{sa_v}^2 x_{ra_u}^2 \right) \over x_{sa_u}^2 x_{ra_v}^2 } \cr 
&= \left(J_1 + J_2 + J_3 +J_4 +J_5 + J_6\right) + \left(J_7 + J_8 + J_9 +J_{10} +J_{11} + J_{12}\right),
}
where we have broken the four-fold sums into the following constituent pieces:
\begingroup
\allowdisplaybreaks
\begin{align*}
J_1 &\equiv \sum_{1 \leq a_u < a_v < r<s \leq K} \Delta_{\langle a_u|s \rangle}\Delta_{\langle a_v|r\rangle}{ \left( x_{rs}^2 x_{a_u a_v}^2  - x_{sa_v}^2 x_{ra_u}^2 \right) \over x_{sa_u}^2 x_{ra_v}^2 }  \cr 
J_2 &\equiv  \sum_{1 \leq a_u < r < a_v<s \leq K} \Delta_{\langle a_u|s \rangle}\Delta_{\langle a_v|r\rangle}{ \left( x_{rs}^2 x_{a_u a_v}^2  - x_{sa_v}^2 x_{ra_u}^2 \right) \over x_{sa_u}^2 x_{ra_v}^2 }  \cr 
J_3 &\equiv  \sum_{1 \leq a_u < r < s<a_v \leq K} \Delta_{\langle a_u|s \rangle}\Delta_{\langle a_v|r\rangle}{ \left( x_{rs}^2 x_{a_u a_v}^2  - x_{sa_v}^2 x_{ra_u}^2 \right) \over x_{sa_u}^2 x_{ra_v}^2 } \cr 
J_4 &\equiv  \sum_{1 \leq r < a_u < a_v<s \leq K} \Delta_{\langle a_u|s \rangle}\Delta_{\langle a_v|r\rangle}{ \left( x_{rs}^2 x_{a_u a_v}^2  - x_{sa_v}^2 x_{ra_u}^2 \right) \over x_{sa_u}^2 x_{ra_v}^2 } \cr 
J_5 &\equiv  \sum_{1 \leq r<a_u<s < a_v \leq K} \Delta_{\langle a_u|s \rangle}\Delta_{\langle a_v|r\rangle}{ \left( x_{rs}^2 x_{a_u a_v}^2  - x_{sa_v}^2 x_{ra_u}^2 \right) \over x_{sa_u}^2 x_{ra_v}^2 }  \cr 
J_6 &\equiv  \sum_{1 \leq r<s <  a_u < a_v  \leq K} \Delta_{\langle a_u|s \rangle}\Delta_{\langle a_v|r\rangle}{ \left( x_{rs}^2 x_{a_u a_v}^2  - x_{sa_v}^2 x_{ra_u}^2 \right) \over x_{sa_u}^2 x_{ra_v}^2 } \,,
\stepcounter{equation}\tag{\theequation}\label{JiDef}
\end{align*}
\endgroup
and
\begingroup
\allowdisplaybreaks
\begin{align*}
J_7 &\equiv \sum_{1 \leq a_u < a_v < s<r \leq K} \Delta_{\langle a_u|s \rangle}\Delta_{\langle a_v|r\rangle}{ \left( x_{rs}^2 x_{a_u a_v}^2  - x_{sa_v}^2 x_{ra_u}^2 \right) \over x_{sa_u}^2 x_{ra_v}^2 }  \cr 
J_8 &\equiv  \sum_{1 \leq a_u < s < a_v<r \leq K} \Delta_{\langle a_u|s \rangle}\Delta_{\langle a_v|r\rangle}{ \left( x_{rs}^2 x_{a_u a_v}^2  - x_{sa_v}^2 x_{ra_u}^2 \right) \over x_{sa_u}^2 x_{ra_v}^2 }  \cr 
J_9 &\equiv  \sum_{1 \leq a_u < s < r<a_v \leq K} \Delta_{\langle a_u|s \rangle}\Delta_{\langle a_v|r\rangle}{ \left( x_{rs}^2 x_{a_u a_v}^2  - x_{sa_v}^2 x_{ra_u}^2 \right) \over x_{sa_u}^2 x_{ra_v}^2 } \cr 
J_{10} &\equiv  \sum_{1 \leq s < a_u < a_v<r \leq K} \Delta_{\langle a_u|s \rangle}\Delta_{\langle a_v|r\rangle}{ \left( x_{rs}^2 x_{a_u a_v}^2  - x_{sa_v}^2 x_{ra_u}^2 \right) \over x_{sa_u}^2 x_{ra_v}^2 } \cr 
J_{11} &\equiv  \sum_{1 \leq s<a_u<r < a_v \leq K} \Delta_{\langle a_u|s \rangle}\Delta_{\langle a_v|r\rangle}{ \left( x_{rs}^2 x_{a_u a_v}^2  - x_{sa_v}^2 x_{ra_u}^2 \right) \over x_{sa_u}^2 x_{ra_v}^2 }  \cr 
J_{12} &\equiv  \sum_{1 \leq s<r <  a_u < a_v  \leq K} \Delta_{\langle a_u|s \rangle}\Delta_{\langle a_v|r\rangle}{ \left( x_{rs}^2 x_{a_u a_v}^2  - x_{sa_v}^2 x_{ra_u}^2 \right) \over x_{sa_u}^2 x_{ra_v}^2 } \,.
\stepcounter{equation}\tag{\theequation}\label{JiDef2}
\end{align*}
\endgroup
Now, observe that switching the dummy variables $r \leftrightarrow a_v$ in $J_2$ turns it manifestly into $-J_1$. Likewise, doing the dummy variable replacement $s \leftrightarrow a_u$ in $J_6$ turns it precisely to $-J_5$, and the switches $r\leftrightarrow a_u, s \leftrightarrow a_v$ turn $J_4$ into $J_3$. Similarly, switching $s\leftrightarrow a_u$ turns $J_{10}$ manifestly into $-J_8$, and $J_{11}$ into $-J_9$. Finally, switching $s\leftrightarrow a_u, r\leftrightarrow a_v$ turns $J_{12}$ into $J_7$. Thus, we conclude $J= 2J_3 + 2J_7$. However, making the dummy variable switch $r \leftrightarrow a_v$ turns $J_7$ into $-J_3$. Thus, $J$ vanishes identically.

\subsection{Proof of~\eno{LHSmRHS}}
\label{APP:CASIMIR}

In this subsection we will show the details of the calculation leading from~\eno{LHSuw}-\eno{RHSuw} to~\eno{LHSmRHS}.
After integer shifting the integral parameters to get all terms in~\eno{LHSuw} to have uniform position dependence as explained below~\eno{RHSuw}, we re-express~\eno{LHSuw} as 
\eqn{}{
LHS&= W_0^{(n)}(x_i) \left( \prod_{i=1}^{n-3} u_i^{\Delta_{\delta_i} \over 2} \right) \!\!\!\!\!\!\! \sum_{\substack{k_1,\ldots,k_{n-3},\\j_{\langle 2|3\rangle},\:j_{\langle 2|4\rangle},\:\ldots,\:j_{\langle n-2|n-1\rangle}}}  
\!\!\!\!\!\!\!\!\!\!\!\!
 \left( \prod_{i=1}^{n-3} u_i^{k_i} \right) \left( \prod_{2\leq r < s \leq n-1} w_{r;s}^{j_{\langle r|s\rangle}} \right)
 \Bigg[\;  c_{(\cdot)} 
 \sum_{i=1}^K m_{\Delta_i}^2 \cr 
  &  -2\: c_{(\cdot)} \sum_{1 \leq r<s \leq K} \left( m_{\Delta_{\langle r|s\rangle}}^2  +   \Delta_{\langle r|s\rangle}\sum_{\ell=1}^{n-2}(\Delta_{\langle r|a_\ell\rangle} + \Delta_{ \langle s|a_\ell\rangle})  - \sum_{\ell=1}^{n-2} \Delta_{\langle r|a_\ell\rangle} \Delta_{\langle s|a_\ell\rangle}   \right)
 \cr
 &  -2 \sum_{\substack{1 \leq r<s \leq K \\ K+1 \leq a_u < a_v \leq n-1}} 
   \left( \widetilde{c}^{r,s; a_u,a_v}_{(k_{s-1}-1, k_s -1, \ldots, k_{a_u-2}-1,j_{\langle r|s\rangle}-1, j_{\langle a_u|a_v\rangle}-1,j_{\langle s|a_u\rangle}+1,j_{\langle r|a_v\rangle}+1) } \right. \cr 
 &\qquad\qquad\qquad\qquad\quad \left. -\:  \widetilde{c}^{r,s; a_u,a_v}_{(j_{\langle s|a_v\rangle}-1, j_{\langle r|a_u\rangle}-1,j_{\langle s|a_u\rangle}+1,j_{\langle r|a_v\rangle}+1)}  \right) \cr 
  &  -2 \sum_{\substack{1 \leq r<s \leq K\\ K+1 \leq a_v < a_u \leq n-1}} 
   \left( \widetilde{c}^{r,s; a_u,a_v}_{(k_{s-1}-1, k_s -1, \ldots, k_{a_v-2}-1,j_{\langle r|s\rangle}-1, j_{\langle a_v|a_u\rangle}-1,j_{\langle s|a_u\rangle}+1,j_{\langle r|a_v\rangle}+1) } \right. \cr 
 &\qquad\qquad\qquad\qquad\quad \left. -\:  \widetilde{c}^{r,s; a_u,a_v}_{(j_{\langle s|a_v\rangle}-1, j_{\langle r|a_u\rangle}-1,j_{\langle s|a_u\rangle}+1,j_{\langle r|a_v\rangle}+1)}  \right) \cr 
  &  -2 \sum_{\substack{1 \leq r<s \leq K\\ K+1 \leq a_u \leq n-1}} 
   \left( \widetilde{c}^{r,s; a_u,n}_{(k_{s-1}-1, k_s -1, \ldots, k_{a_u-2}-1,j_{\langle r|s\rangle}-1, j_{\langle s|a_u\rangle}+1) } -  \widetilde{c}^{r,s; a_u,n}_{(j_{\langle r|a_u\rangle}-1,j_{\langle s|a_u\rangle}+1)}  \right) \cr 
 &  -2 \sum_{\substack{1 \leq r<s \leq K\\ K+1 \leq a_v \leq n-1}} 
   \left( \widetilde{c}^{r,s; n,a_v}_{(k_{s-1}-1, k_s -1, \ldots, k_{a_v-2}-1,j_{\langle r|s\rangle}-1, j_{\langle r|a_v\rangle}+1) } -  \widetilde{c}^{r,s; n,a_v}_{(j_{\langle s|a_v\rangle}-1,j_{\langle r|a_v\rangle}+1)}  \right) 
 \Bigg] ,
}
where we are utilizing the notation~\eno{ctDef}-\eno{cctShort}.

Thus for the above expression to equal~\eno{RHSuw} for all choices of boundary insertion points, each individual term in the sum must equal the conformal Casimir eigenvalue, times the coefficient~\eno{cn}. That is, the proof of~\eno{LHSmRHS} boils down to demonstrating the following non-trivial identity involving the coefficients,
\eqn{ShowId}{
&  c_{(\cdot)} \left[
 \sum_{i=1}^K m_{\Delta_i}^2 
 -2\: \sum_{1 \leq r<s \leq K} \left( m_{\Delta_{\langle r|s\rangle}}^2  +   \Delta_{\langle r|s\rangle}\sum_{\ell=1}^{n-2}(\Delta_{\langle r|a_\ell\rangle} + \Delta_{ \langle s|a_\ell\rangle})  - \sum_{\ell=1}^{n-2}  \Delta_{\langle r|a_\ell\rangle}\Delta_{\langle s|a_\ell\rangle}  \right) \right.
 \cr
 &\quad - m_{\Delta_{\delta_{K-1}}}^2  \Bigg]  -2 \sum_{\substack{1 \leq r<s \leq K \\  \\ K+1 \leq a_u < a_v \leq n-1}} 
   \left( \widetilde{c}^{r,s; a_u,a_v}_{(k_{s-1}-1, k_s -1, \ldots, k_{a_u-2}-1,j_{\langle r|s\rangle}-1, j_{\langle a_u|a_v\rangle}-1,j_{\langle s|a_u\rangle}+1,j_{\langle r|a_v\rangle}+1) } \right. \cr 
 &\quad \left. -\:  \widetilde{c}^{r,s; a_u,a_v}_{(j_{\langle s|a_v\rangle}-1, j_{\langle r|a_u\rangle}-1,j_{\langle s|a_u\rangle}+1,j_{\langle r|a_v\rangle}+1)}  
 + \widetilde{c}^{r,s; a_v,a_u}_{(k_{s-1}-1, k_s -1, \ldots, k_{a_u-2}-1,j_{\langle r|s\rangle}-1, j_{\langle a_u|a_v\rangle}-1,j_{\langle s|a_v\rangle}+1,j_{\langle r|a_u\rangle}+1) } \right. \cr 
 &\quad \left. -\:  \widetilde{c}^{r,s; a_v,a_u}_{(j_{\langle s|a_u\rangle}-1, j_{\langle r|a_v\rangle}-1,j_{\langle s|a_v\rangle}+1,j_{\langle r|a_u\rangle}+1)}  \right) \cr 
 &-2 \sum_{\substack{1 \leq r<s \leq K\\ K+1 \leq a_u \leq n-1}} 
   \left( \widetilde{c}^{r,s; a_u,n}_{(k_{s-1}-1, k_s -1, \ldots, k_{a_u-2}-1,j_{\langle r|s\rangle}-1, j_{\langle s|a_u\rangle}+1) } -  \widetilde{c}^{r,s; a_u,n}_{(j_{\langle r|a_u\rangle}-1,j_{\langle s|a_u\rangle}+1)} \right. \cr  
  &\quad  \left.  +\;\widetilde{c}^{r,s; n,a_u}_{(k_{s-1}-1, k_s -1, \ldots, k_{a_u-2}-1,j_{\langle r|s\rangle}-1, j_{\langle r|a_u\rangle}+1) } -  \widetilde{c}^{r,s; n,a_u}_{(j_{\langle s|a_u\rangle}-1,j_{\langle r|a_u\rangle}+1)}  \right)  \stackrel{!}{=} 0\,,
}
for  $2\leq K \leq n-2$, $n\geq 4$ and all parameters $k_i,j_{\langle \cdot|\cdot\rangle}$.

We can use various symmetry properties of the scaled coefficients $\widetilde{c}$ to simplify~\eno{ShowId}. It is straightforward to check that for $1 \leq r<s \leq K$ and $K+1 \leq a_u <a_v \leq n-1$, 
 \eqn{SymTran1}{
  \widetilde{c}^{r,s; a_u,a_v}_{(j_{\langle s|a_v\rangle}-1, j_{\langle r|a_u\rangle}-1,j_{\langle s|a_u\rangle}+1,j_{\langle r|a_v\rangle}+1)}  &= \widetilde{c}_{(\cdot)}^{r,s;a_v,a_u}  \cr
   \widetilde{c}^{r,s; a_v,a_u}_{(j_{\langle s|a_u\rangle}-1, j_{\langle r|a_v\rangle}-1,j_{\langle s|a_v\rangle}+1,j_{\langle r|a_u\rangle}+1)}  &= \widetilde{c}_{(\cdot)}^{r,s;a_u,a_v} \cr 
    \widetilde{c}^{r,s; a_u,a_v}_{(k_{s-1}-1, k_s -1, \ldots, k_{a_u-2}-1,j_{\langle r|s\rangle}-1, j_{\langle a_u|a_v\rangle}-1,j_{\langle s|a_u\rangle}+1,j_{\langle r|a_v\rangle}+1) }  \cr & \hspace{-5cm}= \widetilde{c}^{r,s; a_v,a_u}_{(k_{s-1}-1, k_s -1, \ldots, k_{a_u-2}-1,j_{\langle r|s\rangle}-1, j_{\langle a_u|a_v\rangle}-1,j_{\langle s|a_v\rangle}+1,j_{\langle r|a_u\rangle}+1) }\,,
  }
 and for $1 \leq r<s \leq K$ and $K+1 \leq a_u \leq n-1$,
 \eqn{SymTran2}{
  \widetilde{c}^{r,s; a_u,n}_{(j_{\langle r|a_u\rangle}-1,j_{\langle s|a_u\rangle}+1)} &= \widetilde{c}^{s,r; a_u,n}_{(\cdot)}   \cr 
  \widetilde{c}^{r,s; n, a_u}_{(j_{\langle s|a_u\rangle}-1,j_{\langle r|a_u\rangle}+1)} &= \widetilde{c}^{r,s; a_u,n}_{(\cdot)}  \cr 
   \widetilde{c}^{r,s; a_u,n}_{(k_{s-1}-1, k_s -1, \ldots, k_{a_u-2}-1,j_{\langle r|s\rangle}-1, j_{\langle s|a_u\rangle}+1) } &= \widetilde{c}^{r,s; n,a_u}_{(k_{s-1}-1, k_s -1, \ldots, k_{a_u-2}-1,j_{\langle r|s\rangle}-1, j_{\langle r|a_u\rangle}+1) }\,.
  }
 These identities are justified at the end of this section.
Using these and the definition~\eno{ctDef}, we can simplify~\eno{ShowId} to
\eqn{ShowIdAgain}{
&  c_{(\cdot)} \left[
 \sum_{i=1}^K m_{\Delta_i}^2 - m_{\Delta_{\delta_{K-1}}}^2 
 -2\: \sum_{1 \leq r<s \leq K} \left( m_{\Delta_{\langle r|s\rangle}}^2  +   \Delta_{\langle r|s\rangle}\sum_{\ell=1}^{n-2}(\Delta_{\langle r|a_\ell\rangle} + \Delta_{ \langle s|a_\ell\rangle})  \right. \right.
 \cr
 &\quad \left. - \sum_{\ell=1}^{n-2}  \Delta_{\langle r|a_\ell\rangle}\Delta_{\langle s|a_\ell\rangle}
 -\sum_{K+1\leq a_u <a_v\leq n}\left( \Delta_{\langle r|a_u \rangle} \Delta_{\langle s|a_v\rangle}+ \Delta_{\langle r|a_v \rangle} \Delta_{\langle s|a_u\rangle}\right)
 \right) \Bigg] \cr 
 &-4 \sum_{\substack{1 \leq r<s \leq K \\  \\ K+1 \leq a_u < a_v \leq n}} 
    \widetilde{c}^{r,s; a_u,a_v}_{(k_{s-1}-1, k_s -1, \ldots, k_{a_u-2}-1,j_{\langle r|s\rangle}-1, j_{\langle a_u|a_v\rangle}-1,j_{\langle s|a_u\rangle}+1,j_{\langle r|a_v\rangle}+1) }  
\stackrel{!}{=} 0\,.
}
Now, using the explicit expressions for the dimensions, one can further simplify~\eno{ShowIdAgain} to~\eno{ShowIdAgain2}. This is straightforward to work out for any given value of $n$ and $K$ (see the ancillary {\tt Mathematica} notebook for calculational details).

\vspace{.5em}
To close this section, let's return to the as yet unproven symmetry transformations~\eno{SymTran1}-\eno{SymTran2}. Consider for example, the first identity in~\eno{SymTran2}. To prove it, it is useful to take a closer look at the coefficients~\eno{cn} and the dimensions $\Delta_{\langle i|j \rangle}$ which can be  visually read off of~\eno{nCombSeries}. Consider the {\it left hand side} first, where we need to study the effect of shifting the integral parameters $j_{\langle r|a_u\rangle} \to j_{\langle r|a_u\rangle}-1,j_{\langle s|a_u\rangle} \to j_{\langle s|a_u\rangle}+1$ in the scaled coefficient $ \widetilde{c}^{r,s; a_u,n}_{(\cdot)}$, for fixed $r,s, a_u$ satisfying $1 \leq r < s \leq K$ and $K+1 \leq a_u \leq n-1$. First assume $r \geq 2$ --- we will come back to the case $r=1$ at the end:
\begin{itemize}
    \item The factor of $\left( \Delta_{1n,2\ldots(n-1)} \right)_{-\sum_{2 \leq \ell_1 < \ell_2 \leq n-1} j_{\langle \ell_1|\ell_2\rangle}}$ in the coefficient~\eno{cn} remains unchanged since the sum over {\it all} $j$-parameters is invariant under this shift.
    
    \item Under the given shift, among other effects discussed below, the coefficient~\eno{cn} picks up an overall factor of \[ {j_{\langle r|a_u\rangle}! \over (j_{\langle r|a_u\rangle}-1)!} {j_{\langle s|a_u\rangle}! \over (j_{\langle s|a_u\rangle}+1)!} = {j_{\langle r|a_u\rangle} \over j_{\langle s|a_u\rangle}+1} = {-\Delta_{\langle r|a_u \rangle}  \over j_{\langle s|a_u\rangle}+1}\,,\]
    where we used $\Delta_{\langle r|a_u \rangle} = -j_{\langle r|a_u \rangle}$ for $r\geq 2$.
    Moreover note that the {\it scaled} coefficient~\eno{ctDef} comes with additional factors of dimensions $\Delta_{\langle r|n\rangle} \Delta_{\langle s|a_u\rangle}$. We will address the first factor shortly, but the other one can be read off of~\eno{nCombSeries} to be,
    \[ \Delta_{\langle s|a_u \rangle}\Big|_{\substack{j_{\langle r|a_u\rangle} \to j_{\langle r|a_u\rangle}-1\\ j_{\langle s|a_u\rangle} \to j_{\langle s|a_u\rangle}+1}} = -j_{\langle s|a_u \rangle}-1\,, \]
     where we have appropriately shifted it as required by the left hand side being evaluated. This will cancel the factor in the denominator of the previous displayed equation.

    \item Looking at the structure of the coefficient~\eno{cn}, the integer shifts may potentially affect certain factors for the choice of the dummy parameter $t+2=a_u$ in~\eno{cn}. However, once again the $\pm 1$ shifts cancel among each other leaving the following factor invariant
    \[  \left(\Delta_{a_u\delta_{a_u-2},\delta_{a_u-1}} \right)_{k_{a_u-2,a_u-1}+\sum_{\ell=2}^{a_u-1}j_{\langle \ell|a_u\rangle}}\Big|_{\substack{j_{\langle r|a_u\rangle} \to j_{\langle r|a_u\rangle}-1\\ j_{\langle s|a_u\rangle} \to j_{\langle s|a_u\rangle}+1}}  = \left(\Delta_{a_u\delta_{a_u-2},\delta_{a_u-1}} \right)_{k_{a_u-2,a_u-1}+\sum_{\ell=2}^{a_u-1}j_{\langle \ell|a_u\rangle}}\,.\]
    On the other hand, when $t+2=r$ or $s$, upon integer shifting as prescribed, one picks up extra overall terms. For $t+2=r$, the following factor, which has been rewritten more suggestively, transforms as,
    \eqn{NewFac}{
    {\Gamma(\Delta_{\langle r|n\rangle})  \over \Gamma(\Delta_{r \delta_{r-1},\delta_{r-2})}}\Bigg|_{\substack{j_{\langle r|a_u\rangle} \to j_{\langle r|a_u\rangle}-1\\ j_{\langle s|a_u\rangle} \to j_{\langle s|a_u\rangle}+1}} &= {\Gamma(\Delta_{r \delta_{r-1},\delta_{r-2}} + k_{r-1,r-2}+\sum_{\ell=r+1}^{n-1}j_{\langle r|\ell\rangle}-1) \over \Gamma(\Delta_{r \delta_{r-1},\delta_{r-2})}} \cr 
    &= {\Gamma(\Delta_{\langle r|n\rangle}-1)  \over \Gamma(\Delta_{r \delta_{r-1},\delta_{r-2})}}   
    =  {1\over \Delta_{\langle r|n\rangle}-1} {\Gamma(\Delta_{\langle r|n\rangle})  \over \Gamma(\Delta_{r \delta_{r-1},\delta_{r-2})}}\,.
    }
    Now recall the extra factor of $\Delta_{\langle r|n \rangle}$ mentioned above, which comes packaged in the scaled coefficient. Under the given shift,
    \[ \Delta_{\langle r|n \rangle}\Big|_{\substack{j_{\langle r|a_u\rangle} \to j_{\langle r|a_u\rangle}-1\\ j_{\langle s|a_u\rangle} \to j_{\langle s|a_u\rangle}+1}} = \Delta_{\langle r|n \rangle}-1\,, \]
    which precisely cancels the factor in the denominator of~\eno{NewFac}.
    Likewise for $t+2=s$, the following factor changes as follows:
    \eqn{NewFac2}{
    {\Gamma(\Delta_{\langle s|n\rangle})  \over \Gamma(\Delta_{s \delta_{s-1},\delta_{s-2})}}\Bigg|_{\substack{j_{\langle r|a_u\rangle} \to j_{\langle r|a_u\rangle}-1\\ j_{\langle s|a_u\rangle} \to j_{\langle s|a_u\rangle}+1}} &= {\Gamma(\Delta_{s \delta_{s-1},\delta_{s-2}} + k_{s-1,s-2}+\sum_{\ell=s+1}^{n-1}j_{\langle s|\ell\rangle}+1) \over \Gamma(\Delta_{s \delta_{s-1},\delta_{s-2})}} \cr 
    &= {\Gamma(\Delta_{\langle s|n\rangle}+1)  \over \Gamma(\Delta_{s \delta_{s-1},\delta_{s-2})}}   
    =  {\Delta_{\langle s|n\rangle}} {\Gamma(\Delta_{\langle s|n\rangle})  \over \Gamma(\Delta_{s \delta_{s-1},\delta_{s-2})}}\,.
    }
\end{itemize}

    Putting everything together, upon evaluting the left hand side of the first identity of~\eno{SymTran2}, we have  recovered  the original {\it unshifted} coefficient~\eno{cn} times an overall factor of $\Delta_{\langle r|a_u\rangle} \Delta_{\langle s|n\rangle}$. This is precisely the right hand side of the first identity of~\eno{SymTran2}.
 
 Returning to the case of $r=1$, the integer parameter $j_{\langle 1|\ell\rangle}$ is undefined, thus one of the shifts becomes vacuous. Consequently, in contrast with the analysis above for $r \geq 2$, the factor of $\left( \Delta_{1n,2\ldots(n-1)} \right)_{-\sum_{2 \leq r < s \leq n-1} j_{\langle r|s\rangle}}$ in the coefficient~\eno{cn} is no longer invariant, but transforms as
    \[ \left( \Delta_{1n,2\ldots(n-1)} \right)_{-\sum_{2 \leq \ell_1 < \ell_2 \leq n-1} j_{\langle \ell_1|\ell_2\rangle}}\Big|_{\substack{ j_{\langle s|a_u\rangle} \to j_{\langle s|a_u\rangle}+1}}  =  {\left( \Delta_{1n,2\ldots(n-1)} \right)_{-\sum_{2 \leq \ell_1 < \ell_2 \leq n-1} j_{\langle \ell_1|\ell_2\rangle}} \over \Delta_{\langle 1|n\rangle} -1}\,.\]
    The denominator above is cancelled by the factor of
    \[ \Delta_{\langle 1|n \rangle}\Big|_{\substack{j_{\langle s|a_u\rangle} \to j_{\langle s|a_u\rangle}+1}} = \Delta_{\langle 1|n \rangle}-1\,, \]
    which comes prepackaged with the scaled coefficient on the left hand side of the identity.
The argument for the cancellation of the factor of $\Delta_{\langle s|a_u \rangle} = -j_{\langle s|a_u\rangle} -1$ is unaffected (the overall minus sign gets cancelled due to the factor of $(-1)^{j_{\langle s|a_u\rangle}}$ in~\eno{cn}), and so is the argument for the appearance of an overall factor of $\Delta_{\langle s|n\rangle}$ since these do not depend on the integer shift rendered vacuous when $r=1$. Finally,  the following factor in the coefficient~\eno{cn} for $t+2 = a_u$, which previously was invariant, instead transforms as
\[  \left(\Delta_{a_u\delta_{a_u-2},\delta_{a_u-1}} \right)_{k_{a_u-2,a_u-1}+\sum_{\ell=2}^{a_u-1}j_{\langle \ell|a_u\rangle}}\Big|_{\substack{j_{\langle s|a_u\rangle} \to j_{\langle s|a_u\rangle}+1}}  = \left(\Delta_{a_u\delta_{a_u-2},\delta_{a_u-1}} \right)_{k_{a_u-2,a_u-1}+\sum_{\ell=2}^{a_u-1}j_{\langle \ell|a_u\rangle}} \Delta_{\langle 1|a_u\rangle}\,.\]
Combining all these observations, once again, the left hand side evaluates to the unshifted coefficient~\eno{cn} times a factor of $\Delta_{\langle 1|a_u\rangle} \Delta_{\langle s|n\rangle} $, which is precisely the expected right hand side for $r=1$. 

The other identities in~\eno{SymTran1}-\eno{SymTran2} can be proven via very similar analyses. In particular, the proofs for the third identity in both~\eno{SymTran1} and~\eno{SymTran2} also go through without further difficulty, since the $k_i$ integer-shifts are uniform across both the left and right hand sides.

\bibliographystyle{ssg}
\bibliography{main} 
\end{document}